\documentclass[11pt,showabstract,showacknowledgments]{iuphd}
\usepackage{cite}
\usepackage{threeparttable}
\usepackage{tikz}

\usetikzlibrary{arrows,shapes,positioning,shadows,trees}

\usepackage{moreverb}
\usepackage{algorithm} % needed for algorithm
\usepackage{algpseudocode}
\usepackage{amsfonts,amssymb} % needed for \mathcal
\usepackage{amsmath} % needed for \align*

\usepackage{adjustbox,lipsum}
\usepackage{graphicx}
\usepackage{caption}
\usepackage{subcaption}
\usepackage{enumerate}
\usepackage{amsmath}
\usepackage{tikz}
\usepackage{color}
\usepackage{multirow}
\usetikzlibrary{shapes.geometric, arrows}
\usepackage{caption}

\usepackage{amsthm}
\usepackage{url} % for url block
\usepackage{footmisc} % for footnotemark symbol

\theoremstyle{definition}

\title{Resource Sharing for Multi-Tenant NoSQL Data Store in Cloud}

\author{Jiaan Zeng}
\date{December 2015}

\committeechair{Beth Plale, Ph.D.}
\readertwo{Martin Swany, Ph.D.}
\readerthree{Judy Qiu, Ph.D.}
\readerfour{David Crandall, Ph.D.}
\readerfive{Atul Prakash, Ph.D.}
\defensedate{December 4th, 2015}

\cryear{2015}
\begin{document}

\maketitle
\acceptancepage
\copyrightpage
\begin{dedication}
\vspace{\fill}
	\begin{center}
		\begin{minipage}{.6\textwidth}
		I dedicate this dissertation to my family.
		\end{minipage}
	\end{center}
\vspace{\fill}
\end{dedication}

\begin{acknowledgments}
First and foremost, I would like to thank my advisor Prof. Beth Plale. She guided me into the research field of data management system and inspired my thinking; she gave me numerous opportunities to explore different directions and identify the research problems. Through working with her, I learned to be a professional researcher. I am thankful to my entire research committee: Prof. Martin Swany, Prof. Judy Qiu, Prof. David Crandall, and Prof. Atul Prakash. Their valuable feedback and professional guidance keep me in the right direction and push me to dive into the details without losing the big picture.

I was fortunate to work with great people in the technical team of the HathiTrust Research Center: Guangchen Ruan, Zong Peng, Milinda Pathirage, Samitha Harshani Liyanage, Miao Chen and Yiming Sun. They continuously work with me on technical problems, give me good advice on paper writing and presentations. In addition, I would like to thank Alexander Crowell and Prof. Atul Prakash at University of Michigan for the collaboration on the data capsule, and Jaimie Murdock and Prof. Colin Allen for applying the data capsule to a real world application. Furthermore, I am grateful to my other colleagues including Yuan Luo, Peng Chen, Quan Zhou, Yu Luo, and Isuru Eranga Suriarachchi at the Data To Insight center. Their support on my projects and research is very helpful. Next I would like to thank Jenny C. Olmes-Stevens, the project manager in the Data To Insight center, for helping me the edits of several papers, and Jodi Stern, the secretary in the Data To Insight center, for travel arrangement and meeting setup. I also appreciate the help from Rob Henderson. He helps me for various system administrative tasks to make some of the research outcome possible.

Finally, I am greatly thankful to my family and friends for their support and understanding, which make this entire process a precious and rewarding experience in my life.
\end{acknowledgments}

%\begin{preface}
%\end{preface}
\begin{abstract}
Multi-tenancy hosting of users in cloud NoSQL data stores is favored by cloud providers because it enables resource sharing at low operating cost. Multi-tenancy takes several forms depending on whether the back-end file system is a local file system (LFS) or a parallel file system (PFS), and on whether tenants are independent or share data across tenants. In this thesis I focus on and propose solutions to two cases: independent data-local file system, and shared data-parallel file system.

In the independent data-local file system case, resource contention occurs under certain conditions in Cassandra and HBase, two state-of-the-art NoSQL stores, causing performance degradation for one tenant by another. We investigate the interference and propose two approaches. The first provides a scheduling scheme that can approximate resource consumption, adapt to workload dynamics and work in a distributed fashion. The second introduces a workload-aware resource reservation approach to prevent interference. The approach relies on a performance model obtained offline and plans the reservation according to different workload resource demands.  Results show the approaches together can prevent interference and adapt to dynamic workloads under multi-tenancy.

In the shared data-parallel file system case, it has been shown that running a distributed NoSQL store over PFS for shared data across tenants is not cost effective. Overheads are introduced due to the unawareness of the NoSQL store of PFS. This dissertation targets the key-value store (KVS), a specific form of NoSQL stores, and proposes a lightweight KVS over a parallel file system to improve efficiency. The solution is built on an embedded KVS for high performance but uses novel data structures to support concurrent writes, giving capability that embedded KVSs are not designed for. Results show the proposed system outperforms Cassandra and Voldemort in several different workloads.
\end{abstract}

\tableofcontents
\listoffigures
\listoftables
\pagestyle{plain}
\startchapters

\chapter{Introduction} \label{title:introduction}
\section{Emerging Characteristics for Big Data Storage}
Science and business today are facing data sets that are growing dramatically in both complexity and volume. In its informal definition, big data consists of large, diverse, and structured or unstructured data. The sheer volume of data increase is predicted to grow exponentially by a factor of 300 from 130 exabytes in 2005 to 40,000 exabytes in 2020 \cite{BigDataSurvey}. Big data applications, whose data and request go beyond a single node's capacity, have begun to revolutionize the underlying storage system. In addition, high availability grows increasingly important as businesses rely on online data services. Furthermore, data are increasingly dynamically generated, coming from various sources with diverse formats and schemas. All these factors taken together suggest distributed storage and flexible data models are the future.

Experience has shown that traditional systems like the relational databases (RDBMS) struggle to handle big data applications as they are difficult and expensive to scale across multiple nodes and have I/O performance that does not meet application requirements \cite{NoSQLSurvey,NoSQLPerf}. In contrast, NoSQL data store (called Not Only SQL) has emerged as an alternate solution to big data storage. NoSQL store distributes and replicates data across multiple nodes. Requests are served by every node in a peer-to-peer fashion, not only increasing overall storage capacity but also the bandwidth. If a node goes down, other nodes can still serve requests for high availability and minimize data loss. Additionally, NoSQL data stores do not impose a rigid structured schema upon the data, thus providing greater flexibility to store the unstructured data. Because of these attracting features, NoSQL data stores have seen a great deal of uptake in both industry \cite{Voldemort,F-HBase,Dynamo} and academia \cite{HTRC,HyperDex}.

With the advent of cloud computing, cloud hosted NoSQL stores have grown in use. Users (called tenants) are willing to move their data infrastructures to the cloud \cite{cloud-future}. Following the spirit of infrastructure-as-a-service, tenants set up NoSQL stores across a set of virtual machines (VMs) that are rented from a cloud provider and billed by a flexible price model, i.e. ``pay-as-you-go'' model. Additionally, many cloud providers offer database-as-a-service to tenants. Such services ease most of the cluster management burdens from tenants. Typical examples include Amazon Dynamo DB \cite{Dynamo,AWSDynamoDB}, Google Cloud Datastore \cite{CloudStore} and Microsoft DocumentDB \cite{DocumentDB}.

For economic reasons though, a cloud hosted NoSQL store is usually used by multiple tenants simultaneously. For example, a database-as-a-service instance like Dynamo DB may be shared by different companies (i.e. tenants). Different departments (i.e. tenants) of an organization may be joint tenants of a single NoSQL store. Generally, there are several advantages of adopting multi-tenancy in various storage services \cite{Multitenancy-Model}. First, multi-tenancy makes management tasks easier. For example, to upgrade the storage service, instead of upgrading multiple service instances, a system admin can update the configurations or code base in a single instance and have the changes immediately available to all tenants. Second, multi-tenancy can yield better resource utilization as the storage service can support dynamic resource allocation which avoids provisioning each tenant with its maximum resources statically. Finally, data sharing is facilitated by multi-tenancy. For these reasons, support for multi-tenancy in cloud hosted NoSQL data store is an important problem.

\section{Multi-tenant NoSQL Data Store}
A NoSQL data store in a cloud environment can be viewed as having a 2-layer architecture: logical view layer and storage layer. The logical view layer presents to tenants a view of the store and a set of APIs for them to interact with the store. The storage layer represents the underlying infrastructure that physically stores and serves the data.

\subsection{Logical View Layer}
The logical view layer determines how tenants see the data and hides the complexities of underlying infrastructures from them. Each tenant has a dedicated view of the store with non-shared data sets. From a tenant's point of view, its workload is run against dedicated resources and is not aware of other tenants' existence. The cloud provider in reality is consolidating each tenant's data into as small an infrastructure as possible to maximize resource utilization. Tenants' data and requests end up co-locating with each other in the underlying infrastructure. Performance interference becomes a concern as isolation of current solutions are flawed.
\begin{figure}[!htbp]
    \centering
    \begin{subfigure}[b]{0.48\textwidth}
        \includegraphics[scale=0.65]{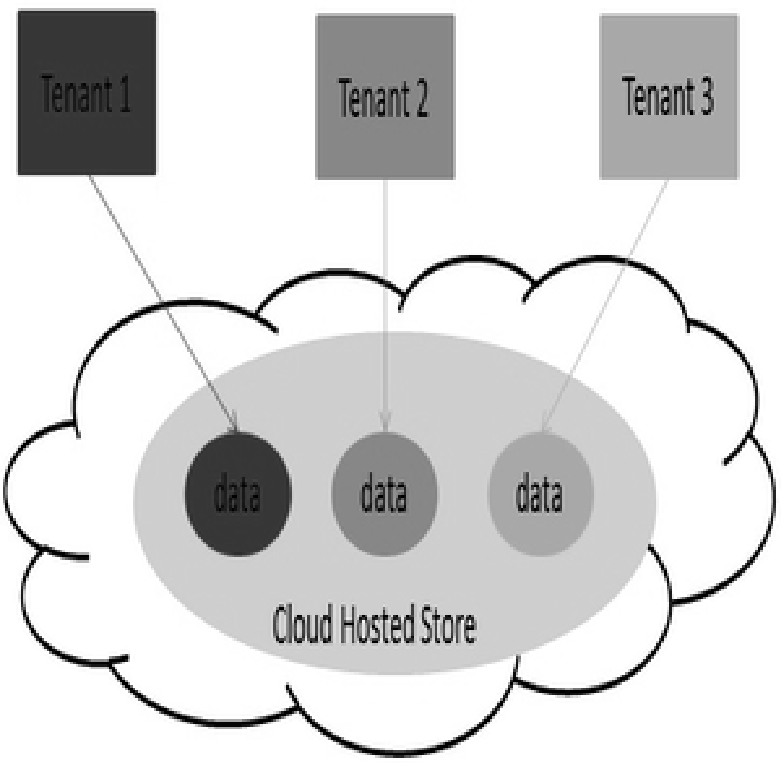}
        \caption{Non-shared data.}
        \label{fig:intro-independent-tenants}
    \end{subfigure}
    ~
    \begin{subfigure}[b]{0.48\textwidth}
        \includegraphics[scale=0.65]{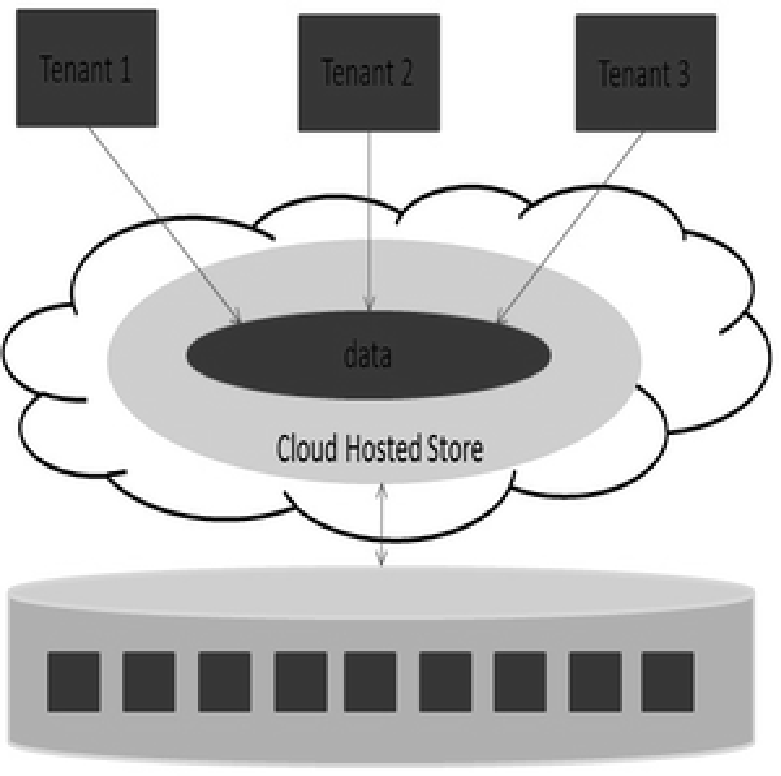}
        \caption{Shared data.}
        \label{fig:intro-cooperative-tenants}
    \end{subfigure}
    \caption{Multi-tenancy in the logical view layer.}
    \label{fig:intro-mult-scenarios}
\end{figure}

On the other hand, tenants may share the same data sets in the store. Tenants may be represented by different components in a pipeline that processes the data. For example, the HathiTrust Research Center (HTRC) \cite{HTRC}, which serves analytical access to nearly 14 million digitized volumes from the HathiTrust digital library, has an ingest component that loads data from remote \textit{rsync} points into Cassandra; a data API component used to read data from Cassandra and serve it to external users and a Solr component indexed data in Cassandra. All these components can be interpreted as tenants which share the same data set stored in Cassandra to form the data infrastructure of HTRC.

It is important to efficiently support the shared data model in a multi-tenancy cloud setting because of the attractiveness of its pay-as-you-go model and ability to bear elasticity.

Chapter \ref{title:background} further describes the models used in multi-tenancy by most data management systems.

\subsection{Storage Layer}
\begin{figure}[!htbp]
    \centering
    \begin{subfigure}[b]{0.48\textwidth}
        \includegraphics[scale=0.65]{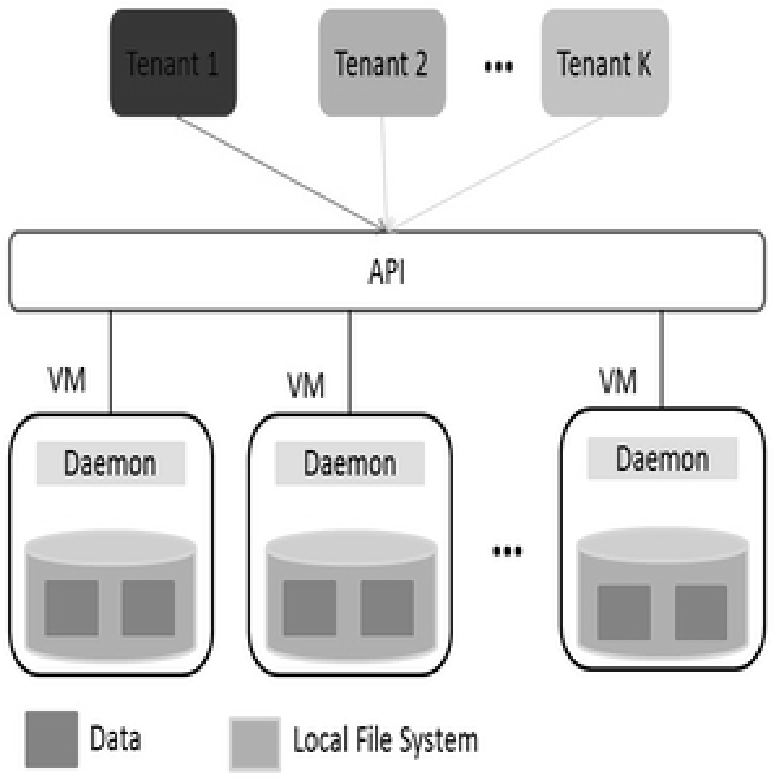}
        \caption{NoSQL over local file system.}
        \label{fig:intro-localfs-store}
    \end{subfigure}
    ~
    \begin{subfigure}[b]{0.48\textwidth}
        \includegraphics[scale=0.65]{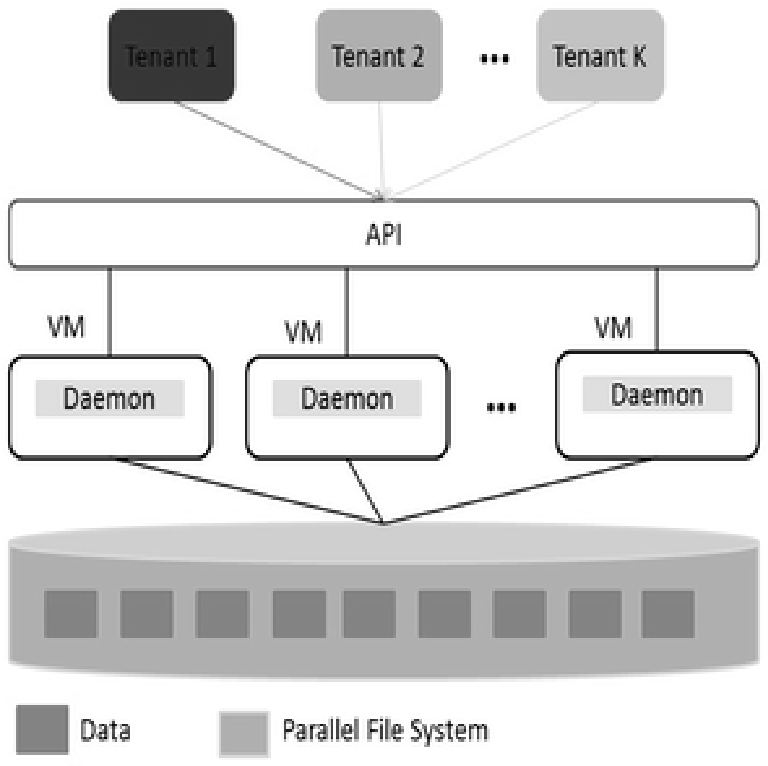}
        \caption{NoSQL over parallel file system.}
        \label{fig:intro-pfs-store}
    \end{subfigure}
    \caption{Different storage layers for NoSQL data store.}
    \label{fig:intro-fs-nosql}
\end{figure}

While the logical view layer serves as the front-end to tenants, the storage layer describes the mechanisms for storing data and handling requests. Figure \ref{fig:intro-localfs-store} shows a NoSQL data store deployed across multiple nodes. Data is stored to the local file system on each node. Because a local file system on a single node may suffer from disk failure and is not scalable, the NoSQL store usually replicates data in a few nodes in the cluster. A persistent daemon service runs on each node in the cluster and communicates with other daemon services to provide a unified view over the local file systems. Various protocols among daemons exist to coordinate daemons' behaviors to support tasks like data replication, failover and so on.

While the local file system has prevailed in the cloud platform for quite some time, the parallel file system (PFS) has begun seeing usage in the cloud in both industry \cite{Lustre-AWS} and academia \cite{Lustre-OpenStack, Lustre-S3}. Originating from the high performance computing (HPC) platform, PFS is a type of clustered file system that partitions data across a dedicated storage node cluster \cite{PFS-Wiki}. PFS provides good scalability, high bandwidth access, and failover, all of which are missing in a local file system solution. PFS serves concurrent access from a number of clients and operates over high-speed networks. It can be mounted to multiple nodes and allows files to be accessed using the same interfaces and semantics as local file systems. Behind the scene, PFS transparently hides the complexities of accessing across different storage nodes, data replication, and fail over from end users. Figure \ref{fig:intro-pfs-store} depicts the scenario of a NoSQL store which is set up over PFS. The daemon service per node gives the illusion that data is stored in local file system while in reality data is transferred to/from PFS transparently. PFS can take over the responsibility of reliable data storage which is important in the NoSQL storage system.

\subsection{Layer Mapping}
Either of the logical views in Figure \ref{fig:intro-mult-scenarios} can be mapped to one of the physical implementations in Figure \ref{fig:intro-fs-nosql}. Usually the NoSQL store relies on the local file system to store data. Thus this dissertation studies the performance interference in the multi-tenancy case using a local file system (non-shared data-LFS). The unawareness of the NoSQL store to the features PFS offers makes it inefficient to run a NoSQL store over PFS today. This dissertation explores the viability of using PFS to support data sharing across tenants in NoSQL data store (shared data-PFS). We leave the investigation of performance interference in NoSQL over PFS for future work.

\section{Research Problems} \label{title:intro-description}
The cloud hosted NoSQL data store has seen a great deal of usage because of its scalability and high availability. Because cloud environments encourage sharing, resource sharing and data reuse across tenants is a growing use case. Multi-tenancy in such a shared environment imposes significant performance challenges on the use of NoSQL store, however. The following are the problems addressed in this dissertation.
\begin{itemize}
  \item Performance interference prevention across tenants in the non-shared data-LFS case
  \item Cost Effective multi-tenant access in the shared data-PFS case
\end{itemize}

\subsection{Performance Interference Prevention}
Tenant data and requests may be consolidated in a single multi-tenant NoSQL store. Due to the co-location, there will be performance interference among tenants caused by resource contention. Interference prevention can be realized from the client perspective when a service level agreement (SLA) is enforced, or can be realized at the server side when fair share is enforced across tenants. This dissertation focuses on providing a non-client centric solution to enforce fair share on the server side. A misbehaved tenant may consume a well-behaved tenant's resources by its workloads, thus degrading the latter's performance. This interference behavior in the multi-tenancy setting is undesirable. The resource reservation approach in Chapter \ref{title:reservation} can potentially be used to enforce SLA for each tenant. We leave the SLA enforcement for future work.

Real world scenarios provide evidence of performance interference. Although Amazon DynamoDB imposes a throughput limit to tenants to prevent the store deing dominated by a few tenants \cite{AWSDynamoDB}, it does not guarantee throughput provision, nor provides fair share among tenants \cite{Pisces,dynamodb-faq}. People from BloomReach Inc. report back-end workloads have a negative impact on front-end workloads' performance in the same Cassandra cluster \cite{bloomreach}. They also report that back-end workloads from different teams (i.e. tenants) may interfere with each other as well \cite{bloomreach}. It is straightforward to address the interference issue by running different tenants' workloads on separate infrastructures, which is the approach BloomReach has taken \cite{bloomreach}. Dedicated hardware provides strong isolation but lowers resource utilization and thus increases cost. Effort is occurring in open source NoSQL projects e.g. Cassandra \cite{Cassandra-URL} and HBase \cite{HBase}, the base of the prototyped systems in this dissertation, to support multi-tenancy \cite{HBase-Quota,Cassandra-Multitenancy}. However, both are still a work-in-progress. Current effort simply schedules requests for tenants in a round robin fashion in Cassandra (the tenant-oriented request scheduling is planned but not implemented yet in HBase) and does not attempt to identify the resource demands of tenants. There is no known performance modeling in the multi-tenant setting. In addition, many of the multi-tenancy features only became available recently and some of them are not merged to the trunk \cite{HBase-RegionServerGrouping} as of the writing of this dissertation.

\subsection{Cost Effective Multi-tenant Access}
Running a distributed NoSQL store over PFS may introduce overheads owning to the un-awareness of the former to the PFS. For example, data may be unnecessarily replicated; extra network trips may be needed to access the PFS because the daemon service delegates all the requests to the back-end file system; and additional overhead may also arise from the data replication and failover protocols, which are unnecessary in the presence of PFS. In addition, the store provider may hold the resources (e.g. VM) even if no request comes, which is not cost effective. That is because if the VMs are shutdown or repurposed, the data stored in the VM's is not accessible anymore. Recently, Greenberg et al. point out the burden and inefficiency of running persistent daemon services for a key-value store in the HPC environment \cite{MDHIM}. We envision PFS will be widely used in the cloud and thus there will be a pressing need to accommodate the features it provides to NoSQL data store.

\section{Contributions}
This dissertation proposes several approaches to address the multi-tenancy issues discussed in Section \ref{title:intro-description} for the shared and non-shared settings in Figure \ref{fig:intro-mult-scenarios}. They are discussed here:

\paragraph{Non-shared data-LFS setting}
I investigate the performance interference in the setting of local file system for storage and no data sharing. An experimental study carried out on Cassandra, a state-of-the-art NoSQL store, shows that a tenant will experience unpredictable performance in terms of throughput when multiple tenants access the store independently. Chapter \ref{title:fairshare} proposes a throughput regulation framework that targets system-wide fairness. Specifically, we adapt and extend the deficit round robin algorithm \cite{fairqueuing} with linear programming as the scheduler to regulate throughput. The solution adaptively changes the scheduling parameters to achieve system-wide fairness. It also protects the throughput of reads in face of scans by splitting a scan operation into small pieces and scheduling them along with reads.

Throughput regulation can be viewed as a form of resource reservation because throughput represents the underlying resource consumption. It assumes every byte delivered to or from the store consumes the same amount of resources. Such an assumption is also used by \cite{Pisces,Cake}. But this assumption does not always hold, especially for workloads having different access patterns and demanding different resources. A workload with a hotspot access pattern may require more cache than a workload with a random access pattern. An equal reservation of cache and disk usage for all tenants may not yield the best result or even fail to provide performance isolation. Chapter \ref{title:reservation} models the impact of various resource demands and proposes a resource reservation framework in HBase to enforce the performance isolation among tenants. The reservation is also elastic in the sense that if a tenant does not use up its reservation, the system is able to reallocate its redundant reservation to tenants in need. Chapter \ref{title:reservation} experimentally evaluates different fair sharing algorithms and tries to quantify the trade-offs between fairness and efficiency. It also quantifies the overhead introduced by the isolation mechanisms as too much overhead can undermine the benefit of fairness.

\paragraph{Shared data-PFS setting}
I explore the feasibility of using PFS for NoSQL data store in a shared data setting. Chapter \ref{title:kvlight} proposes a lightweight key-value store (KVS), a special form of NoSQL data store, that makes use of PFS as the back-end storage and does not require daemon services running in front. Such a feature allows the KVS to be accessed on demand and avoids holding resources when no tenant makes requests. A VM hosting a tenant can be revoked or repurposed once the tenant has completed sending requests. In addition, the responsibility of data reliability is shifted to PFS. The KVS only cares about data organization and serving request which is more lightweight than a traditional KVS. Internally, the proposed KVS is built on an embedded KVS, i.e. Berkeley DB \cite{BDB}, to enable direct file system access for high performance and support the no persistent daemon service feature. Embedded KVS solutions, including Berkeley DB, do not support concurrent writes due to file system locking. The proposed KVS follows the spirit of log structure merge tree \cite{LSM} to organize the data in PFS to support concurrent writes in a distributed environment. To remedy the read deterioration caused by writes, it uses a novel tree-based structure and parallel compaction to efficiently support concurrent reads.

\section{Dissertation Outline}
The remainder of this dissertation is as follows. Chapter \ref{title:background} presents the background information about multi-tenancy model and NoSQL data store. Chapter \ref{title:relatedwork} summaries related work. Chapter \ref{title:fairshare} investigates the performance interference and proposes a throughput regulation mechanism targeting system-wide fairness. Chapter \ref{title:reservation} further studies the performance interference based on the results from Chapter \ref{title:fairshare}, and describes a workload-aware resource reservation mechanism for performance isolation. Chapter \ref{title:kvlight} focuses on building a lightweight key-value store, which is called KVLight, over a parallel file system with novel data structures. Chapter \ref{title:conclusion} concludes with future work.

\chapter{Background} \label{title:background}
Supporting multi-tenancy is important in a storage service like a NoSQL data store. We first present the background of multi-tenancy including its definition and different models. A cloud hosted NoSQL data store usually runs as middleware between user applications and file systems. It consists of a set of service processes running over a raw file system and provides richer data management features than a regular file system. There are many NoSQL systems nowadays with different APIs, data models, and architectures. Thus we present an overview of NoSQL data store from several perspectives including architecture, data distribution, data replication, and resource management, with a focus on three popular NoSQL systems, i.e. Cassandra, HBase, and Berkeley DB \cite{BDB, BDBJE}, which our prototyped systems are based on.

\section{Multi-tenancy Model} \label{title:bg-models}
Multi-tenancy, in its most basic definition, refers to an architecture in which a single software instance serves multiple users, customers, or tenants \cite{Multitenancy}. Multi-tenancy can be found in several different places: network multiplexing \cite{Multitenant-Network}, virtual machine management \cite{Multitenant-VM}, file system sharing \cite{Argon}, job runtime framework \cite{Multitenant-Job}, data management system \cite{Multitenancy-Model} and so on. This dissertation focuses on multi-tenancy in the data management system, particularly NoSQL data store.

\begin{figure}[h]
  \centering
  \includegraphics[scale=0.8]{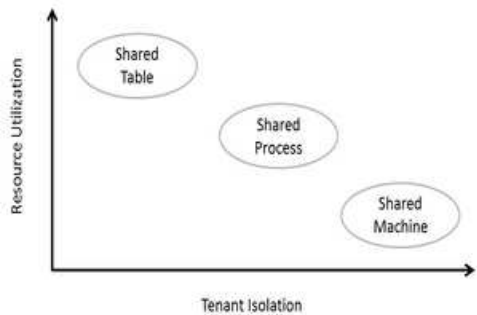}
  \caption{Different multi-tenancy models on storage services.}
  \label{fig:multitenancy-model}
\end{figure}

Generally, in a data management system, the multi-tenancy can be supported in three different models: shared machine, shared table, and shared process \cite{Multitenancy-Model}. Each has different advantages and disadvantages, shown in Figure \ref{fig:multitenancy-model}. In the shared machine model, virtualization technologies like virtual machines are used to host different tenants in a shared machine. It provides the strongest isolation among tenants as tenants are separated by different VMs. On the contrary, the shared table model, which allows different tenants to share the same table, provides the weakest isolation. Data from different tenants are mixed together in the same place, as are the requests.

From the system utilization perspective, the shared machine model sacrifices the most because each tenant is allocated with a VM which holds up some portion of dedicated system resources. The shared table model, however, achieves the highest utilization by aggregating tenant data into a single location and serving without dedicated resources.

The shared process model stays in the middle of aforementioned two models from both the isolation and resource utilization perspective. In the shared process model, each tenant has its own table and enjoys a dedicated view of the storage. Under the hood, the storage service is shared by multiple tenants. Tenants have to share the memory, CPU, I/O, network bandwidth and all other resources allocated to the particular storage service process. Compared with the shared machine model and the shared table model, the shared process model trades a little bit of tenant isolation for better performance and scale. On one hand, it solely relies on throttling and reservation to isolate resource usage across tenants within the same process, which may not be as strong as the shared machine model. On the other hand, it improves resource utilization in the sense that resources may be reused by different tenants and reallocated among tenants. For instance, if a tenant does not use up the memory given, the management system may reallocate the memory to other tenants in need. The shared process model is also superior to the shared table model in terms of flexibility because the shared table model has to provide a unified schema to store data from all different tenants with different formats, which is very difficult and inflexible.

Most research work about interference prevention target the shared process model in regard to provide isolation across tenants because of the advantages mentioned above \cite{Das,A-Cache,Cake,Argon,Argus,Zeng}. Similarly, in this dissertation, we focus on the shared process model for isolation among independent tenants. For tenant consolidation over a shared data set, we use the shared table model as tenants expect to see the same table structure and data in this case.

\section{Overview of NoSQL Data Store}
\begin{figure}[h]
  \centering
  \includegraphics[scale=0.65]{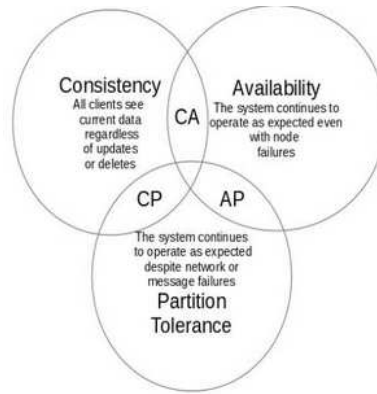}
  \caption[Characteristics of NoSQL store.]{Characteristics of NoSQL store. Source: \url{http://blog.nosqltips.com/2011/04/cap-diagram-for-distribution.html}}
  \label{fig:cap-theorem}
\end{figure}

To deal with the rapid growth of data, NoSQL data stores are designed as modern web-scale databases in mind and the characteristics of schema-free data model, easy replication support, distributed access, simple APIs and eventual consistency \cite{NoSQL-DB-URL}. According to the CAP-theorem \cite{CAP}, conflicts arise among different aspects of a distributed system in terms of three factors: consistency, availability and partition-tolerance. The relationship between factors is shown in Figure \ref{fig:cap-theorem}.

The CAP-theorem postulates that only two of the three factors can be achieved at the same time. For the traditional SQL database, which stresses the ACID properties (Atomic, Consistency, Isolation, and Durability) \cite{CAP}, partition-tolerance or availability is usually sacrificed to honor the consistency. However, for NoSQL data store, because it is designed to be distributed in nature, availability and partition tolerance become critical. Thus most of NoSQL data stores trade in consistency with availability and partition-tolerance. This results in the BASE properties (Basically Available, Soft-state, Eventually consistent) \cite{CAP}.

Next, we investigate NoSQL stores in terms of four aspects: architecture, data distribution, data replication and consistency, and resource management.

\subsection{Architecture}
\begin{figure}[h]
  \centering
  \includegraphics[scale=0.8]{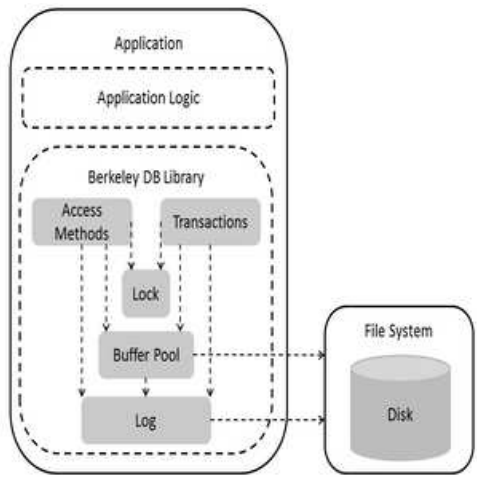}
  \caption{The architecture of Berkeley DB, a single node NoSQL data store.}
  \label{fig:bdb-architecture}
\end{figure}

Generally speaking, a NoSQL store can fall in one of the two categories: single node oriented and multi-node oriented. Single node oriented NoSQL store is designed to work in a single node environment and usually implemented as a library embedded into the application. Unlike multi-node NoSQL stores, which have persistent running daemon services delegate the access to file system through network, single node NoSQL store allows applications to access the file system directly and read or write data without going through a network. In addition, no persistent daemon service is required. Thus it is lighter-weight and performs better than multi-node NoSQL store. Examples include Berkeley DB (BDB) \cite{BDB} and Level DB \cite{leveldb}. Figure \ref{fig:bdb-architecture} displays the architecture in BDB. BDB embeds as a library to an application. It provides a set of APIs for applications to access the data store and carry out transactions. Lock, buffer pool and log are the three main components in BDB. They all run in an application process space. BDB interacts with the file system to store data through standard file system APIs.

However, single node NoSQL store suffers from data loss and is not scalable because all its data is stored in the file system local to the node. Besides, it only allows exclusive writes due to file system locking. Applications in different processes have to take turns to write to the store. In contrast, multi-node NoSQL store (or distributed NoSQL) is designed to work across multiple nodes or even multiple data centers. It partitions and distributes data across nodes with replications, and provides not only reliable data storage but also scalable access as requests are distributed in the cluster. Examples of multi-node NoSQL include HBase, Cassandra, and CouchDB \cite{CouchDB}.

\begin{figure}[!htbp]
   \centering
   \begin{subfigure}[b]{0.8\textwidth}
        \centering
        \includegraphics[scale=0.9]{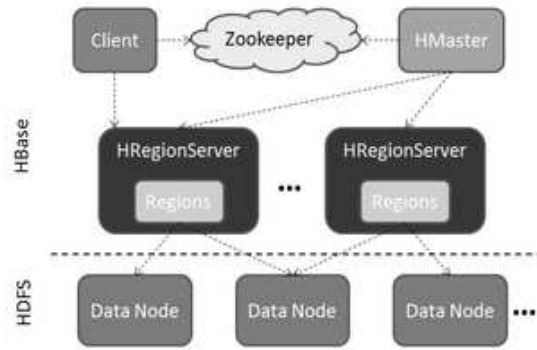}
        \caption{The master-slave architecture of HBase. The \textit{HMaster} monitors and coordinate different \textit{HRegionServer}s, which responde to client for read/write requests directly.}
        \label{fig:hbase-architecture}
    \end{subfigure}
    ~
    \begin{subfigure}[b]{0.8\textwidth}
        \centering
        \includegraphics[scale=0.9]{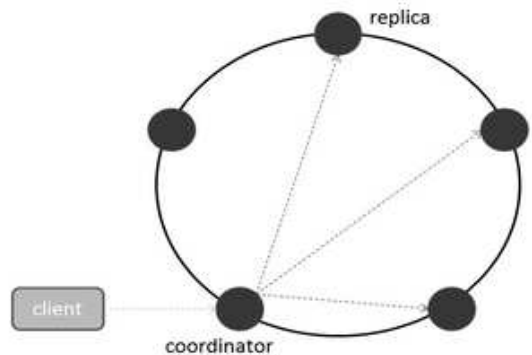}
        \caption{The peer-to-peer architecture of Cassandra. There is no central control point int the system. A client can         connect to any node in the cluster. The connected node serves as a coordinator that forwards the request to the nodes hosting the replicas.}
        \label{fig:cassandra-architecture}
    \end{subfigure}
    \caption{Two main architectures for distributed NoSQL stores.}
    \label{fig:nosql-architecture}
\end{figure}

Most of the distributed NoSQL stores draw heavily from either the master-slave architecture or the peer-to-peer architecture, shown in Figure \ref{fig:nosql-architecture}. The master-slave architecture is used in Google BigTable \cite{BigTable}, and its open source implementations \emph{e.g.} HBase, HyperTable \cite{HyperTable}. The master is responsible for bookkeeping metadata, request routing and coordinating among slaves while the slave is responsible for carrying out the actual workload and responding to clients directly. The peer-to-peer architecture used by Amazon Dynamo \cite{Dynamo} treats each node as an equal peer. There is no central control point in such architecture and thus it can avoid single point of failure. Nodes usually communicate and propagate messages through a gossip protocol. However, due to the peer-to-peer characteristic, it is usually hard to coordinate activities among nodes. Cassandra is an open source implementation of Dynamo. Voldemort and Riak \cite{Riak} are also heavily influenced by Dynamo's design.

In practice, distributed NoSQL store is preferred over a single node NoSQL because of the aforementioned advantages. The main use of single node NoSQL is to serve as a building block for a larger system. For example, Berkeley DB is used as the back-end store in each individual node in Voldemort \cite{Voldemort} and Riak \cite{Riak}. Berkeley DB also sees usage in many other systems \cite{BDB-Usage}.

\subsection{Data Distribution}
\begin{figure}[h]
  \centering
  \includegraphics[scale=0.8]{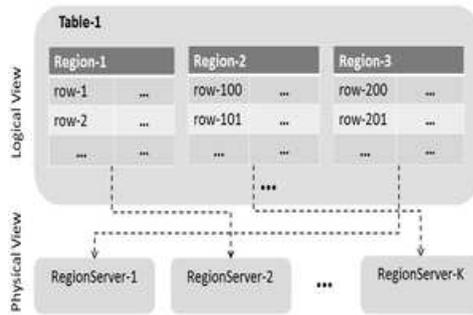}
  \caption{Key-range based data distribution in HBase.}
  \label{fig:key-range-distribution}
\end{figure}

The data distribution scheme describes how data is organized in the underlying file system and distributed across nodes. It determines the way data access works. Different distribution schemes may be suited to different data access patterns. Generally, there are two main distribution approaches for most of the systems: key-range based distribution and hash based distribution. The key-range based distribution has the entire data set sorted according to the order of the key and divides the data set into non-overlapped partitions. A partition represents a range of keys between a minimum key value and a maximum key value. Keys falling within that range go to the corresponding partition. Because keys are sorted, scan queries can be answered very easily and quickly. HBase uses the key-range based distribution to distribute data. Figure \ref{fig:key-range-distribution} shows a distribution scenario. In HBase, data is stored in a table as rows and sorted by the key \emph{i.e.} the row key. As the data grows, the table will be split into several pieces called \textit{Region}s. HBase sysadmin can specify what the splitting policy may be for the store. After the splitting, each region consists of a bunch of sorted rows and is the basic unit of data distribution. Regions are distributed across \textit{RegionServer}s running in each node in the cluster. The HMaster holds a request routing table which describes the mapping between a key to a region and its server. The client interacts with the HMaster first to figure out which RegionServers to connect for its key requests. Then the client communicates to the RegionServers directly for data access. Although HBase can support efficient scan operations, it may suffer from load imbalance situation as some key ranges may be accessed more frequently than other key ranges.

\begin{figure}[h]
  \centering
  \includegraphics[scale=0.8]{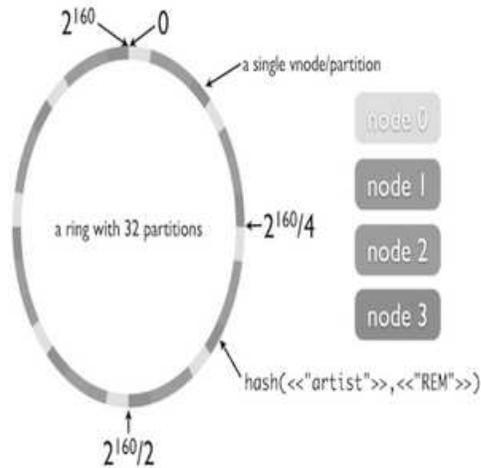}
  \caption[The principal of consistent hashing.]{The principal of consistent hashing. It is used widely in the peer-to-peer architecture \emph{e.g.} Dynamo, Cassandra, Riak, and so on. Source: \url{http://www.paperplanes.de/2011/12/9/the-magic-of-consistent-hashing.html}}
  \label{fig:consistent-hashing}
\end{figure}

The hash based distribution uses a hash function to randomly hash keys to different servers. The randomness may avoid the load imbalance occurring in key-range distribution. In practice, consistent hashing is usually preferred because it makes adding and removing machines in the system easy \cite{ConsistentHashing}. Specifically, the output values of the hash function form a ring. The same hash function is applied to all data and mapped to positions in the ring. Each individual node in the cluster maps itself to a position in the ring as well and has knowledge about other nodes' positions. A node is responsible for storing all the keys that fall in the range between this node and its predecessor node. A client can send requests to any of the nodes. A node receiving a request acts as a coordinator that either reads or writes data to its local storage, or forwards the request to another proper node based on the output value of the hash function. To deal with various node capacities and data skew, a node may be mapped to multiple positions in the ring to take more data if it has a larger capacity or further randomize the mapping in the case of data skew. Figure \ref{fig:consistent-hashing} shows an example of 4 nodes mapped into multiple positions in the ring. Each position is responsible for a range of keys. Although hash based distribution can avoid load imbalance, it makes serving scan operation very difficult as data is not sorted in the backend.

Cassandra by default uses consistent hashing to partition the data. It uses a MD5 hashing algorithm to hash the row key to a big integer. The output of MD5 is guaranteed to follow balanced distribution even if the input keys do not show an even distribution. It follows the steps described above to handle requests. Voldemort and Riak also use consistent hashing to distribute data and follow similar request handling steps.

Finally, Berkeley DB supports both the key-range based (in the form of a B-Tree \cite{BDB}) and the hash based data organization. But it does not distribute data across nodes as it is a single node oriented NoSQL data store. It simply stores data into different files.

\subsection{Data Replication and Consistency}
Distributed NoSQL data store usually replicates data to provide highly reliable and concurrent data access. Different replication policies may be applied and lead to different consistency levels. Notice that Berkeley DB is not concerned with data replication and consistency as it only stores data locally.

HBase's replication policy and consistency inherit from the Hadoop Distributed File System (HDFS) which is the underlying storage for HBase. HDFS replicates a block, its basic data unit, into two other nodes in the cluster by default. If one of the three nodes is down, the other replicas can still serve the requests. HDFS maintains strong consistency for the data access. A write to HDFS returns only when all the replicas are written successfully. Thus any reads after the write can get the latest value. To update a record, the new value will be appended to the write-ahead log file stored in HDFS and saved in the \textit{MemStore} (an in-memory structure). Reads which follow can query the \textit{MemStore} to get the latest value.

In contrast to HBase's rigid and singular model of replication and consistency, Cassandra, the representative of a peer-to-peer system, has a flexible scheme. Similar to HBase, it replicates a piece of data to two other nodes. Cassandra allows applications to pick replication policies such as ``Rack Unaware'' which randomly places the replicas in the cluster, ``Rack Aware'' which tries to place replicas in different racks in the same data center and ``Datacenter Aware'' which intends to place replicas in different data centers for greater reliability. Cassandra by default follows eventual consistency but allows different consistency levels to exist in the system. It uses the quorum to manage the replicas. The application can specify the condition of a successful operation as the number of replicas responds. For example, a quorum with value ``ONE'' can have a write return when only one replica is written successfully and other replicas are still being written. A quorum with value ``ALL'' requires a read to return only when all the replicas respond.

\subsection{Resource Management}
NoSQL data store usually has multiple resources involved to serve requests. For example, CPU for serialization and de-serialization, memory for caching and buffering, disk for reading and writing, network for transferring, and etc. Configuration about how resources are allocated plays an important role in performance in some cases.

Single node NoSQL functions as an embedded library and manages the memory as well as disk usage all by itself. Berkeley DB manages the memory used for reads caching and writes buffering. It also controls the way to generate the data files on disk, e.g. how big a file is, when to flush a file, and etc.

Compared with the monotonic resource management of single node NoSQL, distributed NoSQL store usually separates the management into layers. In HBase, HDFS stores data and manages the disk resource, while the HBase daemon service controls the CPU, cache and network resources. In Cassandra, the \textit{RPC} layer controls the request scheduling while the \textit{StorageProxy} layer takes care of the actual reading and writing.

\chapter{Related Work} \label{title:relatedwork}
Multi-tenancy hosting of users in cloud NoSQL data stores is favored by cloud providers because it enables resource sharing at lower operating cost. As discussed in Chapter \ref{title:background}, the models of multi-tenancy can be realized in three different abstractions: shared machine, shared table and shared process \cite{Multitenancy-Model}. Xiong et al. \cite{Xiong} allow tenants to set up database instances within VMs that share a single host. The usage of VM by tenants uses the shared machine model where tenants share the same set of hosts. Salesforce.com \cite{Salesforce}, provided as software-as-a-service (SaaS), uses the shared table model where different tenants share the same set of database tables. Amazon EC2 \cite{EC2} provides infrastructure-as-a-service (IaaS) by allowing tenants to create virtual machines (VM) in shared hosts. Amazon DynamoDB service \cite{AWSDynamoDB} and Relational Cloud \cite{RelationalCloud} expose themselves as platform-as-a-service (PaaS), and follow the shared process model where tenants share the same data store process. Next, we discuss the multi-tenancy in the shared service case and the shared data case.
%In general, the shared machine model provides the strongest isolation but introduces significant overheads. The shared table model is the most efficient but hard to be applied to data with different formats. The shared process model stays in the middle of the two above -- it avoids many of the overheads in the shared machine model but provides a much stronger isolation than the shared data model has.

\section{Storage Service Sharing}
The shared process model is usually preferred to serve tenants with non-shared data \cite{Multitenancy-Model}, because it provides reasonable isolation without imposing too much overhead as discussed in Section \ref{title:bg-models}. Therefore, a majority of work \cite{Das,A-Cache,SQLVM,Cake,Argon,Pisces} about storage service sharing targets the shared process model. This section first summarizes multi-tenancy support in different storage services classified as file system, relational database and NoSQL store. Then it surveys the literature of resource scheduling approaches used for performance isolation.

\subsection{Storage Services}
\paragraph{File system}
Wachs et al. use a time-quanta-based disk scheduling approach with cache space partitioning for performance insulation among applications running on a single file server \cite{Argon}. We adopt the idea of partitioning cache and disk for tenants but coordinate the partitioning over these two resources through a constraint optimization model rather than treating the resource independently. In addition, we focus on a distributed store which is more complicated than a single node server. Due to the complexity, the underlying storage system is sometimes treated as a black box as responses coming out from it vary in several aspects e.g. latency, size, etc. Gulati et al. use a feedback-based approach on a black-box storage system to dynamically adjust the number of IOs issued to the storage system with observed latency as feedback \cite{PARDA}. We employ the idea of feedback-based scheduling, but model all the tenant behaviors in a unified constraint optimization problem and globally adjust the scheduling parameters instead of having separate adjustments per tenant.
%Gulati et al. tune the scheduling parameters including the request batch size and the length of outstanding request queue based on the scheduler to trade efficiency versus fairness \cite{IOShare}.

\paragraph{Relational Database}
Narasayya et al. propose an abstract of resource reservation called \textit{SQLVM} on resources such as CPU, I/O and memory for tenant performance isolation and focus on I/O scheduling \cite{SQLVM}. Das et al. present a CPU scheduling approach in SQLVM to reserve CPU usage for CPU interference prevention \cite{Das}. While our approach uses resource reservation as well, we target multiple resources as a whole instead of treating them individually, because various resources may not be entirely independent. With regard to multiple resources, Soundararajan et al. propose a multi-resource allocator to dynamically partition the database's cache and its storage bandwidth so as to minimize request latency for all the tenants \cite{Multi-Resource}. However, different from \cite{Multi-Resource}, we intend to provide fairness across tenants. Additionally, we attempt the partitioning in a distributed NoSQL store with hierarchical architecture instead of a monolithic RDBMS. Walraven et al. utilize a central scheduler to dispatch requests to different back-end RDBMS in a multi-tier web application \cite{Walraven}. Our work relies on each node to enforce the resource reservation instead of a central scheduler.

\paragraph{NoSQL Data Store}
Pisces \cite{Pisces} uses partition placement, replica selection, and fair queuing to provide multi-tenant fair share in terms of throughput in Membase, a memory-based NoSQL store with hash partitioning. Like Pisces, our work also targets system-wide fair share. In addition, we also adapt the deficit round robin algorithm \cite{DRR} for scheduling. Unlike Pisces, our target storage abstraction has disk, memory and network resources involved, and is much more complicated than the memory-based store Pisces uses. Also unlike Pisces, our approach distinguishes different resource demands from different workloads. Besides, our approach does not assume a static load distribution. It can dynamically adjust tenant weights to achieve system-wide fairness. A-Cache \cite{A-Cache} divides the block cache space in HBase and limits a tenant's cache activities within the cache space it is assigned to resolve the cache interference among tenants. Our experiments show that the cache partition by itself does not resolve interference in some cases.
%In contrast, we target disk-based NoSQL stores with consistent hashing. Our performance isolation mechanisms run over a storage abstraction which is much more complicated than the memory back-end where Pisces runs, because a request may go through the cache, network stack, and file system, and have other nodes involved as well. As a result, we use a feedback-based scheduling approach, which is similar to \cite{PARDA}, to schedule when requesting and adjust the scheduling parameters when responding in a asynchronous way, while Pisces runs these two steps synchronously.

The open source community has made efforts on supporting performance isolation for multi-tenancy in HBase and Cassandra \cite{HBase-Quota,Cassandra-Multitenancy}, although in a limited way. In HBase, to prevent a tenant from taking over the entire store, quota management is used \cite{HBase-Quota}. It enforces the maximum number of tables and the maximum number of regions a namespace i.e. a tenant can create in the store. Furthermore, to prevent the interference caused by data and request co-location, HBase uses the concept of region server grouping \cite{HBase-RegionServerGrouping}. It instructs the load balancer to assign regions such that a region server only serves a particular set of regions from a particular set of tenants. Despite the significant amount of efforts on supporting multi-tenancy in HBase, it is still a work-in-progress. Many of the multi-tenancy features only became available recently and some of them are not merged to the trunk yet \cite{HBase-RegionServerGrouping}. In addition, the current multi-tenancy support does not consider resource utilization of workloads and uses a static way to dispatch data and requests \emph{e.g.} the region server grouping approach. As our experiments show in chapter \ref{title:reservation}, ignoring resource demands from workloads may lead to low utilization or even failure of isolation. Our work in HBase can shed some light of the future development of multi-tenancy support.

Compared with HBase, the support of multi-tenancy in Cassandra is lessened \cite{Cassandra-Multitenancy}. Cassandra uses a simple weighted round robin algorithm to schedule requests from different tenants. Such a scheduler fails to prevent the domination because it keeps looking for pending requests. An ill-behaved tenant can use a large number of threads to send requests and take up most of the scheduling chances easily. Thus multi-tenancy support in Cassandra is an on-going effort -- there are still many open tickets \cite{Cassandra-Multitenancy}. Our work in chapter \ref{title:fairshare} extends Cassandra in terms of multi-tenancy support in a scheduling approach with a feedback control loop and some adaptive control mechanisms.

\subsection{Resource Scheduling}
Resource scheduling can regulate the resource usage to avoid domination i.e. some tenants dominate the use of resources. These kinds of resources include CPU, cache, disk, network bandwidth, etc. We use resource scheduling in this dissertation to enforce fair share. We adapt the deficit round robin algorithm \cite{DRR} because of its simplicity and effectiveness. However, we present related work of resource scheduling as some of it could strengthen the results in future work. We first present the general scheduling algorithms classified as the virtual-time-based approach and the quanta-based approach. Then we discuss the usage of scheduling algorithms in two typical scenarios: reservation and proportional share. Most work on resource scheduling focus on either enforcing resource reservation \cite{mclock,A-Cache,Das}, or providing proportional share \cite{Pisces,Argon,PARDA,XenSchedulers}.

\paragraph{Scheduling Algorithms}
Generalized processor sharing (GPS) is an idealized scheduler and achieves perfect fairness with the assumption that tenants' traffic is fluid \cite{GPS}. However, in real world scenarios, resource schedulers can only approximate the behavior of GPS due to the discretionary nature of computer \cite{GPS}. There are two categories of approximation: virtual-time-based approximation and quanta-based approximation. The virtual-time-based approximation estimates a request's start time and finish time as virtual time, and uses them as scheduling criterions. Fair queuing scheduler (FQ) assumes request time is linear to the size of data delivered \cite{FQ}. Weighted fair queuing scheduler (WFQ) extends FQ by considering weights in the estimation of finish time \cite{GPS}. Both FQ and WFQ pick a task with the smallest finish time as the next task to run. \cite{Network} presents a comparison of different fair queuing algorithms. The quanta based approximation does round robin scheduling to schedule tasks according to the resource quanta allocated. Weighted round robin (WRR) scheduler allocates quanta based on tenant weights. It works for fixed size tasks but struggles with variable size tasks because it requires an estimation of mean task size \cite{WRR}. Deficit round robin (DRR) is a variation of WRR in the sense that it approximates GPS without knowing the mean size of tasks \cite{DRR}. In each scheduling round, DRR schedules tasks according to a tenant's quanta. Remaining quanta will be accumulated to the next scheduling round. Due to its simplicity and low time complexity as shown in \cite{Pisces, Das}, we use DRR in our resource scheduling. We also empirically compare DRR with WFQ, a virtual-time-based scheduler, in Chapter \ref{title:reservation}.
%On the contrary, start-time fair queuing scheduler (SFQ) schedules tasks in order of their start time not finish time which can handle tasks with variable \cite{SFQ}.

\paragraph{Reservation}
mClock uses reservation and limitation to mitigate I/O interference across VMs running on the same hypervisor \cite{mclock}. Its virtual-time-based scheduling approach statically allocates I/O resources, which may cause the storage capacity to be under utilized. In contrast, our reservation is elastic and adaptively changes according to workloads. A-Cache reserves the block cache space for tenants to protect hotspot oriented workload \cite{A-Cache}. Our experiments show that our framework proposed is able to resolve interference in some cases where A-Cache fails because it only considers block cache. Das et al. calculate the deficit of CPU reservation and propose a variant of deficit round robin algorithm (DRR) \cite{DRR} for elastic CPU reservation \cite{Das}. However, the elastic approach simply boosts the reservation to a fixed percentage and is not flexible enough for workloads with dynamic resource demands. Our elastic approach adjusts the reservation proportionally according to the actual throughput a tenant yields, and is able to reallocate resources when a tenant's workload changes its resource demands. SQLVM reserves IOPS (IO operations/second) for each tenant \cite{SQLVM}. It employs the virtual-time-based scheduling approach in \cite{mclock} and translates the IOPS to a deadline for each tenant to guide the scheduling. Narasayya et al. study a page replacement algorithm for fair sharing the buffer pool memory in a RDBMS \cite{multitenant-memory}. The deadline oriented approach and memory sharing technique can potentially be used in our framework to enforce service-level objectives specified by tenants.

\paragraph{Proportional Share}
Pisces \cite{Pisces} adapts the deficit round robin algorithm \cite{DRR} for throughput regulation and intends to achieve Dominant Resource Fairness (DRF) \cite{DRF}. We argue that the physical resources reflected from bytes read and bytes written are not independent and violate a fundamental assumption of DRF. Unlike Pisces, our approach asynchronously updates tenant credit account per request instead of synchronously because the disk-based NoSQL store may have a longer delay than a memory-based store Pisces uses. Wachs et al. use a time-quanta-based scheduling for fair sharing the disk \cite{Argon}. A tenant withholds the disk until the given time expires even though the tenant may not fully use the disk at that time frame, which leads to low disk utilization. Gulati et al. use the FAST-TCP algorithm to detect congestion and provide fair share on a black box storage system \cite{PARDA}. Our framework in HBase could adapt this approach in the HDFS scheduling level. Our schedulers use credits to represent the chances of scheduling. The virtualization hypervisor Xen \cite{XenSchedulers} also uses the notion of credit to schedule VCPU, a virtual CPU mapped to a physical core. However, unlike our schedulers where credits are only an approximation of the requests' resource consumption, the CPU usage of a VCPU is represented by the credits in the Xen scheduler directly. Additionally, a VCPU can yield its usage of the host CPU due to I/O blocking. A request in DRR cannot yield itself once it is scheduled.

\section{Data Sharing}
This dissertation targets the key-value store (KVS), a specific form of NoSQL data store, in the shared data case. KVLight, the proposed system, uses the shared table model for multi-tenancy, which is the most efficient way of sharing data with the same format among tenants as discussed above.

\subsection{File System and Key-Value Store}
There have been efforts on integrating parallel file system (PFS) in the cloud \cite{Lustre-AWS,Lustre-OpenStack,Lustre-S3,IUJetStream}. We focus on building a storage layer over parallel file system to better utilize its features. Yin et al. compare the performance between a parallel file system and a KVS in terms of throughput \cite{pfs-kvs}. Ren et al. propose a file system that utilizes an embedded KVS to manage the file system metadata \cite{tablefs}. KVLight is a KVS that is built on top of a PFS and realizes its data reliability and high scalability features.

\subsection{Key-Value Store As A Library}
Single node KVS (S-KVS) usually embeds itself as a library to the application for high performance access in a single node environment. Berkeley DB uses B-Tree or Hash to organize the key-value pairs \cite{BDB}, while LevelDB stores data in files in a logical tree level \cite{leveldb}. RocksDB \cite{RocksDB} extends LevelDB in terms of multi-cores utilization, multi-threaded compaction, and so on. Systems like NVMKV \cite{NVMKV}, FlushStore \cite{flushstore} focus on building S-KVS over flash storage with suitable data structures. KVLight extends the usability of S-KVS for concurrent writes in a distributed environment where traditional S-KVS does not support.

A distributed KVS over multiple nodes (M-KVS) is usually adopted in real world scenarios to handle concurrent access. Several M-KVSs use S-KVS as the building block because it is lightweight and has high performance. Both Dynamo \cite{Dynamo} and Voldemort \cite{Voldemort} use Berkeley DB as their default back-end storage. Riak builds its own S-KVS called Bitcask \cite{bitcask} but allows LevelDB as one of its back-end options \cite{Riak}. Most M-KVS require persistent running servers and assume the underlying file system is not reliable. KVLight is designed to support concurrent access workloads but can be accessed on-demand without persistently running servers. MDHIM \cite{MDHIM} is a recently developed KVS that also provides on-demand access without running persistent servers in a distributed environment through MPI and a S-KVS i.e. LevelDB. However, unlike MDHIM, KVLight does not require applications to run a MPI cluster with fixed number of processes due to the static data partition.

The idea of running a storage system as a library or ``serverless'' system is also employed in file systems. PLFS is a library file system which optimizes an application's data layout for the underlying file system \cite{plfs}. DeltaFS embeds the file system metadata server as a library in an application to remove the centralized metadata server bottleneck \cite{deltafs}. They are both orthogonal and complementary to KVLight.

\subsection{Compaction Management}
Most KVS achieve high write throughput through log-based write. The proposed KVS in chapter \ref{title:kvlight} also follows the same spirit. Updates are appended to log files rather than applied in-place. As a result, a read has to consult several files generated by writes to get the data. A background procedure called compaction runs periodically to remedy the read deterioration caused by writes \cite{HBase-Book,Cassandra-Book,CompactionManagement}. BigTable's compaction merges a fixed number of data files whose sizes are the smallest into one single file \cite{BigTable}. Cassandra and HBase also adopt such an approach. This can reduce the number of files a read request needs to consult. Although the number of data files decreases, a read request may still have to linearly scan several data files because the key range of each file overlaps with others. LevelDB uses a level compaction to compact files level by level \cite{LevelDBImpl}. It divides the key space into disjoint key ranges in separate files. Files in the same level will have disjoint key ranges. In a compaction, a file in level-L and files in level-(L+1) with overlapped keys are merged into a new file in level-(L+1). Therefore, a read request can be answered without linear scanning multiple files. The disjoint partitioning is not directly applicable in our system because we need to accommodate concurrent writes from different processes. Instead, we propose a tree-based compaction strategy to approximate the level compaction. Additionally, we also allow several compactions to run in parallel to speed up the entire remedy process. This is similar to the multi-threaded compaction in RocksDB \cite{RocksDB} which only applies in S-KVS. Ahmad et al. propose a compaction management framework that offloads the compaction on a dedicated server to lower the impact on actual workloads and uses a cache pre-fetching scheme to avoid the performance penalty from offloading \cite{CompactionManagement}. We adopt the offloading idea and leave the cache usage in future work.

\chapter{Multi-Tenant Fair Share in NoSQL Stores} \label{title:fairshare}
As discussed in Section \ref{title:introduction}, often for economic reasons, a NoSQL data store will be shared by multiple tenants. The independent tenants may be from a single organization or from different organizations. Tenant workloads operate on disjoint data sets and a tenant should not see the impact of the workloads of other users.  In a cloud environment, tenants often want to treat the entire storage system as a black box that can scale on demand, while in reality their data sets are usually co-located and cause resource contentions \cite{Pisces}. Thus a critical goal of multi-tenancy is fair sharing across tenants. In a fair-shared system, fairness is achieved in the sense that a tenant gets her share of the system no matter what other tenants do, i.e., a well-behaved tenant should not experience any impact from misbehaved tenants. Fair share is also the foundation of providing different service level agreements to different tenants. Especially in a cloud environment, some tenants are willing to pay more for a larger share of system resources.

Fair share can be realized at two different levels: 1) at the infrastructure level where fairness is guaranteed by directly scheduling physical resources (disk, CPU, network, etc.) or 2) at the application level where fairness is guaranteed by application level scheduling \emph{e.g.}, Hadoop fair scheduler \cite{FairScheduler} schedules based on task slots and Moab scheduler \cite{Moab} schedules based on compute nodes. For traditional storage systems, fair share is usually provided at the infrastructure level and involves a single physical resource on a single machine \cite{Multi-Resource}, \emph{e.g.}, disk bandwidth, CPU, network bandwidth, etc. Ensuring fair sharing in a distributed NoSQL data store is more challenging because NoSQL data stores engage multiple types of resources: cache memory, CPU, disk.  It is difficult to maintain a scheduler and queues for each type of physical resource because these schedulers need to cooperate to serve requests. In addition,  the use of a coordinator node in Dynamo style NoSQL stores, \emph{e.g.}, Cassandra, involves multiple nodes to handle one request. The coordinator node takes requests and forwards the requests to the nodes where the data is located. Such features require the fair scheduling algorithms to coordinate among nodes. Therefore, our proposed scheduling approach for ensuring fair share in a multi-tenancy NoSQL store is done at the application level.

In this chapter, we propose a novel approach to provide fair share across multiple tenants for NoSQL data stores, especially for Cassandra \cite{Cassandra-URL,Cassandra}. The approach is designed to work in a distributed manner -- cooperation among nodes is taken into account to provide system-wide fairness (global fairness) instead of single node fairness (local fairness). Furthermore, the system provides fair share for read operations by preventing the head-of-line blocking impact of scan operations. In summary, this chapter makes the following contributions:
\begin{itemize}
  \item A framework that employs a feedback control loop to monitor and schedule requests;
  \item A scheduler based on the adaption and extension of deficit round robin algorithm \cite{fairqueuing} with linear programming;
  \item Adaptive control approaches to provide global fairness instead of local fairness and protect reads from scans;
  \item Experimental results show effectiveness of our system.
\end{itemize}

\section{Fairness Experiments in Cassandra}
\begin{figure}[htbp]
   \begin{subfigure}[b]{0.22\textwidth}
        \centering
        \includegraphics[scale=0.20]{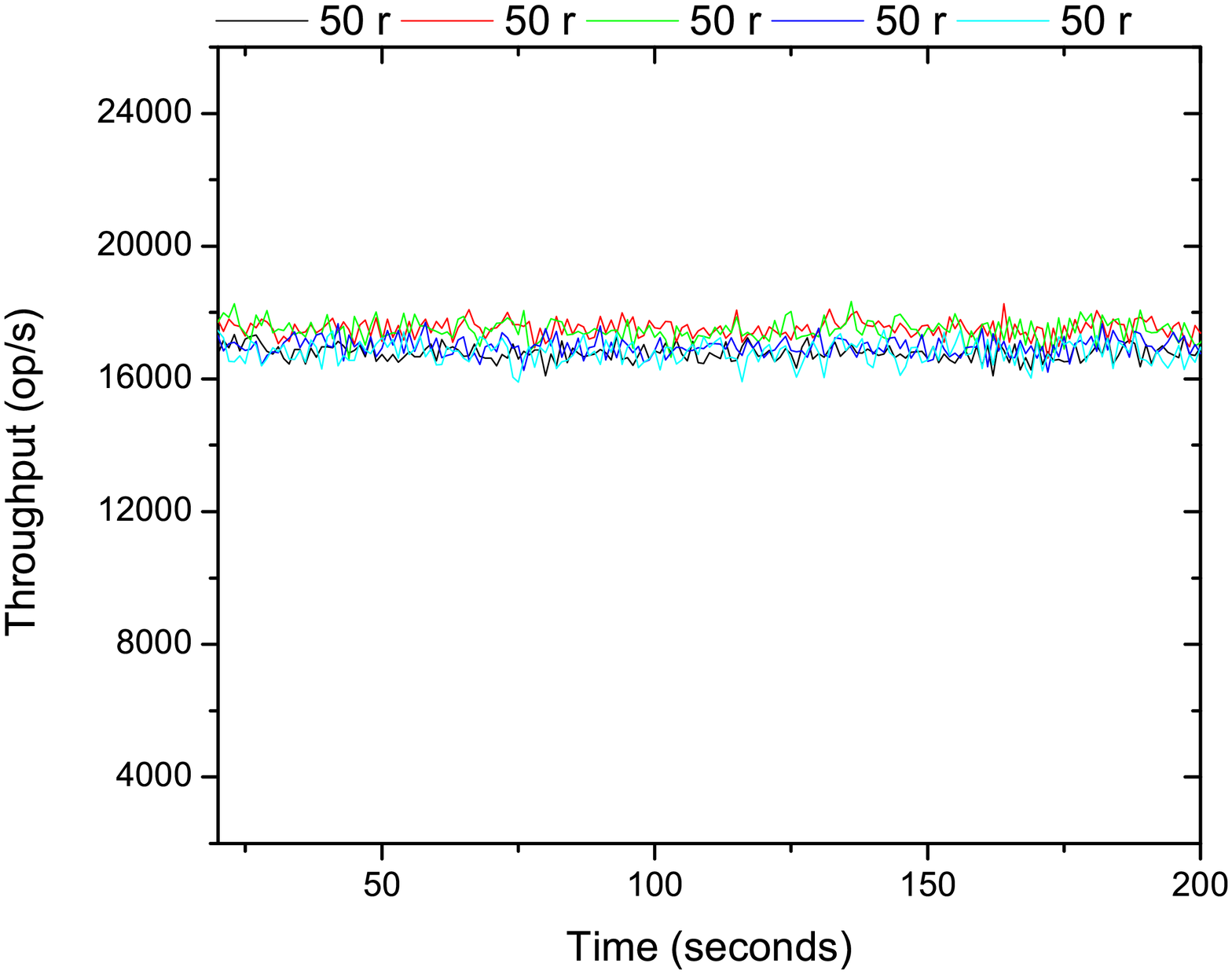}
        \caption{}
        \label{subfig:fairness-motivation:samerate}
    \end{subfigure}
    ~~~~~~~~~~
    \begin{subfigure}[b]{0.22\textwidth}
        \centering
        \includegraphics[scale=0.20]{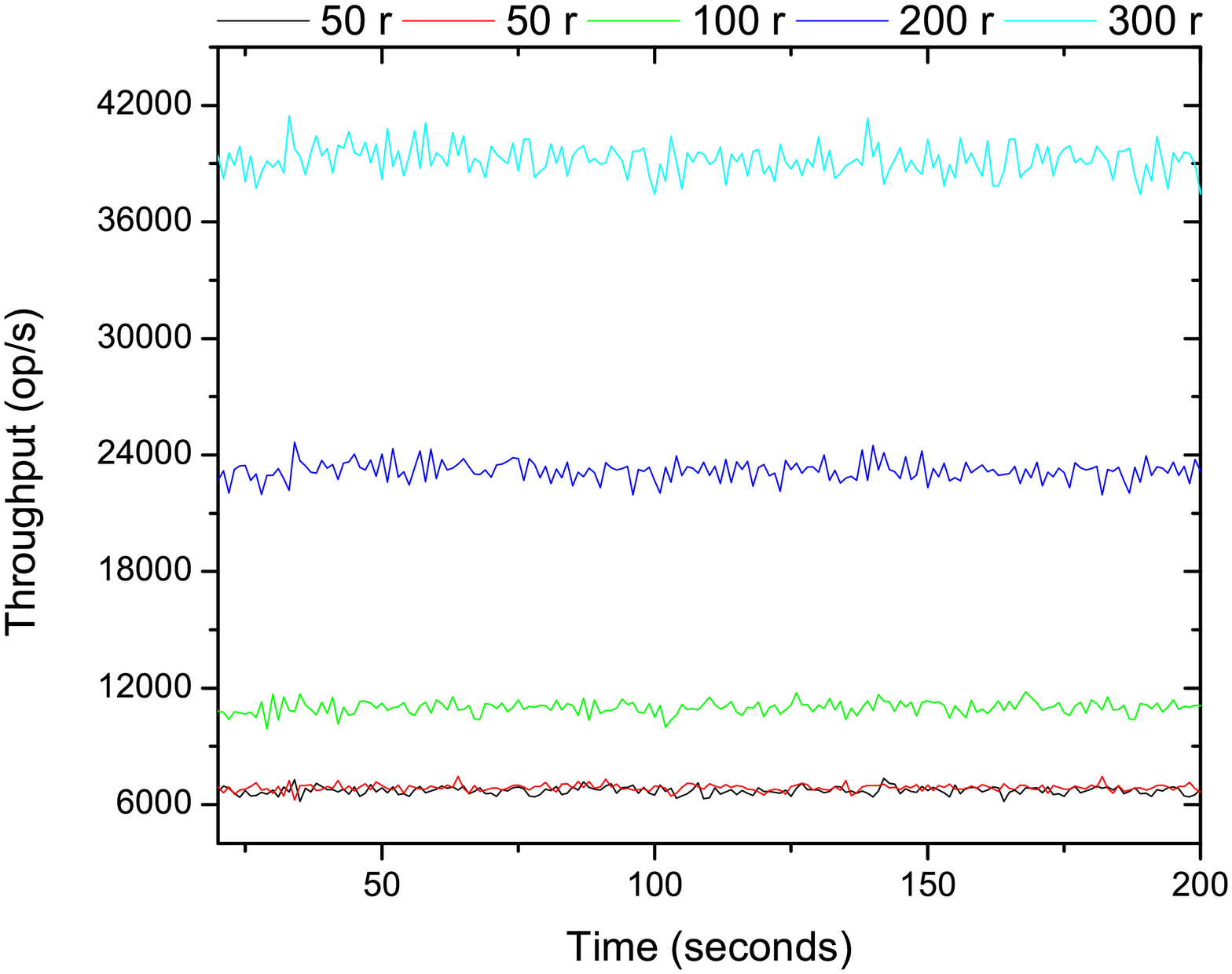}
        \caption{}
        \label{subfig:fairness-motivation:diffrate}
    \end{subfigure}
    ~~~~~~~~~~
    \begin{subfigure}[b]{0.22\textwidth}
        \centering
        \includegraphics[scale=0.20]{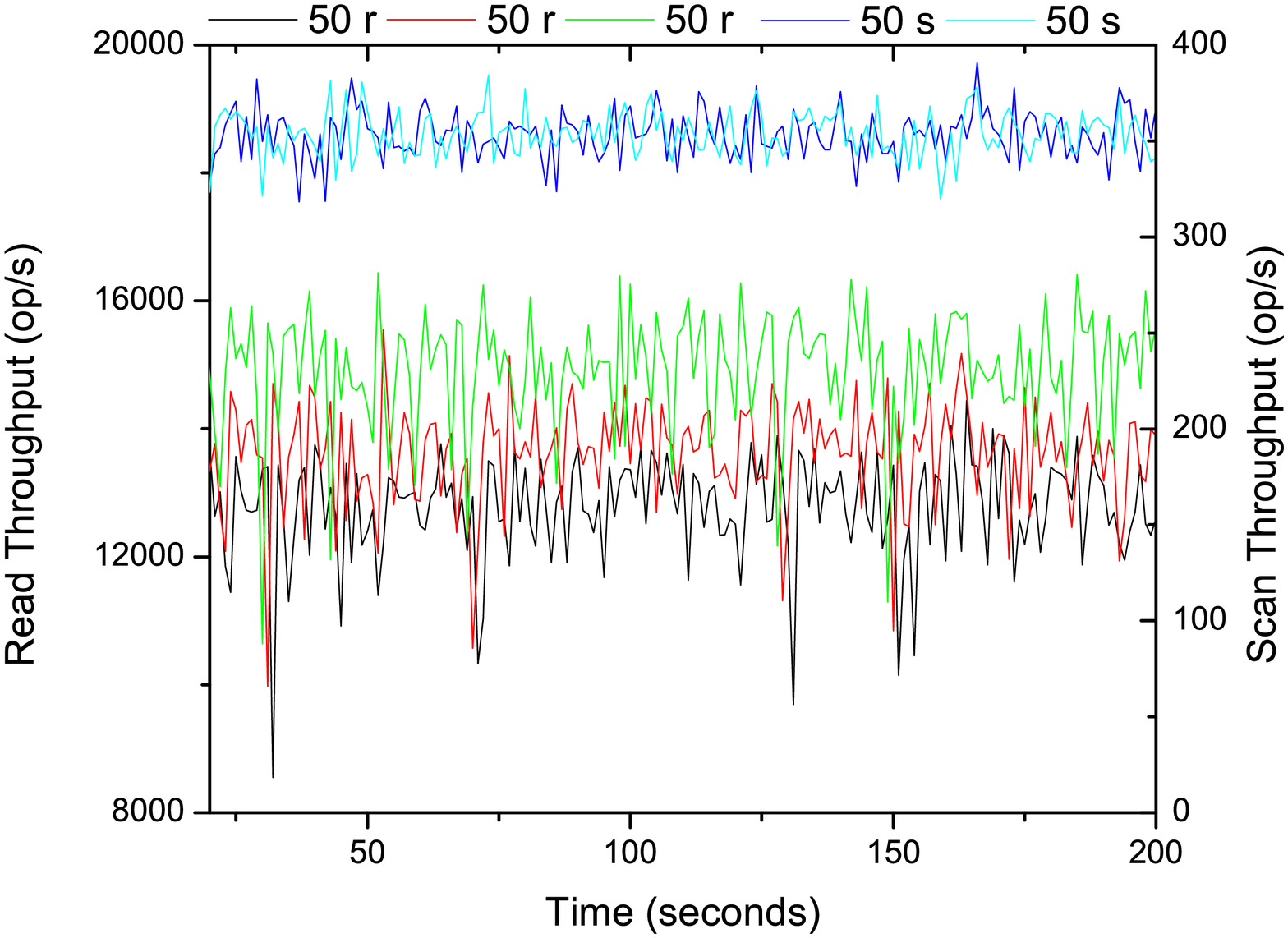}
        \caption{}
        \label{subfig:fairness-motivation:difftype}
    \end{subfigure}
    \caption[Throughputs of different workloads in a shared Cassandra.]{Throughputs of different workloads in a shared Cassandra. Each tenant's throughput is represented by a line of a different color. The legend shows the number of threads a tenant uses and what requests it is sending, e.g.,``50 r" means the tenant uses 50 threads to send read requests, ``50 s" means the tenant uses 50 threads to send scan requests. (a) Tenants using the same number of threads can fair share. (b) Unfairness occurs when tenants use different \# threads. (c) Unfairness occurs when read operations coexist with scan operations. }
    \label{fig:fairness-motivation}
\end{figure}
We motivate the need of fair share support in Cassandra as follows. We use Yahoo Cloud Storage Benchmark (YCSB) \cite{YCSB} to generate the workloads and simulate multiple tenant access. We set up a 9-node Cassandra cluster and allocate 5 additional nodes to run the YCSB benchmark clients. Each tenant owns 1,000,000 rows, each row size is 1.2 KB, the row is pre-loaded by YCSB in a table that is independent from other tenants. A YCSB client connects to multiple nodes in a Cassandra cluster through multiple TCP connections. Each TCP connection is managed by one thread which can be observed from the server side as well because Cassandra launches one thread per connection to handle requests. Like \cite{Pisces,Walraven,MeT}, we measure \textit{operation throughput} i.e. operations per second (op/s) for each tenant to represent a tenant's share on the system. For more details of the experiment setup, please see section \ref{subtitle:fairshare-evaluation}. Figure \ref{fig:fairness-motivation} plots the throughput as a function of time.

In the first experiment, Figure \ref{subfig:fairness-motivation:samerate} demonstrates that all tenants see roughly the same throughput at 17,000 op/s which we interpret to mean that tenants each receive a fair share of Cassandra under this scenario. In the second experiment, Figure \ref{subfig:fairness-motivation:diffrate} shows tenants with 50 threads see throughput of about 6,000 op/s while the tenant with 300 threads sees throughput that is 7 times higher (about 40,000 op/s). Compared to the first experiment, throughput for the tenants with 50 threads drops 60\%. Our supposition is that the large drop in performance to the 50-thread tenant is because Cassandra's resource allocation is done based on the size of the workload that the client generates. This test clearly shows that in a shared Cassandra cluster, a tenant's throughput can be influenced by other tenants' demands, i.e., the number of threads. Figure \ref{subfig:fairness-motivation:difftype} delineates the result of the third experiment. The read-only tenants' throughput oscillates dramatically and is worse than the one in Figure \ref{subfig:fairness-motivation:diffrate}. Even with the same number of threads, the read-only tenants do not get similar throughput to each other. We attribute the difference to the impact of the scan operation because it makes the system suffer from head-of-line blocking. In summary, the fair share among tenants of Cassandra depends on tenant's workload. Both the number of threads a tenant uses and types of requests a tenant sends can lead to unfairness in Cassandra when multiple tenants were present.

In the first experiment, Figure \ref{subfig:fairness-motivation:samerate}, all tenants run a read-only workload that reads one row per request. Tenants are configured at 50 threads each, which saturates the throughput of our Cassandra cluster.  We can tell from Figure \ref{subfig:fairness-motivation:samerate} that all tenants see roughly the same throughput at 17,000 op/s which we interpret to mean that tenants each receive a fair share of Cassandra under this scenario. In the second experiment, we configure a unique number of threads for each tenant: 50, 50, 100, 200, and 300 threads for the five tenants.  We rerun the read-only workload to see if tenants again see similar throughput. Figure \ref{subfig:fairness-motivation:diffrate} shows that tenants with 50 threads see throughput of about 6,000 op/s while the tenant with 300 threads sees throughput that is 7 times higher (about 40,000 op/s). Compared to the first experiment, throughput for the tenants with 50 threads drops 60\%.  Our hunch is that the large drop in performance is because Cassandra's resource allocation is done based on the size of the workload that the client generates.  A 300-thread client generates requests faster than a 50 thread client; Cassandra sees this and adjusts resources.   This test clearly shows that in a shared Cassandra cluster, a tenant's throughput is influenced by other tenants' demands. Tenants with more threads deprive tenants with fewer threads' share of the cluster and thus lead to unfairness from the perspective of the smaller tenant. In the third experiment, we configure 50 threads per tenant. Three of the tenants run the same read-only workload as in the first experiment (read-only tenant), while the other two tenants run a scan-only workload that scans 200 rows per request (scan-only tenant). Figure \ref{subfig:fairness-motivation:difftype} delineates the result. The read-only tenants' throughputs oscillate dramatically and are worse than the ones in Figure \ref{subfig:fairness-motivation:diffrate}.

In summary, the fair share among tenants of Cassandra depends on tenant workloads. We demonstrate that both number of threads and request type can lead to unfairness in Cassandra when multiple tenants were present. Our fairness control approach targets these two factors to provide fair share for multi-tenancy.

\section{Request Scheduling}
\begin{figure}[h]
  \centering
  \includegraphics[scale=0.8]{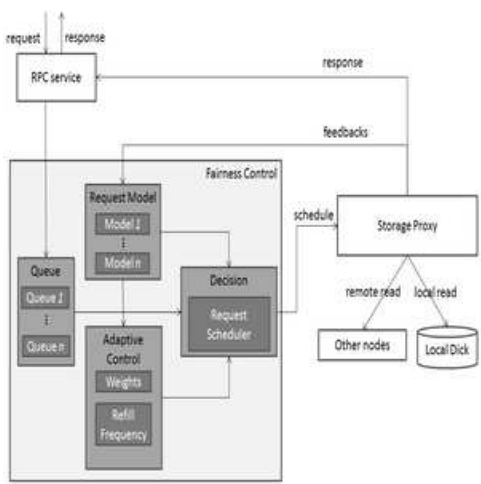}
  \caption{Architecture.}
  \label{fig:fairness-architecture}
\end{figure}

Two architectural components of Cassandra are involved in request handling: an \textit{RPC} service that exchanges information across the network, and a \textit{StorageProxy} service for reading and writing data from all replicas, maintaining certain consistency levels, detecting and handling failures.  This is shown in Figure \ref{fig:fairness-architecture}. Our proposed approach adds a third component that resides between RPC and StorageProxy and treats the latter as a black box. Such a design decouples our scheduler from the underlying storage technology, making it easier to adapt to different situations e.g., using solid state drive instead of hard disk as the storage media.

Our proposed fairness control scheduler, working at the application level, uses feedback collected from responses returned from the StorageProxy to guide scheduling.  As such, it does not track the number of physical resources a request consumes, a solution that is more modular.  The fairness control scheduler is designed with four major pieces: 1) Queues that hold tenants' requests; 2) Request Scheduler that schedules requests to the StorageProxy, 3) Request Models that collect different metrics from a response as feedbacks to support scheduling; 4) Adaptive Controller that adaptively changes some of the scheduling parameters e.g., weights. Each tenant has its own queue and request model. When a request arrives at a coordinator node, the RPC service puts it into the corresponding tenant's queue. The scheduler schedules a request from queues to the StorageProxy for processing. When a response is returned, the fairness control scheduler collects metrics, e.g., number of bytes read, number of requests in last second, etc., as feedback from responses and serves them as input to the Request Model.

The intuition behind request scheduling for fair share is this:  \textit{if a tenant has consumed more resources in the past, its upcoming requests will get fewer opportunities to be scheduled}. This draws from the Moab scheduler \cite{Moab} where future jobs have fewer chances to be scheduled if past jobs consume more resources. We use the deficit round robin (DRR) \cite{DRR} algorithm to schedule requests because of its simplicity and effectiveness, and linear programming to model the aforementioned intuition. DRR creates a credit account giving some number of initial credits to each tenant. Upon a request, DRR removes credits from the tenant's credit account based on the size of the resource consumption of the request. A request will not be scheduled if the tenant's credit account has insufficient credits. The scheduler then does round robin scheduling among the tenants. A tenant's credit account is refilled when it is its turn to be scheduled again in the round robin circuit. We adapt and extend DRR for request scheduling as follows.

First, the scheduler uses bytes read and written as delivered by the StorageProxy as an indirect means to measure the physical resource consumption. It assumes the existence of a linear function that could combine bytes read from and bytes written to the StorageProxy, which allows us to estimate and quantify underlying physical resources consumption. Dominant resource fairness (DRF) \cite{DRF} could be used here, and this is what Pisces does \cite{Pisces}. However, we argue that the physical resources reflected from bytes read and bytes written are not independent which violates a fundamental assumption of DRF. We thus leave the usefulness of DRF for future work. Second, DRR requires that the resource consumption of a request is known in advance. This is impossible for a read operation because Cassandra does not know how many bytes a read operation requests before it processes the request. Thus we will predict the size of an upcoming read operation using a simple averaging of bytes in a sliding window of bytes of previous read operations. Other approaches could be used here, for example linear regression. Third, to provide fair share within a certain time frame, we refill a tenant's credit account only when all tenants' credit accounts run dry. The number of credits a tenant has indicates how big its chance is to be scheduled in in the current round of DRR and should be related to her previous resource consumptions, i.e., the more resources she consumes, the less opportunities her requests will get scheduled. We use linear programming to implement this idea and describe it below.
\begin{table}[h]
    \centering
    \caption{Notations.}
    \label{tab:fairness-notations}
    \begin{tabular}{|l|l|}
      \hline
      \textbf{Notations} & \textbf{Description} \\
      \hline
      $n$ & Number of tenants \\
      \hline
      $b_{i}$ & Resources consumed by tenant $i$ since last refill\\
      \hline
      $B_{i}$ & Resources estimated to be consumed by tenant $i$ \\
      \hline
      $x_{i}$ & Credits assigned to tenant $i$ \\
      \hline
      $m_{i}$ & Scalar to translate credits to resources for tenant $i$ \\
      \hline
      $w_{i}$ & Weight for tenant $i$ \\
      \hline
      $M$ & Total credits for all tenants \\
      \hline
    \end{tabular}
\end{table}

Notations in Table \ref{tab:fairness-notations} are used for discussion. $b_{i}$ is the sum of bytes read and written in current implementation. We express $B_{i}$ in a linear equation in equation \ref{eq:bytesdeliver}. $m_{i} \times x_{i}$ is the resource tenant $i$ can have if $x_{i}$ credits are assigned to her. Notice that $b_{i}$ is collected as feedbacks and $m_{i}$ is a positive constant.
\begin{equation} \label{eq:bytesdeliver}
B_{i} = B(x_{i}) = b_{i} + m_{i} \times x_{i}, i = 1,2,\ldots,n
\end{equation}

The optimization objective is to achieve max-min fairness which is expressed in equation \ref{eq:lp} where $\{x_{i}\}$ are control variables. By solving equation \ref{eq:lp}, we can have $\{x_{i}\}$ that enforce max-min fairness among tenants. Meanwhile, if $b_{i}$ is large, then $x_{i}$ will be small because of the optimization objective and constraints in equation \ref{eq:lp}.
\begin{equation} \label{eq:lp}
    \begin{aligned}
        %\text{max} \quad
        \max_{x_{1}, \dots, x_{n}} \quad
        \min_{x_{1}, \dots, x_{n}}{\{B(x_{i})\}}& \\
        \text{s.t.} \quad
        \sum_{1 \leq i \leq n}x_{i} &= M \\
        x_{i} &\geq 0, i = 1,2,\ldots,n
    \end{aligned}
\end{equation}

To convert equation \ref{eq:lp} to the standard form in linear programming, we introduce an auxiliary variable $z$ and rewrite equation \ref{eq:lp} to equation \ref{eq:standardlp}.
\begin{equation} \label{eq:standardlp}
    \begin{aligned}
        \max_{x_{1}, \dots, x_{n}} \quad
        z \\
        \text{s.t.} \quad
        -(b_{i} + m_{i} \times x_{i}) &\leq -z \\
        \sum_{1 \leq i \leq n}x_{i} &= M \\
        x_{i} &\geq 0, i = 1,2,\ldots,n
    \end{aligned}
\end{equation}

We show that equation \ref{eq:standardlp} has optimal solutions. First, equation \ref{eq:standardlp} has at least one feasible solution, e.g., $x_{1}=M$, $x_{i}=0, i = 2,\ldots,n$. Second, $z$ is bounded as shown in equation \ref{eq:upperbound}. Since $z$ is a convex function and bounded, and feasible solutions exist for constraints, an optimal solution exists for equation \ref{eq:standardlp} according to \cite{LP-Book}.
\begin{equation} \label{eq:upperbound}
0 \leq z \leq \max{\{b_{i} + m_{i} \times x_{i}\}} \leq \max{\{b_{i} + m_{i} \times M\}}
\end{equation}

Besides even tenant share, our system allows a system admin to set weights for each tenant to have weighted tenant share. Each tenant is given weight $w_{i}$. A tenant with a larger weight gets more share of the system. We extend our model in equation \ref{eq:standardlp} to consider different weights for different tenants. To avoid confusions, we use variable $u$ and rewrite the linear model in equation \ref{eq:weightedlp}. The proof of existence of optimal solution is similar to the one for equation \ref{eq:standardlp}.
\begin{equation} \label{eq:weightedlp}
    \begin{aligned}
        \max_{x_{1}, \dots, x_{n}} \quad
        u \\
        \text{s.t.} \quad
        -(b_{i} + m_{i} \times x_{i}) &\leq -u \times w_{i} \\
        \sum_{1 \leq i \leq n}x_{i} &= M \\
        \sum_{1 \leq i \leq n}w_{i} &= 1 \\
        x_{i} &\geq 0, i = 1,2,\ldots,n  \\
        w_{i} &\geq 0, i = 1,2,\ldots,n
    \end{aligned}
\end{equation}

We present the scheduling algorithm in Algorithm \ref{alg:lp-ex-drr}.
\begin{algorithm}[!htbp]
    \caption{Request Scheduling Algorithm}\label{alg:lp-ex-drr}
    \begin{algorithmic}[1]
        \State $credit_{i}$ is the credits in tenant $i$'s credit account.
        \State $est_{i}$ is the estimation of a request resource demand for tenant $i$.

        \Procedure{Schedule}{}
            \For{each tenant $i$}
            \State $est_{i}$ $\leftarrow$ RequestRcEstimation(tenant $i$)
            \While{$m_{i} \times credit_{i} \geq est_{i}$}
                \State Take a request if tenant $i$'s queue is not empty.
                \State $credit_{i} \leftarrow credit_{i} - est_{i} / m_{i}$
            \EndWhile
            \EndFor
            \If{RefillCredits() is true}
                \State $\{x_{i}\}$ = LPModel($\{b_{i}\}$, $\{w_{i}\}$), $i=1,2, \dots, n$
                \State Assign credit $x_{i}$ to tenant $i$, $i=1,2, \dots, n$
            \EndIf
        \EndProcedure

         \Procedure{RefillCredits}{}
            \If{all tenants' credit accounts run dry}
                Return True
            \Else
                Return False
            \EndIf
        \EndProcedure
    \end{algorithmic}
\end{algorithm}

\section{Adaptive Control Mechanisms}
The adaptive control mechanisms include the local weight adjustment approach to provide system-wide fair share and scan operation splitting to avoid head-of-line blocking for read operation.

\begin{figure}
    \centering
    \begin{subfigure}[b]{0.4\textwidth}
        \includegraphics[scale=0.3]{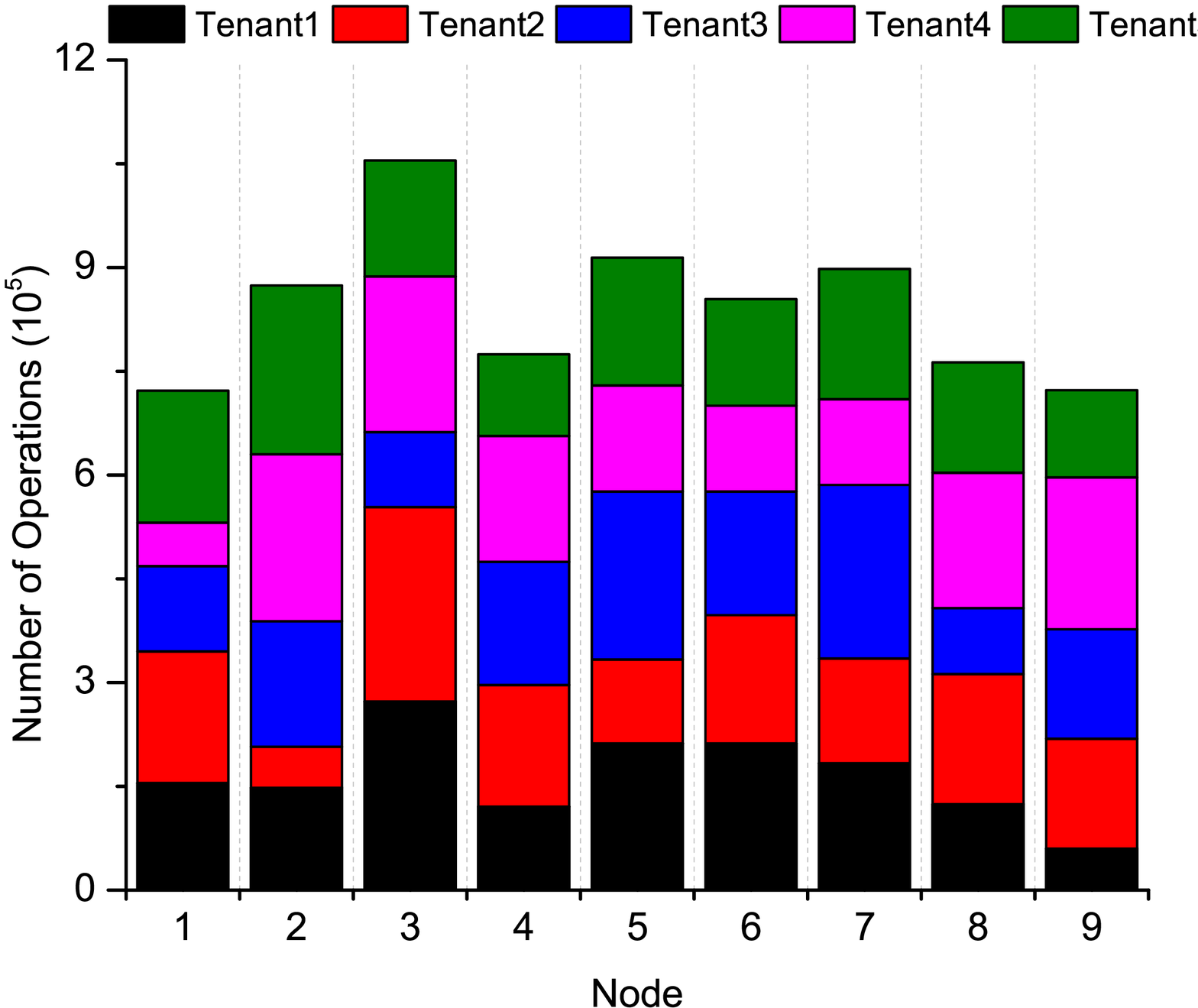}
        \caption{Cumulative throughput distribution.}
        \label{subfig:global:operations}
    \end{subfigure}
    ~~~
    \begin{subfigure}[b]{0.4\textwidth}
        \includegraphics[scale=0.3]{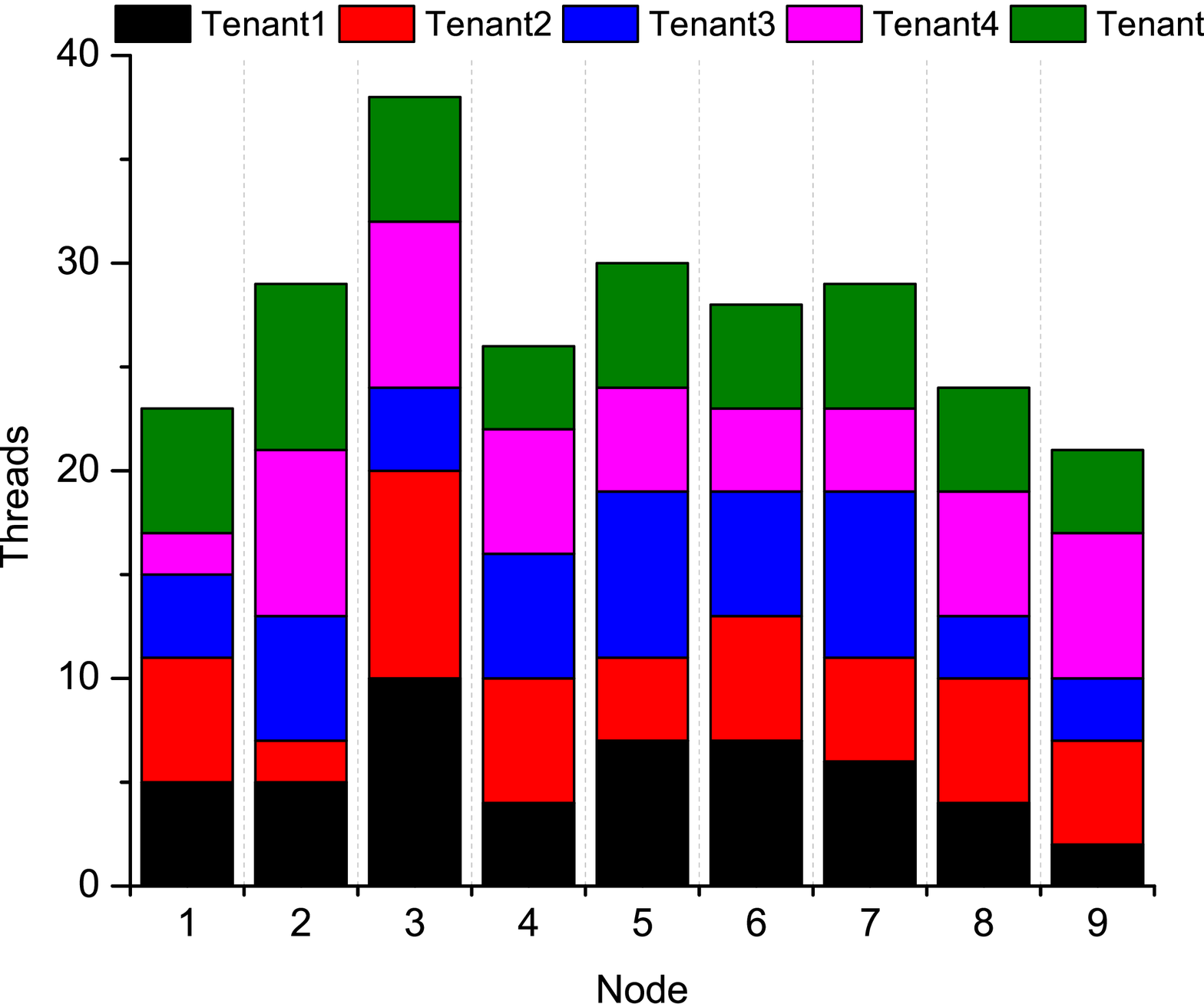}
        \caption{Thread counts distribution.}
        \label{subfig:global:threads}
    \end{subfigure}
    \caption{Distributions of cumulative throughput and thread count over Cassandra nodes.}
    \label{fig:global-local}
\end{figure}

\paragraph{Local weight adjustment}
The scheduling approach presented above focuses on providing fair share in a single node. We call such fairness \textit{local fairness} and system-wide fairness as \textit{global fairness}. It is easy to show local fairness results in global fairness. However, in a distributed environment, local fairness is not a sufficient condition for global fairness and can lead to inefficiency. We demonstrate this through example: for the experiment in Figure \ref{subfig:fairness-motivation:samerate}, we plot the distribution of cumulative throughput over nodes for each tenant within a certain time frame in Figure \ref{subfig:global:operations}. It shows that although tenants have global fairness, i.e., their throughputs are about the same as Figure \ref{subfig:fairness-motivation:samerate} shows, they actually receive different throughputs on different nodes. For instance, tenant \#1 has the lowest throughput in node \#9 and the highest throughput in node \#3, while tenant \#3 has the lowest throughput in node \#3 and the highest throughput in node \#5. If we enforce local fairness for tenant \#1 in node \#9, then throughputs of other tenants are constrained unnecessarily.

To investigate the causes, we plot the distribution of number of threads each tenant uses to connect over nodes in Figure \ref{subfig:global:threads}. We see that a tenant's cumulative throughput on a node is proportional to its number of threads connected to the same node. This matches our expectation that the number of threads is related to throughput.  \textit{In summary, Figure \ref{fig:global-local} suggests that 1) global fairness can be achieved without local fairness; 2) number of threads can be used as a hint to regulate throughput.}

We thus propose a local weight adjustment approach to achieve global fairness. Algorithm \ref{alg:global-control} presents the approach. The idea is to recalculate the weight a tenant should have on a particular node based on the ratio of its thread count on that node to its total thread count on all the nodes (line 10 to 15). Then for each node, we can calculate the new weight assignment based on the tenants' new credits (line 16 to 20). In this step, the maximum credits a node has would be changed and different from each other, although the total credits a tenant has within the system is not changed. Finally, the algorithm disseminates the new weight assignment along with the new maximum credit for each node (line 21 to 24).
\begin{algorithm}[!htbp]
    \caption{Local Weight Adjustment Algorithm}\label{alg:global-control}
    \begin{algorithmic}[1]
        \State $total$ is the total credits for the entire system.
        \State $n$ is the number of tenants, $m$ is the number of nodes.
        \State $c_{i,.}$ is the total credits for tenant $i$. $c_{i,j}$ is credit for tenant $i$ on node $j$.
        \State $d_{i,.}$ is the total thread count for tenant $i$. $d_{i,j}$ is the thread count for tenant $i$ on node $j$.
        \State $w_{i,j}$ is the weight for tenant $i$ on node $j$. $W_{i}$ is the weight for tenant $i$ initially.

        \Procedure{LocalWeightAdjustment}{}
            \For{each tenant $i$}
                \State $c_{i,.}$ $\leftarrow total \times W_{i}$
                \For{each node $j$}
                \State $c_{i,j} \leftarrow c_{i,.} \times (d_{i, j}/d_{i,.})$
                \EndFor
            \EndFor

            \For{each node $j$}
                \For{each tenant $i$}
                \State $w_{i,j} \leftarrow c_{i,j}/\sum_{1 \leq i \leq n}c_{i,j}$
                \EndFor
            \EndFor

            \For{each node $j$}
                \State Disseminate $w_{i,j}, i=1,\dots,n$ and
                \State $\sum_{1 \leq i \leq n}c_{i,j}$ to node $j$.
            \EndFor
        \EndProcedure
    \end{algorithmic}
\end{algorithm}

\paragraph{Interference between read and scan}
We have demonstrated that the presence of scan operations will influence throughput of read operations. But read operations and scan operations often coexist in the same system. \textit{MapReduce} style workloads frequently use scan operations while front-end workloads perform read operations. The read operations have to wait for physical resources held by the scan operations, a problem known as head-of-line blocking.

One solution is to use preemption to allow waiting operations to deprive physical resources from current operation temporarily. However, not all physical resources are preemptable, e.g.,network bandwidth and disk I/O. In Cassandra, once an operation is scheduled to process, it is difficult to suspend that operation because it may trigger operations on other nodes. Instead of exploring how to suspend an ongoing operation temporarily, we use a simple approach, which is used in other systems as well \cite{Jeff,Cake}. That is to split a scan operation into small pieces such that each piece does not lead to head-of-line blocking. The scheduler then schedules small pieces of scan operations along with read operations so that a read operation can get more chances to be scheduled. The results from these small pieces need to be merged before handing back to the client. This approach trades performance of scan with performance of read because the execution of a scan operation is interleaved with the execution of read operation and the number of disk seeks is increased due to multiple small pieces. The size of a piece is negatively correlated to the performance of read, and positively correlated to the performance of scan. As the size increases, the throughput of scan goes up and the throughput of read goes down, vice versa. Automatically tuning the number of pieces a scan operation is chopped, based on different workload scenarios, is an ongoing work.

\section{Evaluation} \label{subtitle:fairshare-evaluation}
We use computation resources of FutureGrid \cite{FG} in the evaluation of our system. Each node has 2 Intel(R) Xeon(R) 2.93 GHz CPUs, 25 GB memory and a 800 GB local disk. Nodes are connected with InfiniBand. Using a 9-node cluster, we install a modified version of Cassandra 1.2.4 equipped with the fairness control scheduler in every node retaining the default Cassandra settings. On the client side, we use YCSB \cite{YCSB} to generate the workloads and use 5 additional nodes to run the clients to simulate 5 tenants accessing Cassandra. Each tenant stays in a separate node, so interference on the client side is avoided. Each tenant has its own \textit{Keyspace} with one \textit{ColumnFamily}. We pre-load 1,000,000 rows into each tenant's \textit{Keyspace}. The row size is 1.2 KB and replication factor is set to 3. The consistency level is set to 1. The tests are run with a read-only workload and a read-write workload. For the former, all tenants send read operations only. For the read-write workload, some tenants send read operations while other tenants send an even mix of read operations and write operations. We configure the target throughput of the read-only workload to be a large number so that a tenant can send as many read operations as possible. The target throughput of the read-write workload is configured to be 25,000 op/s so as to avoid disk saturation and dramatic increase of data size.

We use operation throughput, ops/second (op/s), to represent a tenant's share of the system. We compute min-max ratio of throughput as follows.
\begin{equation}
rate_{minmax}=\min{\{throughput_{i}\}}/\max{\{throughput_{i}\}}, i = 1, 2, \ldots, n,
\end{equation}
where $throughput_{i}$ is the throughput of tenant $i$, among all tenants as the fairness metric. To get a stable throughput, we report the results after a ramp-up time of 20 seconds. We first evaluate the overall performance of our system by running different workloads on different data sets. Then we study the effectiveness of local weight adjustment. Finally, we test the system with read and scan mixed workload.

\subsection{Overall Performance}
\begin{figure}[!htbp]
    \centering
    \begin{subfigure}[b]{0.4\textwidth}
        \includegraphics[scale=0.3]{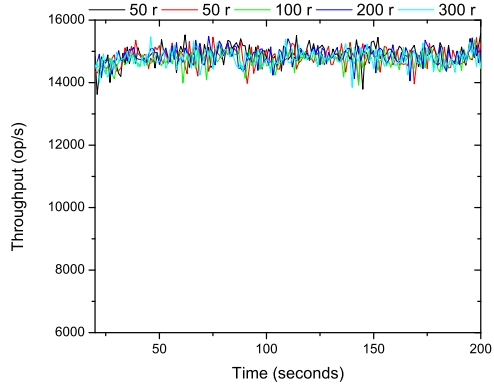}
        \caption{Fixed size data, even share, 16\% throughput decrease}
        \label{subfig:evaluation:fixequal}
    \end{subfigure}
    ~~~
    \begin{subfigure}[b]{0.4\textwidth}
        \includegraphics[scale=0.3]{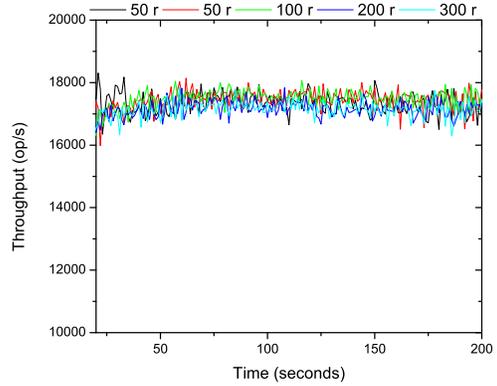}
        \caption{Variable size data, even share, 15\% throughput decrease}
        \label{subfig:evaluation:varequal}
    \end{subfigure}
    ~
    \begin{subfigure}[b]{0.4\textwidth}
        \includegraphics[scale=0.3]{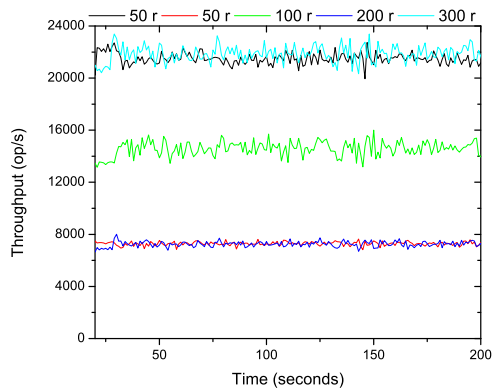}
        \caption{Fixed size data, weighted share, 16\% throughput decrease}
        \label{subfig:evaluation:fixdiff}
    \end{subfigure}
    ~~~
    \begin{subfigure}[b]{0.4\textwidth}
        \includegraphics[scale=0.3]{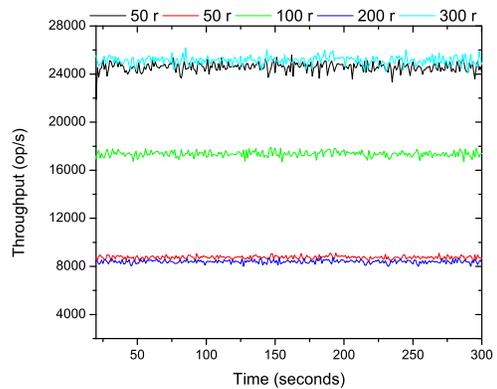}
        \caption{Variable size data, weighted share, 16\% throughput decrease}
        \label{subfig:evaluation:vardiff}
    \end{subfigure}
  \caption{Fair share for read-only workload on data sets with different tenant share configurations.}
  \label{fig:rperformance}
\end{figure}

\begin{figure}[!htbp]
    \centering
    \begin{subfigure}[b]{0.4\textwidth}
        \includegraphics[scale=0.3]{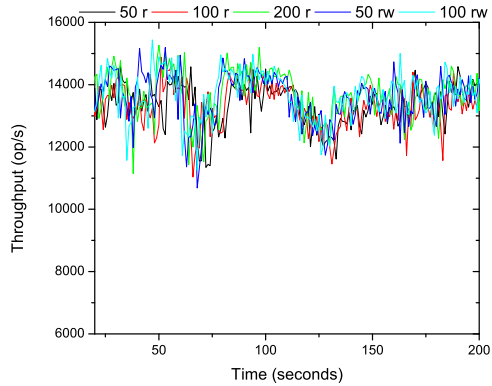}
        \caption{Fixed size data, even share, 25\% throughput decrease}
        \label{subfig:rwevaluation:fixequal}
    \end{subfigure}
    ~~~
    \begin{subfigure}[b]{0.4\textwidth}
        \includegraphics[scale=0.3]{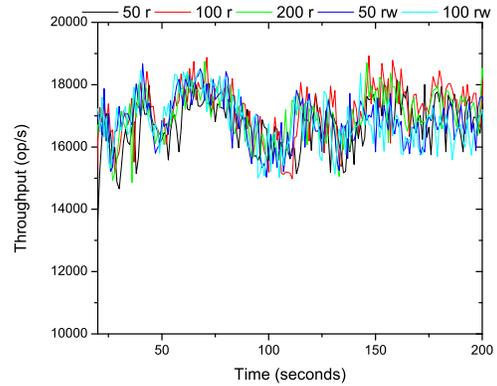}
        \caption{Variable size data, even share, 24\% throughput decrease}
        \label{subfig:rwevaluation:varequal}
    \end{subfigure}
    ~
    \begin{subfigure}[b]{0.4\textwidth}
        \includegraphics[scale=0.3]{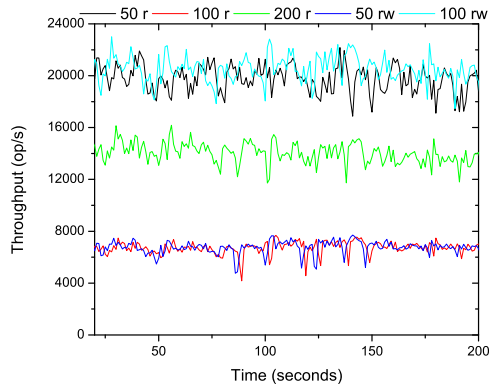}
        \caption{Fixed size data, weighted share, 25\% throughput decrease}
        \label{subfig:rwevaluation:fixdiff}
    \end{subfigure}
    ~~~
    \begin{subfigure}[b]{0.4\textwidth}
        \includegraphics[scale=0.3]{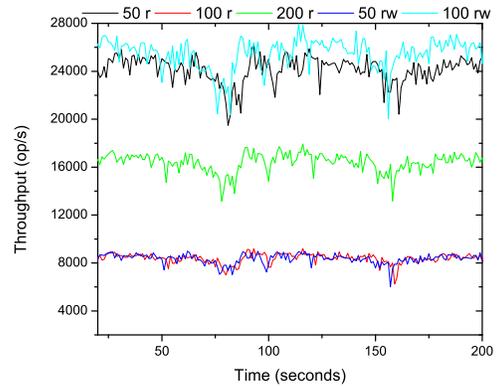}
        \caption{Variable size data, weighted share, 23\% throughput decrease}
        \label{subfig:rwevaluation:vardiff}
    \end{subfigure}
  \caption{Fair share for read-write workload on data sets with different tenant share configurations.}
  \label{fig:rwperformance}
\end{figure}

We begin by assessing whether our system can provide fair share when tenants use different numbers of threads. Then we test if the system can differentiate tenants based on their weight configurations. To test the system thoroughly, in addition to the fixed size data set, we use YCSB to pre-load 1,000,000 rows, whose row size varies from 100B to 1.2 KB uniformly for each tenant. Figure \ref{fig:rperformance} and \ref{fig:rwperformance} plot the throughputs as a function of time. The letter in the legend stands for the workload a tenant runs. For instance, ``50 r" means the tenant uses 50 threads to send read operations, while ``50 rw" means the tenant uses 50 threads to send read as well as write operations. The percentage of throughput decrease measures the performance degradation comparing with the vanilla system. The throughput represents the system throughput and is measured as aggregated average throughput from all tenants. Table \ref{tab:minmaxratio} summarizes the min-max ratio for different runs.

We run the read-only workload for even tenant share i.e., weight 0.2 for each tenant in this case. We configure the number of threads as 50, 50, 100, 200, and 300 for each tenant respectively. Figure \ref{subfig:evaluation:fixequal} and \ref{subfig:evaluation:varequal} show the results. As reported from Table \ref{tab:minmaxratio}, the mean min-max ratio is close to 1, i.e., 0.95 (std. 0.02) for fixed size data and 0.94 (std. 0.02) for various size data. Therefore, our system can provide fair share in either fixed size or variable size data given that one tenant demands more (e.g., the one with 300 threads) than another tenant does (e.g., the one with 50 threads).
\begin{table}[h]
    \centering
    \caption{Fairness of workloads on across data sets.}
    \label{tab:minmaxratio}
    \begin{tabular}{|l|l|l|l|}
      \hline
      \textbf{Workload} & \textbf{Data Size} & \textbf{Mean $rate_{minmax}$} & \textbf{Std. $rate_{minmax}$} \\
      \hline
      Read-Only & Fixed & 0.95 & 0.02 \\
      \hline
      Read-Only & Variable & 0.94 & 0.02 \\
      \hline
      Read-Write & Fixed & 0.90 & 0.04 \\
      \hline
      Read-Write & Variable & 0.91 & 0.03 \\
      \hline
    \end{tabular}
\end{table}

Additionally, we run the same read-only workload with the same threads setting as above, but configure weighted tenant share. Specifically, the weight for each tenant is configured as 0.3, 0.1, 0.2, 0.1, and 0.3. The corresponding number of threads is 50, 50, 100, 200, and 300 for each tenant. The purpose is threefold. First, we test if tenants' throughputs could be differentiated based on their weights. Second, we evaluate if the number of threads has an impact on the differentiation by configuring tenants with 50 and 300 threads to both have weight 0.3. Third, we want to see if the tenant with weight 0.2 will see similar throughput compared to the one in even tenant share. Figure \ref{subfig:evaluation:fixdiff} and \ref{subfig:evaluation:vardiff} depict the results. It is clear that 5 tenants are classified into 3 categories and tenants' throughputs are roughly proportional to 3:2:1 which is equal to their weights ratio. Note that the tenant with 50 threads gets similar throughput as the tenant with 300 threads does, showing that the fairness control scheduler can eliminate the impact of a tenant's thread count in differentiation. In addition, the tenant with weight 0.2 gets similar throughput compared to the one in even tenant share. That verifies the system can preserve a tenant's throughput in either even or weighted tenant share.

Next we rerun the tests with the read-write workload. Three tenants are configured to send read requests with 50, 100, and 200 threads respectively (read-only tenants). The other two tenants are configured to send an evenly mixed workload of reads and writes with 50 and 100 threads respectively (read-write tenants). Figure \ref{subfig:rwevaluation:fixequal} and \ref{subfig:rwevaluation:varequal} report the throughputs of even tenant share, while figure \ref{subfig:rwevaluation:fixdiff} and \ref{subfig:rwevaluation:vardiff} display the throughputs of weighted tenant share. The percentage of throughput decrease comparing with vanilla Cassandra is also reported in each case. Compared to throughputs in Figure \ref{fig:rperformance}, the throughputs in Figure \ref{fig:rwperformance} are similar in the sense that tenants get proportional throughputs to their weights no matter how many threads they have. However, the performance degrades in terms of throughput decreases, min-max ratio also drops and oscillates. We attribute the performance degradation to the inability of isolating the performance of read operations from the performance of write operations. Write operations in Cassandra have read operations to read more \textit{MemTables} and \textit{SSTables} for the same tenant. This triggers the \textit{compaction} procedure \cite{Cassandra}, which requires many disk I/O and influences other tenants' throughputs. Similar situation also happens to Bigtable like systems \cite{BigTable}. To isolate writes from reads, a performance model that can predict the impact of internal writes is required. That is given a set of writes, the model should be able to predict the amount of internal writes. We leave such a performance modeling and the isolation between read and write for future work.

Finally, we quantify the throughput degradation caused by the fair share scheduler. Throughput degradation is measured from the client side. This work does not extend to server side profiling to discern the precise location and type of throughput degradation. We first measure degradation in scenarios having interference, i.e. the scenes in Figure \ref{fig:rperformance} and \ref{fig:rwperformance}. The percentage of aggregated throughput decrease varies by scenarios. The read-only workloads experience about 16\% degradation while the read-write workloads have about 25\% degradation. We think the reason is the inaccurate estimate of resources consumed in read-write mixed workloads imposes unnecessary constraints on requests. We further study the degradation in a fair share scenario by having all tenants run the read-only workload and the read-write workload respectively with 50 threads per tenant. We observe about 9\% throughput decrease for both workloads. We think the reasons for smaller degradation compared with interference scenarios are two folds. First, the 50-threads workloads do not fully utilize the system as evidenced by smaller aggregated throughput generated. Second, smaller thread count incurs fewer contentions to the scheduler.

The main reason behind these degradations, we believe, is the number of credits used in the scheduling. Intuitively, the number of credits is directly proportional to throughput, but inversely proportional to fairness. The more credits are given, the fewer constraints the scheduler will impose, which results in higher throughput. Meanwhile, however, fairness decreases because the scheduler has less control over throughput regulation. We speculate that increasing 15\% of the credits given might help to lower the overhead without hurting too much of fairness based on the observations that read-only workloads experience about 15\% overhead and fairness (i.e. the min-max ratio) is very close to 1. We experimentally study the impact of the number of credits on overhead and fairness in the follow-up work in Chapter \ref{title:reservation}, and present a near-optimal setting of the credit to avoid high overhead without sacrificing too much fairness. It is an open problem to profile the server side system to better tune the credit parameter.

%Finally, we assess the overhead of the scheduler by having all tenants run the read-only workload and the read-write workload respectively with 50 threads per tenant. We compare the throughputs with those generated by a vanilla Cassandra. We observe 7\% $\sim$ 9\% throughput decrease due to the fairness control scheduler. From this we conclude that our system provides fair share among multiple tenants with multiple number of threads and differentiates tenants based on their weight configuration.

\subsection{Effectiveness of Adaptive Control Mechanisms}
\begin{figure}[h]
  \centering
  \includegraphics[scale=0.4]{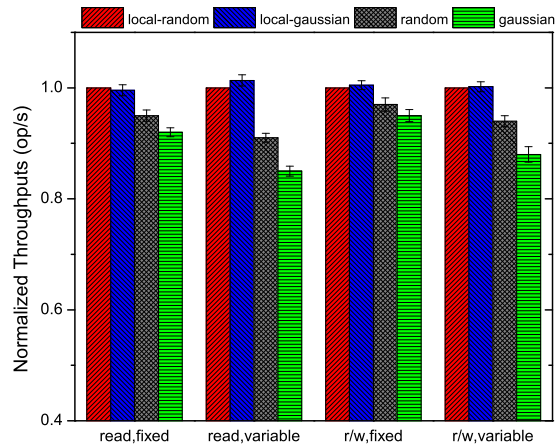}
  \caption[Improvement in throughput through local weight adjustment.]{Improvement in throughput through local weight adjustment. ``local-random" and ``local-gaussian" mean the local weight adjustment is applied to random and gaussian thread distribution while ``random" and ``gaussian" mean no local weight adjustment.}
  \label{fig:throughput}
\end{figure}

\paragraph{Local weight adjustment}
We compare the throughputs of the read-only workload as well as the read-write workload with local weight adjustment to the ones without it. Two thread distributions are tested. The first one is random distribution where each tenant thread randomly picks a server node to connect to, while the second one is gaussian distribution where each tenant thread picks a server node based on a pre-defined gaussian distribution. Figure \ref{fig:throughput} presents the normalized throughputs. When the local weight adjustment is in place, the bars with red and blue colors have similar heights on fixed size data and variable size data. Therefore, the local weight adjustment can handle different thread distributions. When there is no local weight adjustment present, the gray bar is higher than the green bar which means the random distribution gets more throughputs than the gaussian distribution does in our system. Additionally, the throughput with the local weight adjustment is up to 8\% higher for random distribution and 15\% higher for gaussian distribution than the throughput without it respectively. We attribute the throughput improvement to the ability that the local weight adjustment can redistribute the weights based on a global view of tenants' demands and avoid unnecessary constraints.

\paragraph{Interference between read and scan}
\begin{figure}[h]
  \centering
  \includegraphics[scale=0.3]{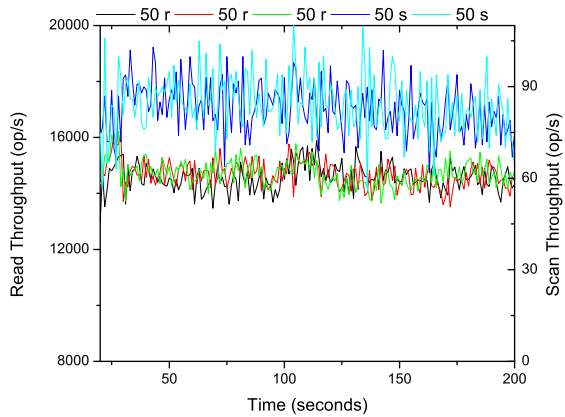}
  \caption{Fair share protects the throughput of the read-only workload from scan operations.}
  \label{fig:readscan}
\end{figure}

We test the effectiveness of our approach under read + scan workloads. The test uses 3 tenants who send read operations (read tenants) while the 2 remaining tenants send scan operations (scan tenants). A scan operation scans 200 rows per request. All tenants run 50 threads. The scheduler splits a scan request into 5 rows per request.  The results, summarized in Figure \ref{fig:readscan}, show throughput of read tenants at close to 15,000 op/s which is similar to the throughput in Figure \ref{subfig:evaluation:fixequal}. In addition, the throughput oscillation is much smaller than shown in Figure \ref{subfig:fairness-motivation:difftype}. This demonstrates that our approach can preserve the read tenant's throughput under a mixed load of read tenants and scan tenants.

\section{Summary}
In this chapter, we examined the fairness in NoSQL data stores under multi-tenancy, with a focus on the Cassandra NoSQL store. We propose extensions, both methodologically and through a prototype implementation, of ensuring fairness that employs a deficit round robin algorithm with linear programming to schedule tenants' requests. We adaptively adjust tenants' weights on each node to improve throughput. Finally we protect the throughput of read operations in face of scan operations by splitting one scan operation into small pieces and scheduling them along with read operations.

Future study is of the impact of writes on reads so as to isolate read and write performance. Besides, statistical machine learning may be effective in predicting future resource consumption and detect slow tenants. Additionally, different resource models that either combine resource types in a non-linear function or use the dominant resource fairness approach may improve fairness further. Finally, it appears beneficial to extend Cassandra's gossip protocol to integrate a more robust leader selection algorithm for local weight adjustment.

\chapter{Workload-Aware Resource Reservation for Multi-Tenant NoSQL Stores} \label{title:reservation}
Resource reservation is a common approach to avoiding performance interference among tenants. The basic idea is to dedicate a portion of a resource to a tenant for its use. Chapter \ref{title:fairshare} uses throughput regulation to provide fair access among tenants. Such an approach can be viewed as a special case of resource reservation -- throughput represents the underlying actual resource consumptions and is treated as a ``resource'' for each tenant.

As workloads usually have multiple resources involved e.g. memory for caching, CPU for serialization or deserialization, disk for reading or writing data, a tenant needs to acquire a reservation on each resource. But reservations are not all alike: a workload that has a hotspot access pattern may require more cache than does a workload with a random access pattern. An equal reservation of cache and disk usage for both workloads will not yield the best result. So reservations have to be based on workload characteristics, also called workload-aware reservation.

A workload-aware reservation becomes more complicated if a workload bears dynamics i.e. a workload changes its access pattern during the access, which requires the system to be able to adjust accordingly. In addition, the distributed nature of NoSQL stores makes the workload-aware reservation more difficult. For a typical NoSQL store, a request is sent to one node which may contact several other nodes to fetch the data. It is complicated to have a coordination among different resources and nodes.

Previous research on preventing performance interference does so by simplifying the scenario, either by considering a single resource \cite{Krebs,Walraven,Das,A-Cache} (\emph{e.g.} CPU, cache), or representing multiple resources consumption as a single ``virtual resource'' consumption \cite{Pisces,Zeng}. Similarly, work in Chapter \ref{title:fairshare} uses throughput to approximate the underlying resource consumption of each tenant and regulates the throughput to provide fair access. Ignoring various resource demands that workloads have could lead to low resource utilization as the system imposes unnecessary constraints to tenants and even failure of preventing interference.

Therefore, we propose Argus (the 100-eyed watchman in Greek mythology), a workload-aware resource reservation framework that targets multiple resource reservations and aims to prevent performance interference, in terms of fair throughput violation, in NoSQL stores. Specifically, Argus focuses on cache and disk reservations. It enforces the cache reservation by splitting the cache space among tenants. It approximates the disk usage by the throughput of a distributed file system and uses a request scheduler to enforce throughput reservation. Argus models the workload-aware reservation as a constrained optimization and uses the stochastic hill climbing algorithm to find the proper reservation according to various workloads' resource demands. We applied the idea of Argus to HBase \cite{HBase}, a state-of-art NoSQL data store. In summary, this chapter makes the following contributions:
\begin{itemize}
  \item Quantitative evidences for existence of interference in HBase;
  \item Mechanisms to enforce reservation on both cache and disk resource under multi-tenancy;
  \item Offline performance model and stochastic hill climbing algorithm to discover a near-optimal resource reservation plan;
  \item Experimental results that show our system successfully prevents interference across tenants.
\end{itemize}

\section{Analysis of Interference} \label{subtitle:interference-analysis}
NoSQL data stores are typically deployed across multiple nodes for enhanced availability and performance. Data are represented as rows and distributed across nodes. We motivate our approach by showing that NoSQL data stores can suffer from performance interference when multiple tenants access simultaneously; and even that a reservation for a single resource can fail to prevent interference in some cases.

\subsection{Setup}
As \cite{Pisces,A-Cache,Zeng} show, multi-tenant performance interference could occur in various NoSQL stores. In this chapter, we study HBase \cite{HBase}, a popular NoSQL store. HBase is an open source implementation of Google BigTable \cite{BigTable}. It abstracts the data partition and distribution to a distributed file system i.e. HDFS \cite{HDFS} and runs on top of it. HBase follows the master-slave design: the \textit{HMaster} on the master node is responsible for coordinating and monitoring slaves nodes activities; the \textit{HRegionServers} on the slave nodes handle client requests directly. The HRegionServer exchanges data with HDFS in the unit of a block and implements a block cache equipped with a LRU replacement algorithm to avoid HDFS access. Thus HBase can be viewed as a two-level hierarchy storage system and provides a clean separation between different resource managements in different levels: HDFS manages the disk resource while HBase itself takes care of the caching and CPU usage.

We use the Yahoo Cloud Storage Benchmark (YCSB) \cite{YCSB} to simulate multi-tenant access. We set up a 28-node HBase cluster with block cache size set to 1,200 MB per node. We preload 80,000,000 rows for each tenant, where a row is about 1.2 KB. Additional details on the experiment setup are given in Section \ref{subtitle:reservation-evaluation}. We run two YCSB clients to simulate two tenant access. The clients are run on 2 additional nodes to avoid interference on the client side.

We define and name several workloads with different access patterns below to test HBase in a multi-tenant setting. Each Get request fetches one row per request.
\begin{enumerate}
  \item \textit{Uniform}: Series of Get requests that retrieve any data with equal probability from the table.
  \item \textit{Extreme Hotspot (ExHot)}: Series of Get requests that retrieve a small portion of the data in the table. The requested data is small enough to fit into cache entirely.
  \item \textit{Regular Hotspot (Hot)}: Shows hotspot pattern but the data requested cannot fit into cache entirely.
\end{enumerate}

\subsection{Interference Experiments}
We define several metrics by which performance is measured: operation throughput, throughput violation, cache occupancy, and HDFS throughput. Similar to \cite{Walraven,Pisces,Zeng,MeT}, we measure the \textit{operation throughput}, i.e. operation per second (ops/sec), from the client side to reflect each client's share of the system. Similar to \cite{Das}, to quantify the interference, we calculate the \textit{throughput violation} as $violation_{i}=(baseline_{i}-throughput_{i})/baseline_{i}$, where $baseline_{i}$ and $throughput_{i}$ are the baseline throughput and actual throughput of tenant $i$ respectively. The baseline throughput is observed when the cluster is dedicated for such a workload while the actual throughput is recorded when the cluster is shared by multi-tenant workloads. To investigate the underlying resource consumptions, we take advantage of HBase's level design and break down a request's resource consumption into the usage of cache and disk since simple key-value pair access is not CPU intensive. The cache usage is measured as the cache occupancy i.e. the ratio between current cache size a tenant takes to the total cache size. The disk usage is difficult to directly measure because a request may involve the disks in a few other nodes. Thus we approximate the disk usage as HDFS throughput from the abstract of disk access on multiple nodes which HDFS provides. The implementation details of tracking the cache occupancy and HDFS throughput are discussed in Section \ref{subtitle:enforcement}.

\paragraph{Baseline} To measure interference, we first establish the baseline for the selected workloads. Table \ref{tab:workparameters} summarizes the different parameters used in YCSB for the workloads. The records column is the range of the records that will be accessed. For hotspot workloads, there will be lots of repeated records access as there are only 0.2 and 3 million out of 80 million records accessed. To get a stable throughput, we report the results in Table \ref{tab:baseline} after a ramp-up time of 300 seconds, after which the throughput tend to stay stable especially for workloads with hotspot. The throughput as well as the HDFS throughput is averaged over a 800 seconds period.
\begin{table}[h]
    \centering
    \caption{Workload parameters.}
    \label{tab:workparameters}
    \begin{tabular}{|l|l|l|l|}
      \hline
      \textbf{Workload} & \textbf{Key Distribution} & \textbf{Records} \\
      \hline
      Uniform Get & Uniform & 80 million \\
      \hline
      Extreme Hotspot Get & Zipfian & 0.2 million \\
      \hline
      Regular Hotspot Get & Zipfian & 3 million \\
      \hline
    \end{tabular}
\end{table}

\begin{table}[h]
    \centering
    \caption{Baseline throughput for different workloads.}
    \label{tab:baseline}
    \begin{tabular}{|l|l|}
      \hline
      \textbf{Workload} & \textbf{Throughput (ops/sec)} \\
      \hline
      Uniform Get & 892.32 \\
      \hline
      Extreme Hotspot Get & 19853.53 \\
      \hline
      Regular Hotspot Get & 2030.76 \\
      \hline
    \end{tabular}
\end{table}

\paragraph{Interference on different resources}
\begin{figure}[!htbp]
   \centering
   \begin{subfigure}[b]{0.45\textwidth}
        \includegraphics[scale=0.3]{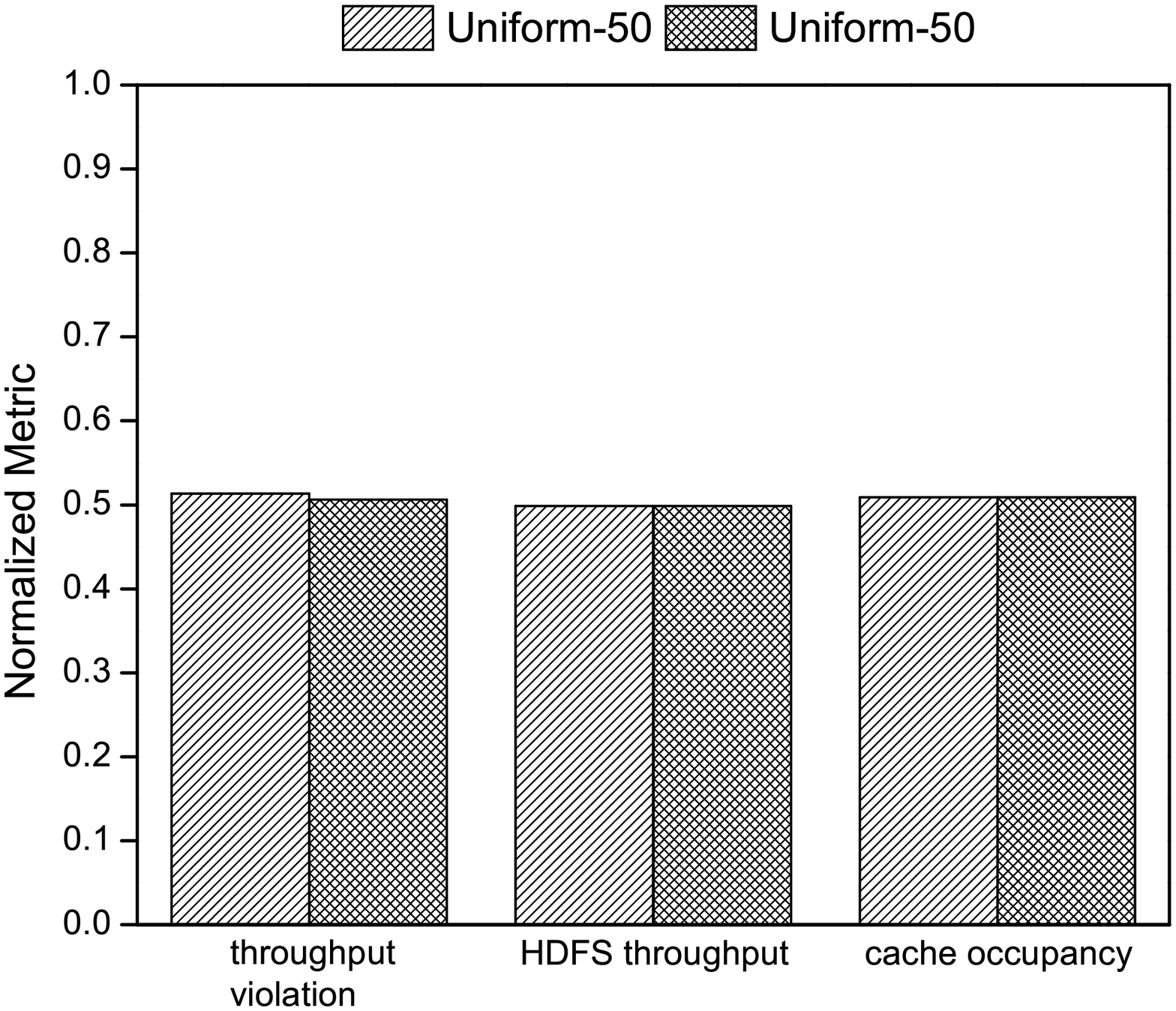}
        \caption{}
        \label{subfig:interference:fair}
    \end{subfigure}
    ~~~
    \begin{subfigure}[b]{0.45\textwidth}
        \includegraphics[scale=0.3]{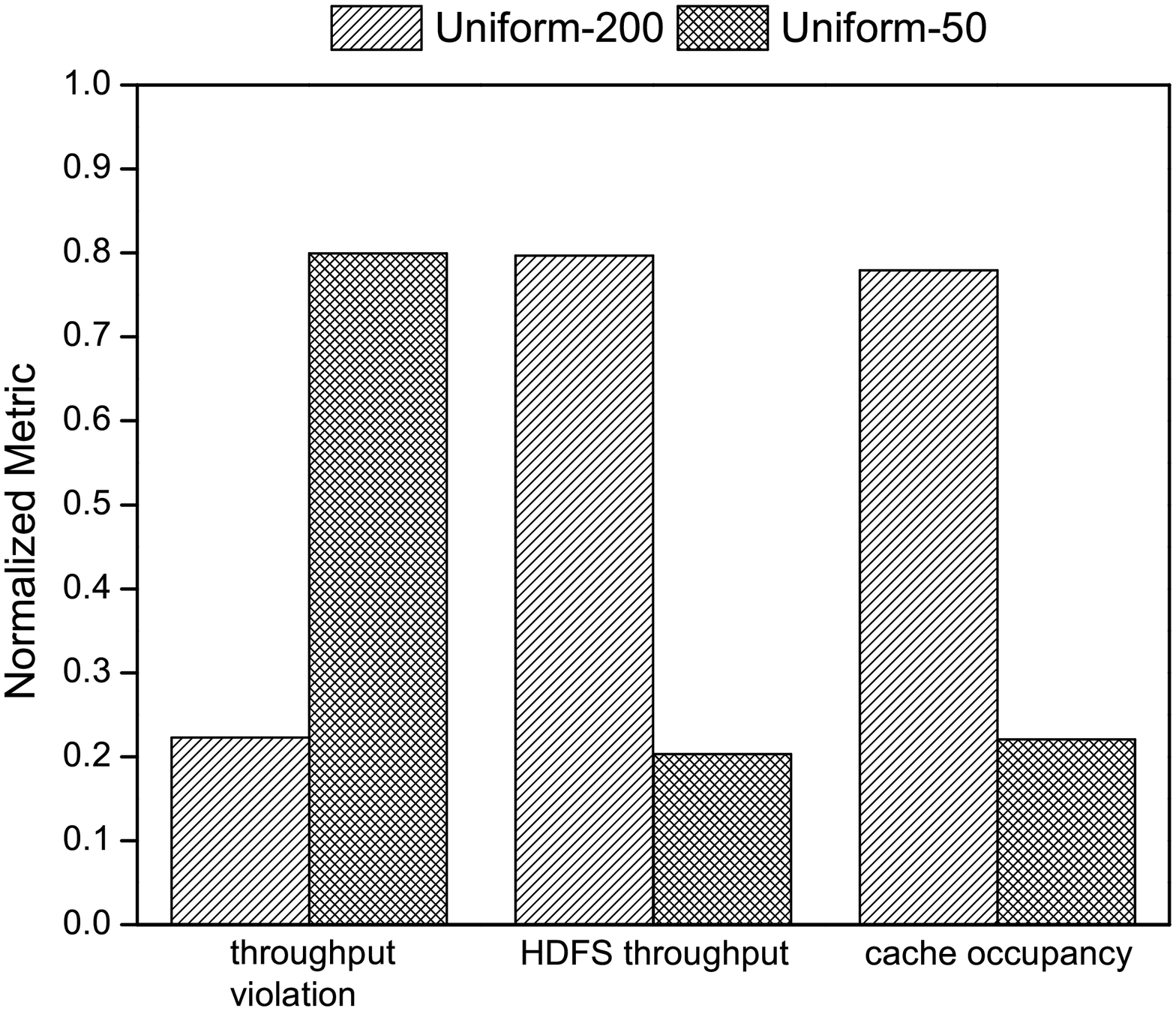}
        \caption{}
        \label{subfig:interference:bigthreads}
    \end{subfigure}
    ~
    \begin{subfigure}[b]{0.45\textwidth}
        \includegraphics[scale=0.3]{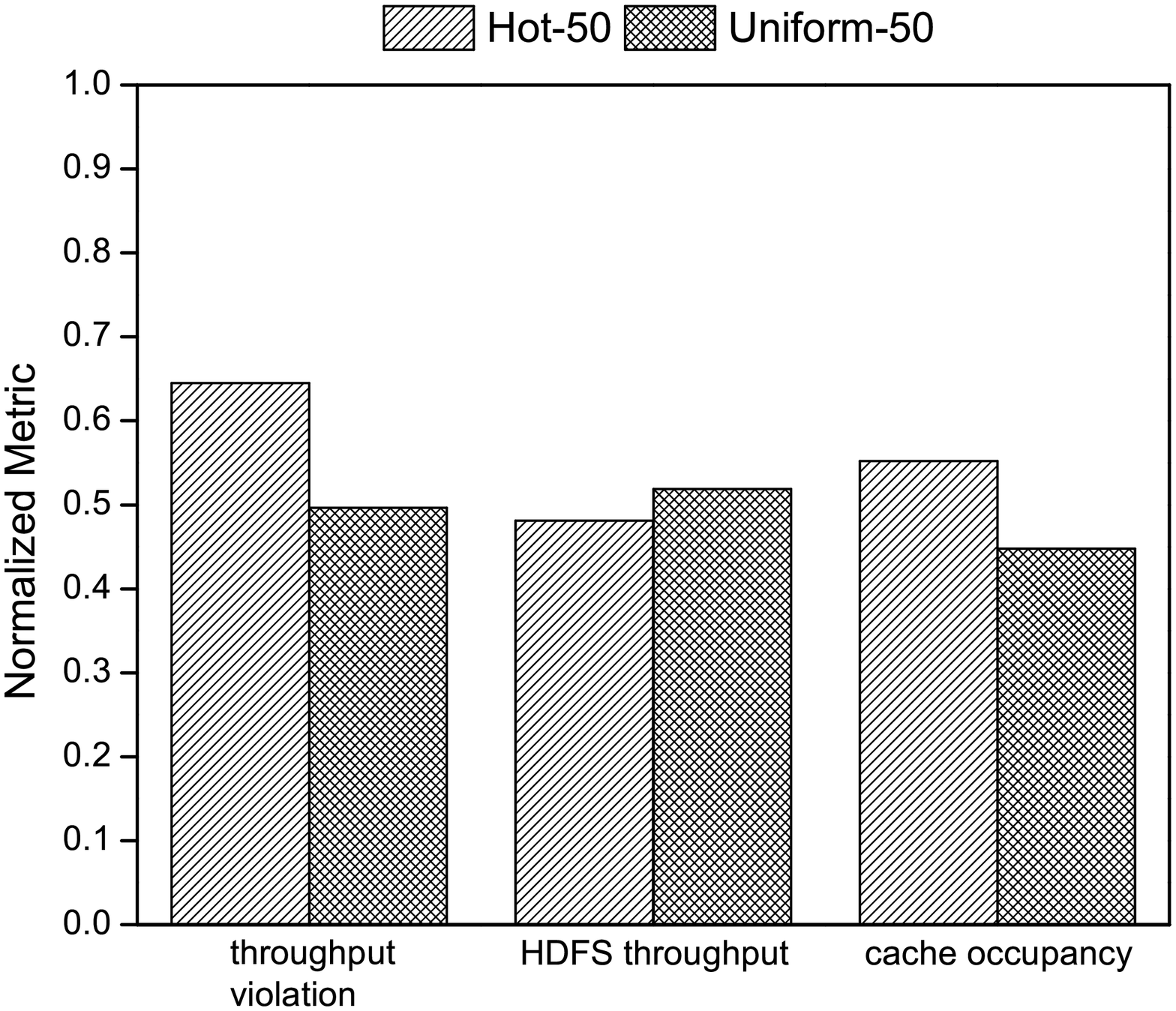}
        \caption{}
        \label{subfig:interference:hotreg}
    \end{subfigure}
    ~~~
    \begin{subfigure}[b]{0.45\textwidth}
        \includegraphics[scale=0.3]{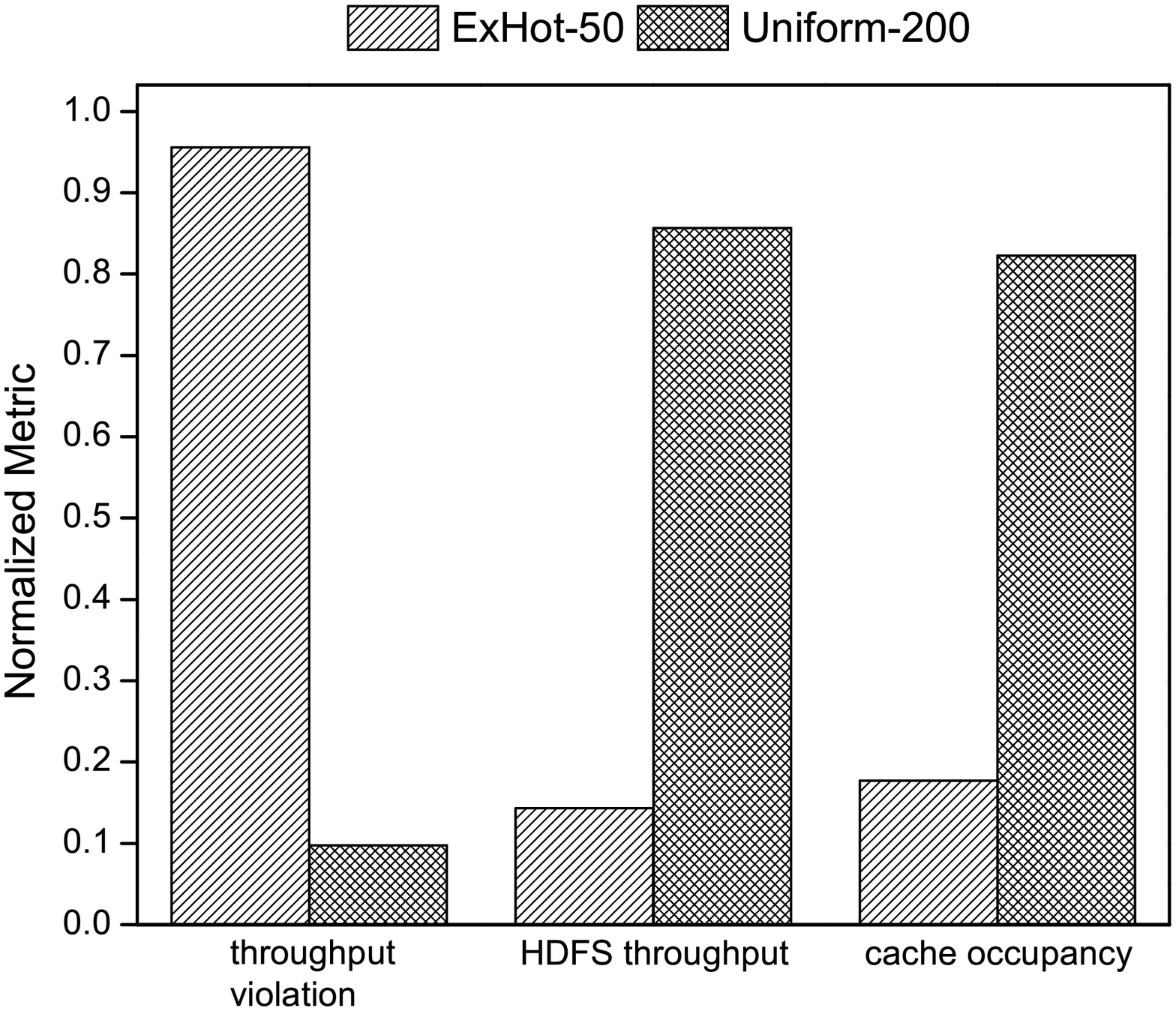}
        \caption{}
        \label{subfig:interference:exhotbig}
    \end{subfigure}
    \caption[Normalized metrics show interference occurs in different resources.]{Normalized metrics show interference occurs in different resources. We measure client throughput, HDFS throughput and cache occupancy for each tenant. The legend shows the workload type and number of threads it uses. For example, ``Uniform-200'' means uniform workload sent by 200 threads, ``Hot-50'' stands for regular hotspot workload sent by 50 threads, and ``ExHot-50'' means extreme hotspot workload sent by 50 threads.}
    \label{fig:interference}
\end{figure}

We next conduct several experiments that mix the workloads shown in Table \ref{tab:workparameters} to investigate the interference. For each experiment, we measure throughput violation, normalized HDFS throughput and cache occupancy. Figure \ref{fig:interference} plots the results.

In Figure \ref{subfig:interference:fair}, two tenants run the uniform workload with 50 threads. They see similar throughput violation which we interpret as fair access between these tenants. They also have similar HDFS throughput as well as cache occupancies. In Figure \ref{subfig:interference:bigthreads}, tenant \#1 uses 200 threads to run the Uniform workload while tenant \#2 still uses 50 threads. We observe that the throughput violation of tenant \#2 is about 4 times higher than tenant \#1's. Similarly, tenant \#1's HDFS throughput and cache occupancy are about 4 times higher than tenant \#2's. We believe tenant \#1 is able to take resources from tenant \#2 by launching more threads to send requests. \textit{HBase does not prevent throughput interference among tenants which use different thread number.}

In Figure \ref{subfig:interference:hotreg}, tenant \#1 runs the regular hotspot workload and tenant \#2 runs the uniform workload. Both use 50 threads. Throughput violation of tenant \#1 goes above 60\% while tenant \#2's is only 50\%. Furthermore, tenant \#1 receives similar HDFS throughput as tenant \#2 and tenant \#1's cache occupancy is only 10\% higher than tenant \#2's. This indicates tenant \#2 may take some cache space from tenant \#1 which causes tenant \#1 to read from HDFS and thus degrades its performance. \textit{HBase fails to isolate resources among workloads with different resource demands.}

Figure \ref{subfig:interference:exhotbig} shows the results where tenant \#1 runs the extreme hotspot workload with 50 threads and tenant \#2 runs the uniform workload with 200 threads. The throughput violation of tenant \#1 exceeds 90\%. Its cache occupancy is only about 20\% which explains why its throughput drops significantly. In contrast, tenant \#2 only suffers 10\% throughput violation because its HDFS throughput and cache occupancy are about 5 times higher than tenant \#1's. Compared with Figure \ref{subfig:interference:hotreg}, tenant \#1's throughput is less even it is supposed to read more data from the cache. \textit{HBase's incapability of isolating resources is magnified when different resource demands and thread number coexist for workloads.}

We conclude from the experiments above that 1) the number of threads a tenant uses to connect to HBase and the data access pattern e.g. hotspot can lead to performance interference; 2) interference could occur in different resources in HBase, e.g., cache, disk, or both. 3) cache occupancy and HDFS throughput can indeed reflect workload's resource demands.

\paragraph{Single resource reservation}
A common way to prevent interference is to reserve resources so that a tenant is guaranteed a certain amount of resources. Owning to its simplicity, single resource reservation e.g. bytes delivered, CPU usage, and cache usage, has been used by many people \cite{Pisces,Zeng,Das,A-Cache}. In this section, we study two single resource reservation approaches.
\begin{figure*}[!htbp]
    \centering
    \begin{subfigure}[b]{0.45\textwidth}
        \includegraphics[scale=0.3]{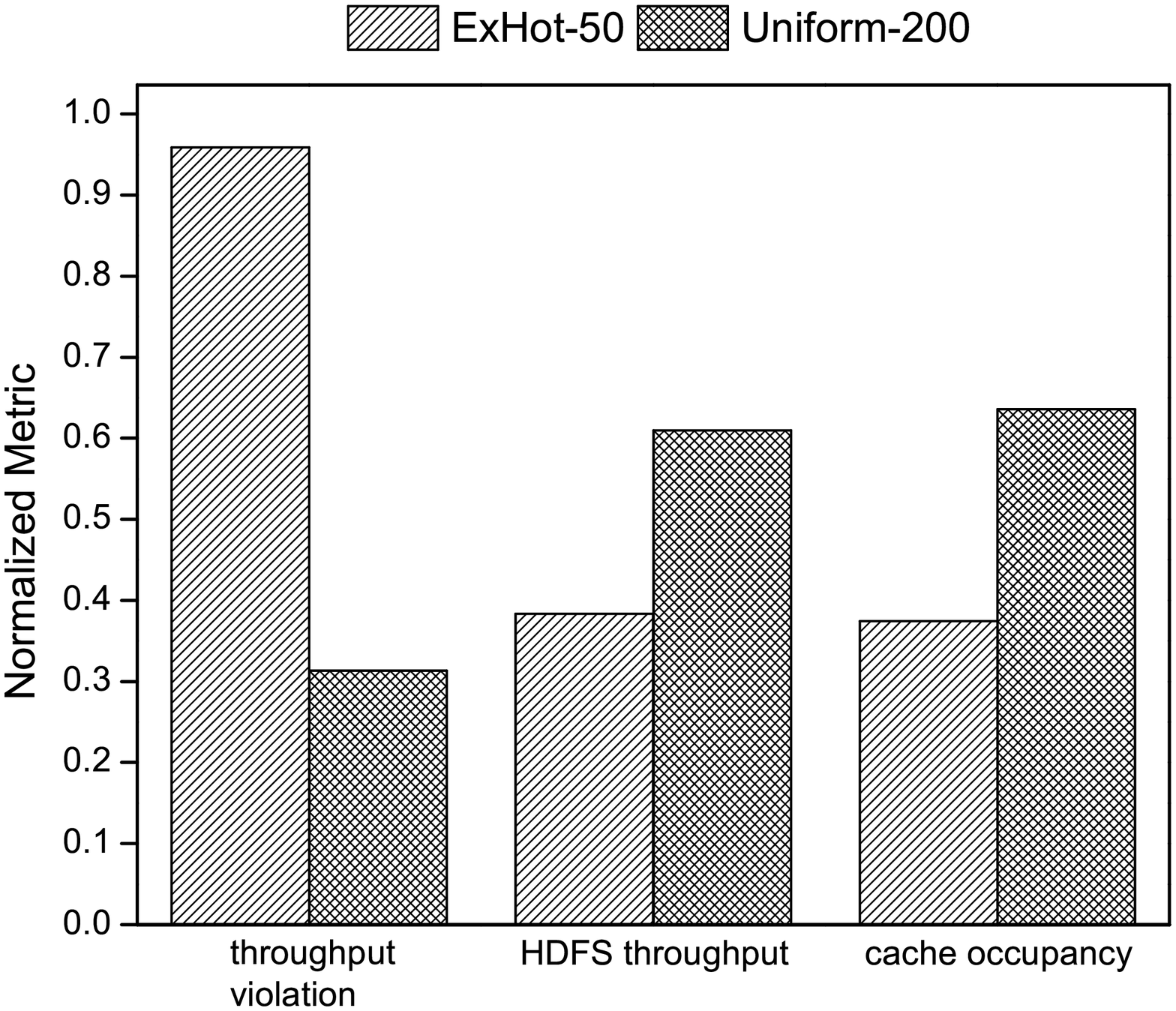}
        \caption{}
        \label{subfig:throughputpartition}
    \end{subfigure}
    ~
    \begin{subfigure}[b]{0.45\textwidth}
        \includegraphics[scale=0.3]{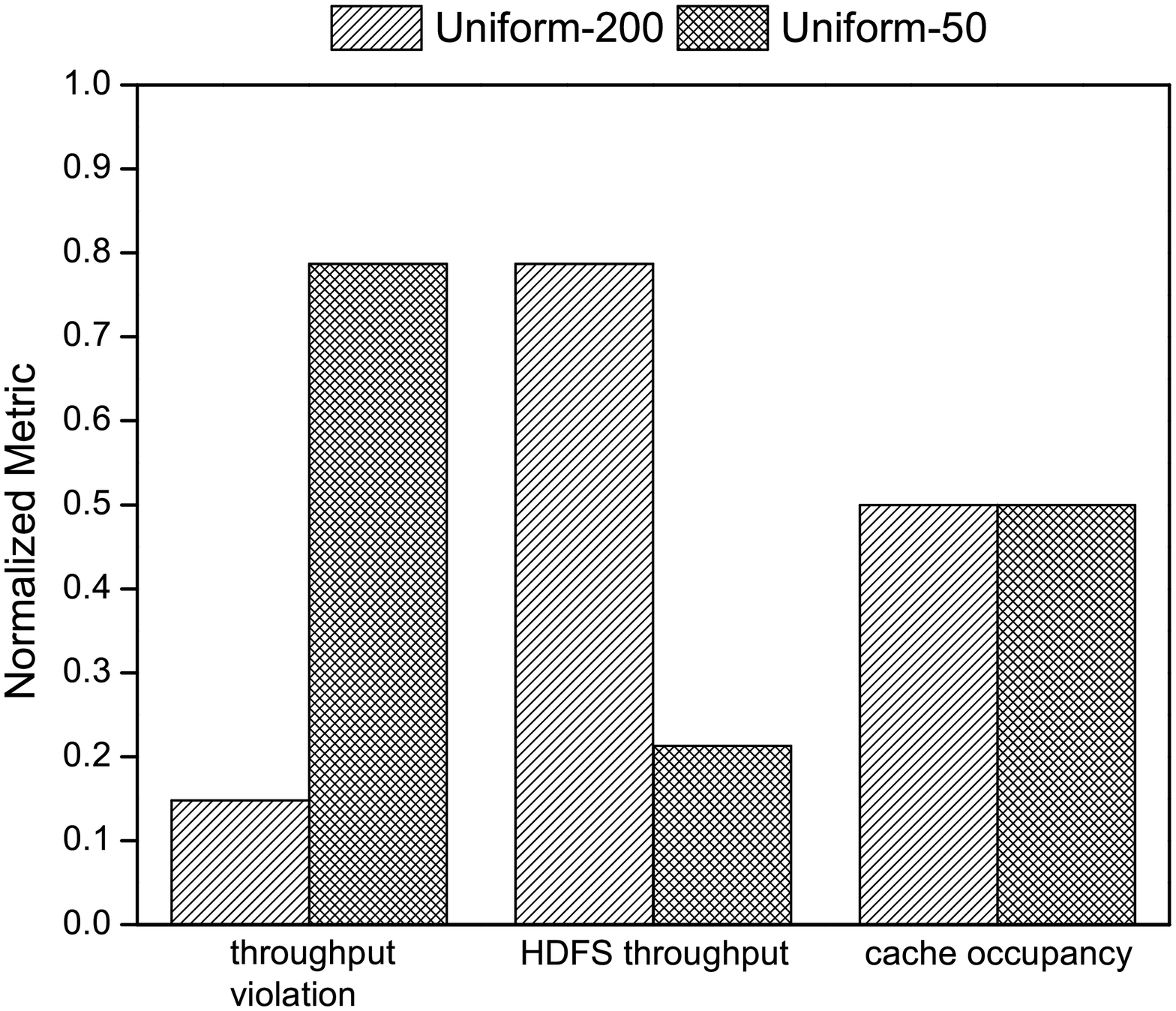}
        \caption{}
        \label{subfig:cachepartition}
    \end{subfigure}
    \caption{Single resource reservation fails to prevent interference for workloads with different resource demands.}
    \label{fig:singlerc}
\end{figure*}

Similar to \cite{Pisces,Zeng}, we use the bytes delivered from HBase as a ``virtual resource'' to represent the underlying resource consumption. Such an approach imposes a maximum number of bytes HBase can deliver to a tenant over a certain period. This approach is able to provide fair access in the case of Figure \ref{subfig:interference:bigthreads}. In this experiment, tenant \#2 runs a uniform workload with 200 threads and tenant \#1 runs an extreme hotspot workload with 50 threads. Figure \ref{subfig:throughputpartition} shows the result. Although both tenants see similar throughput, tenant \#1 suffers more than 90\% throughput decrease when compared with its baseline. Tenant \#2 also has less HDFS throughput and cache occupancy due to the throughput regulation. Therefore, resource approximation using bytes delivered lowers tenant \#2's share but fails to increase tenant \#1's because it cannot identify the cache and disk consumption.

In the second experiment, we divide up the block cache space into half for two tenants so as to provide strong isolation in the cache space as suggested in \cite{A-Cache}. Tenant \#1 uses 200 threads and tenant \#2 uses 50 threads to run the uniform workload simultaneously. Figure \ref{subfig:cachepartition} displays the results. Although both tenants share the cache space equally, tenant \#2 experiences about 80\% throughput violation because tenant \#1 uses more threads to send requests faster and thus deprives tenant \#2's HDFS throughput. Therefore for workloads that are not cache sensitive, cache space isolation does not prevent interference.

In summary, single resource reservation e.g. bytes delivered in the first experiment and cache in the second experiment did not prevent interference across tenants because it ignored the actual resource demands of workloads. The results motivate us to develop a workload-aware reservation approach that targets multiple resources.

\section{Resource Reservation} \label{subtitle:enforcement}
\begin{figure}[h]
  \centering
  \includegraphics[scale=0.9]{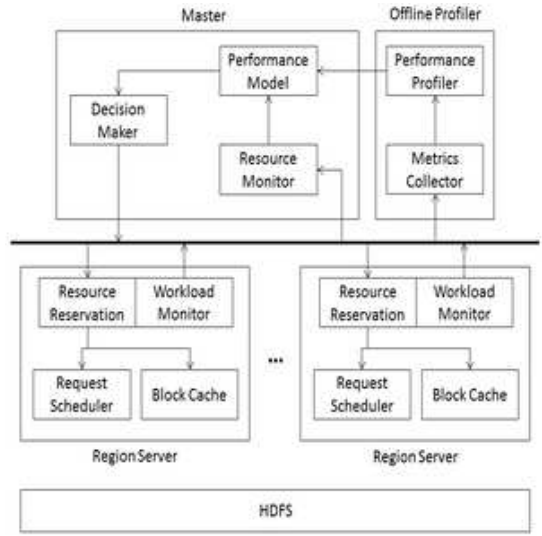}
  \caption{Architecture.}
  \label{fig:architecture-argus}
\end{figure}

We design and implement Argus, a workload-aware resource reservation framework, to prevent performance interference across tenants. Argus is built on HBase's master-slave architecture (see Figure \ref{fig:architecture-argus} for details). The Master collects the resource info from different slave nodes and makes wise resource reservation decisions. The RegionServer serves as an executer to enforce any reservation plans decided by the Master. The disk access to HDFS is controlled by the request scheduler in the RegionServer. The cache access in the RegionServer is enhanced with multi-tenancy support.

Inside a RegionServer, the workload monitor module collects workloads' performance metrics, and the resource reservation module describes current resource reservation policy. Disk reservation is approximated by HDFS throughput reservation as discussed in \ref{subtitle:interference-analysis}. The request scheduler is used to enforce the HDFS throughput reservation given. The block cache shipped with vanilla HBase is made resource reservation aware to support cache reservation as well. The Master has three pieces: 1) Resource monitor that aggregates the workload information from all the RegionServers; 2) Performance model that takes workload information and estimates performance; 3) Decision maker that takes advantage of the performance model to find an optimum resource reservation policy and sends it to all the RegionServers. The performance model relies on an offline profiler that uses linear interpolation to predict the performance.

Given a resource reservation policy, it is critical to enforce resource reservation in each individual node as it provides the basis for cluster-wide resource reservation. As discussed in Section \ref{subtitle:interference-analysis}, we focus on two resources: block cache and disk, both of which will be discussed in the following sections.

\subsection{Block Cache Reservation} \label{subtitle:cache-reservation}
The cache reservation is used to provide strong isolation in the cache space for tenants. In our current prototype, we divide up the entire block cache space into partitions and limit a tenant's cache activities to the cache partition to which it is assigned. Unlike A-Cache \cite{A-Cache} that replaces HBase's default cache replacement, we apply the built-in LRU cache replacement in HBase to the cache partition of each tenant because the cache replacement in HBase has been improved to prioritize the eviction based on the times the blocks are reused. Although strict cache reservation can provide strong isolation, it can result in poor cache utilization if some tenants may not use up their reservations. We classify such situations into two categories: some tenants change their access patterns e.g. they change from hotspot access to random access; some tenants slow down their request rates. We discuss the details and present the solutions in Section \ref{subtitle:elastic-reservation}.

\subsection{Disk Reservation}
Owing to the reasons mentioned in Section \ref{subtitle:interference-analysis}, we use HDFS throughput to approximate the disk usage. We design the request scheduler in the RegionServer instead of in the Hadoop Distributed File System (HDFS) because many of the file systems including HDFS are not designed with multi-tenancy in mind: multi-tenancy enforcement is carried out by the application built on top of the file system.

As stated in Chapter \ref{title:relatedwork}, there are two types of scheduling approaches that approximate the generalized processor sharing (GPS) model \cite{GPS} to provide fair sharing. One is virtual time based approximation and the other one is quanta based approximation. To understand which approach may work well in the context of multi-tenancy in HBase, we study weighted fair queuing (WFQ) \cite{GPS}, which is a virtual time based scheduler, and deficit round robin (DRR) \cite{DRR}, which is a quanta based scheduler. We experimentally compare these scheduling approaches in terms of fairness and efficiency.

\subsubsection{A. Approaches for Request Scheduling}
Each tenant is assigned a queue to hold its requests. WFQ schedules requests among queues according to their finish time. To lower the computation cost of estimating request time, the notion of virtual time is used to order requests. Each request is tagged with a virtual start time and a virtual finish time. Equations \ref{eq:virStart} and \ref{eq:virFinish} show how they are calculated \cite{FairQueuing-Wiki,WeightedFairQueuing}, where $S_{i}^{n}$ and $F_{i}^{n}$ are the virtual start time and the virtual finish time for the $n^{th}$ request of tenant $i$ respectively, $v(t)$ is the virtual time for real time $t$, $L_{i}^{n}$ is the size of the $n^{th}$ request and $r_{i}$ is the share of tenant $i$.
\begin{equation} \label{eq:virStart}
S_{i}^{n} = max(v(t), F_{i}^{n-1})
\end{equation}

\begin{equation} \label{eq:virFinish}
F_{i}^{n} = S_{i}^{n} + L_{i}^{n} \times r_{i}
\end{equation}

The virtual start time is the maximum of current virtual time and the virtual finish time of the last request. The virtual finish time is based on an estimate of how long the request will take. The estimate assumes a linear relationship between request length and virtual time. The complexity of WFQ in each scheduling round is $O(log(n))$ as it needs to select the request with the smallest virtual finish time from $n$ queues in a min-heap.

DRR is a variant of weighted round robin \cite{WRR} that uses quanta (sometimes called tokens or credits) to throttle requests. DRR associates each tenant with a credit account. To schedule a request, the scheduler takes some credits off from the tenant's account according to the size of the request. Eventually a tenant's credit account will exhaust and need to be refilled. There are two refill strategies: refill the accounts periodically; refill when tenants are either exhausted i.e. not enough credits or inactive i.e. no pending requests. Periodic refill can improve utilization as it does not need to wait until other tenants meet the refill criterions.

\subsubsection{B. Comparison}
The goal of the evaluation is to assess the fairness and efficiency of various scheduling approaches in the context of throughput reservation. We implemented WFQ and DRR as the request scheduling approach in HBase respectively. DRR requires some adaptions and extensions in the context of HBase. First, we have the scheduler interpret the credits as the bytes read from or written to HDFS. It assumes there is a linear function that translates the credits to the underlying resources usage, mainly disk access, in HDFS. Second, upon the arrival of a read request, the scheduler does not know how much data will be read from HDFS. The scheduler simply uses an average size over a sliding window, which has the bytes read of 10 previous requests, as the bytes needed for upcoming requests. We also use this as a prediction in WFQ. Krebs et al. discuss more advanced prediction options \cite{Krebs}.

The evaluation environment is the same as the one in Section \ref{subtitle:interference-analysis}. A 28-node HBase cluster is used. Two YCSB clients run on two additional nodes to send uniform read-only workloads. One uses 50 threads while the other one uses 200 threads. Target throughput is set as a large number to allow the client to send as many requests as possible. Since we focus on disk throughput, we disable the block cache in HBase to eliminate its impact. Similar to \cite{Das}, to quantify fairness, we use the \textit{Jain index} (\textit{J-index}) defined in equation \ref{eq:j-index} where $v_{i}$ is the throughput violation of tenant $i$ and can be expressed as $v_{i} = (b_{i} - t_{i})/b_{i}$; $b_{i}$ is the baseline throughput; and $t_{i}$ is the observed throughput. The baseline is established when the workload is run delicately in the cluster. \cite{vfair} also measures fairness through comparing the actual throughput to the baseline throughput. The value of $J$ varies between $1$, where the violation of each tenant is the same, to $1/n$, where one tenant gets the largest $v_{i}$ while other tenants' $v_{i}=0$. The denominator of equation \ref{eq:j-index} will never be zero as long as there are competitions among tenants. If all tenants run slower than they should be, there is no need to impose constraints through the scheduling.
\begin{equation} \label{eq:j-index}
J(v_{1}, v_{2}, \dots, v_{n}) = \frac{(\sum_{1 \leq i \leq n}{v_{i}})^{2}}{n \times \sum_{1 \leq i \leq n}{v_{i}^{2}}}
\end{equation}

Besides fairness, we also consider efficiency which is defined as the average of throughput violation for all tenants in equation \ref{eq:efficiency}. The larger value of J-index and E indicates better fairness and higher efficiency respectively.
\begin{equation} \label{eq:efficiency}
E(v_{1}, v_{2}, \dots, v_{n}) = 1 - \frac{\sum_{1 \leq i \leq n}{v_{i}}}{n}
\end{equation}

Table \ref{tab:scheduling-comparison} summaries the values of the J-index and E for WFQ, DRR and a no scheduling approach. For fairness, the no scheduling approach yields the worst fairness (lowest J-index) in face of tenants running workloads with different thread numbers. DRR outperforms WFQ. We think it is because WFQ assumes requests with the same size take the same time to be processed which does not hold in our experiments. In fact, we observed a large time variant for requests with the same size. \cite{Multi-Resource} also evidences that such a variant in a single node file system setting leads to failure of fairness enforcement. For efficiency, the no scheduling approach has the highest value while DRR has the lowest number. We attribute that to the constraints the scheduler imposes. Section \ref{subtitle:micro-argus} presents more details about the tradeoff of fairness and efficiency for DRR. With the above results, we can conclude that DRR is able to provide stronger resource isolation which results in better fair access than WFQ does. In the rest of this paper, we focus on the usage of DRR.
\begin{table}[!htbp]
    \centering
    \caption{Comparison of different scheduling approaches.}
    \label{tab:scheduling-comparison}
    \begin{tabular}{|c|c|c|c|c|}
      \hline
      \textbf{Metric} & \textbf{NoSchedule} & \textbf{WFQ} & \textbf{DRR} \\
      \hline
      \emph{J-Index} & 0.708 & 0.874 & 0.996 \\
      \hline
      \emph{Efficiency} & 0.513 & 0.493 & 0.481 \\
      \hline
    \end{tabular}
\end{table}
%\begin{table}[h]
%    \centering
%    \caption{Statistics of request latency. A tenant sends 100,000 read requests with 50 threads. The minimum latency, maximum latency and average latency are reported.}
%    \label{tab:request-time-statistic}
%    \begin{tabular}{|c|c|c|}
%      \hline
%      \textbf{Minimum} & \textbf{Maximum} & \textbf{Average} \\
%      \hline
%      5.497 ms & 423.651 ms & 82.043 ms \\
%      \hline
%    \end{tabular}
%\end{table}

\begin{algorithm}
    \caption{Request Scheduling Algorithm}\label{alg:fs-scheduling}
    \begin{algorithmic}[1]
        \State $credit_{i}$ is the current credits in tenant $i$'s credit account.
        \State $est_{i}$ is the estimation of bytes a request reads from or written to HDFS for tenant $i$.
        \State $actual_{i}$ is the actual bytes a request reads from or written to HDFS for tenant $i$.

        \Procedure{Schedule}{}
            \For{each tenant $i$}
            \State $est_{i}$ $\leftarrow$ BytesEstimation(tenant $i$)
            \While{$credit_{i} \geq est_{i}$ and tenant $i$'s queue is not empty}
                \State $credit_{i} \leftarrow credit_{i} - est_{i}$
                \State $est_{i}$ $\leftarrow$ BytesEstimation(tenant $i$)
            \EndWhile
            \EndFor
        \EndProcedure

        \Procedure{Refund}{}
            \If{request is served from cache}
                \State $credits_{i} \leftarrow credit_{i} + est_{i}$
            \Else
                \State $credits_{i} \leftarrow credit_{i} + (est_{i} - actual_{i})$
            \EndIf
        \EndProcedure

        \Procedure{Refill}{}
            \State Redistribute credits assignment if slow tenants exist.
        \EndProcedure
    \end{algorithmic}
\end{algorithm}

To integrate the block cache into DRR, we introduce a refund procedure that refunds credits later if the request can be served from cache. The refund procedure can also refund positive credits if the amount of bytes is overestimated or negative credits if it is underestimated. Algorithm \ref{alg:fs-scheduling} describes the adaption of DRR. The \textit{Schedule} procedure runs in the background to schedule requests from different tenants' queues in a round robin fashion. The \textit{Refund} procedure is invoked when a request finishes. The \textit{Refill} procedure refills tenants' credit accounts periodically and boosts some of the tenants' credits if necessary. The details of it will be discussed in Section \ref{subtitle:elastic-reservation}.

\subsection{Elastic Reservation} \label{subtitle:elastic-reservation}
Neither the cache reservation or disk reservation would be efficient if some tenants did not use up their resources reserved, because both reservations are applied statically without any elasticity. Static resource holding will lead to inefficiency as some of them may be idle and cannot be used by other tenants in need.

There are two cases when a tenant does not use up its reservation. One is when its access pattern does not need much of the resource reserved. For example, a random access workload does not need cache very much, neither does a hotspot workload need disk resource. Therefore, reservation has to consider workload resource demands, i.e. workload-aware. We will present the solution of workload-aware reservation in Section \ref{subtitle:planning}. The other one is when a tenant slows down its throughput (called slow tenants). To deal with such a situation, we redistribute the resources. Specifically, redundant resources from the slow tenants will be taken away and distributed evenly among tenants that are in need.

We establish an expected throughput as a reference to detect if a tenant slows down. We obtain the baseline throughput by running the workload in a dedicated manner. Then the expected throughput is calculated by dividing the baseline throughput with the number of tenants. We assume both the size of cache and the number of credits reserved are linear to the throughput. In the current prototype, a tenant will give away 10\% of its cache and credits to other tenants in need if it slows down every 10\%. Busy tenants will share evenly the cache and credits given away. The \textit{Refill} procedure in Algorithm \ref{alg:fs-scheduling} implements the aforementioned reallocation. In the prototype, it runs every 2 seconds to refill the credit accounts and determines if cache and credit adjustments are needed. If there are slow tenants, it will adjust the credits and notify the cache module for cache resizing accordingly. To deal with the case where a slow tenant may bump up its throughput later, we allow slow tenants to retain the same credit amount they had before the credit redistribution even they may not need them. The cache reservation is reset periodically (3 minutes in the current prototype) and runs with equal reservation for a short period (30 seconds in the current prototype) so as to give slow tenants a chance to increase the throughput. A more accurate way of detecting slow tenants as well as reallocating cache and credits among tenants is in the future work.

\section{Reservation Planning} \label{subtitle:planning}
We have described the mechanisms used to reserve resources. We also described the cases where some tenants may not use up their reservations. The elastic reservation approach mentioned in Section \ref{subtitle:elastic-reservation} adjusts reservation in a monotonic way and is not suitable to handle the case where tenants have different resource demands. Because resource usages are not independent, \emph{e.g.} increasing cache allocation may decrease the disk usage and vice versa, reservation of this kind requires a model that reflects the dependency between different resources. In this section, we discuss the reservation planning used to decide how much resource to reserve for each tenant according to its demands dynamically.

\subsection{Problem Formalization}
\begin{figure*}[!htbp]
   \begin{subfigure}[b]{0.32\textwidth}
        \includegraphics[scale=0.25]{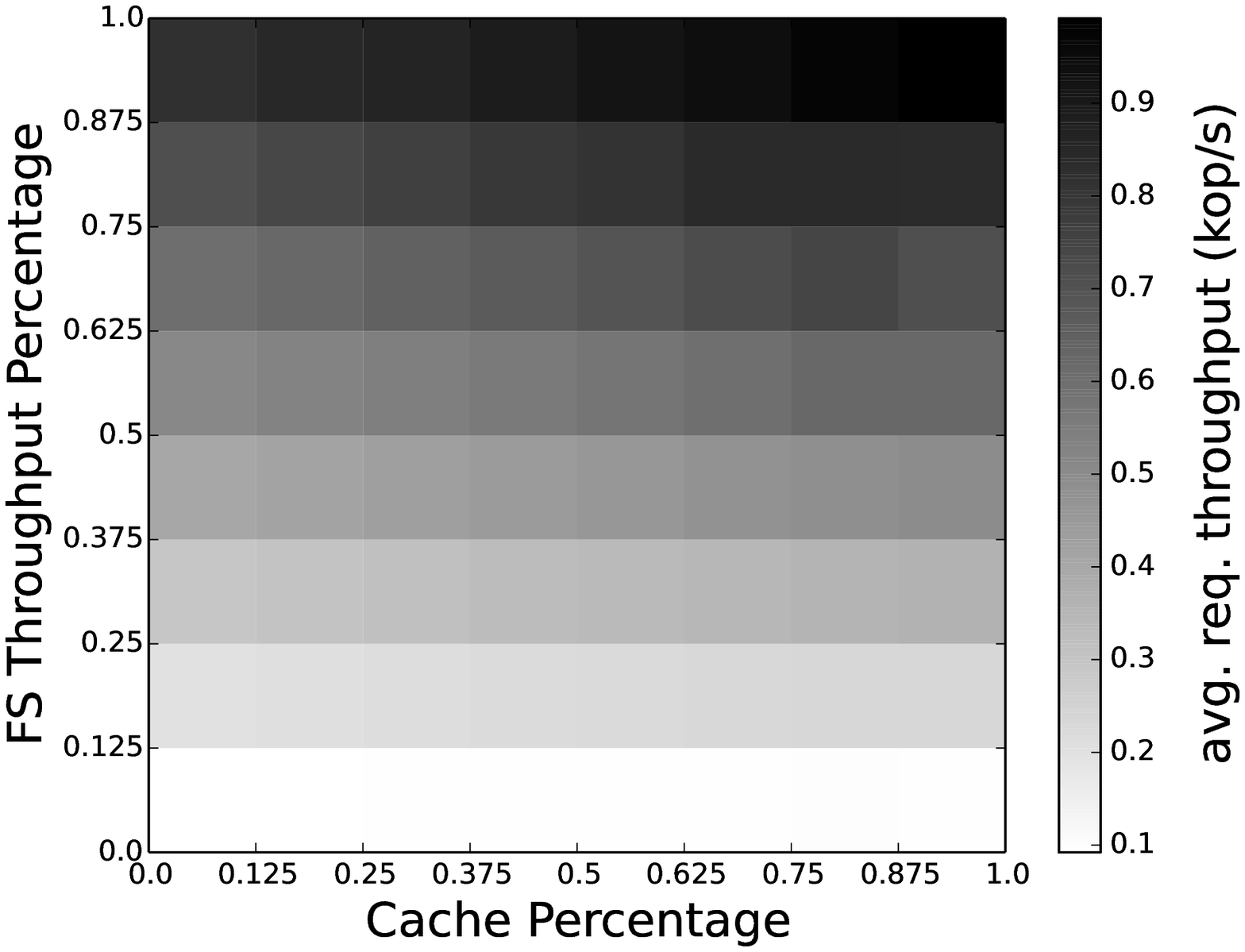}
        \caption{Uniform workload.}
        \label{subfig:reg-heat}
    \end{subfigure}
    ~
    \begin{subfigure}[b]{0.32\textwidth}
        \includegraphics[scale=0.25]{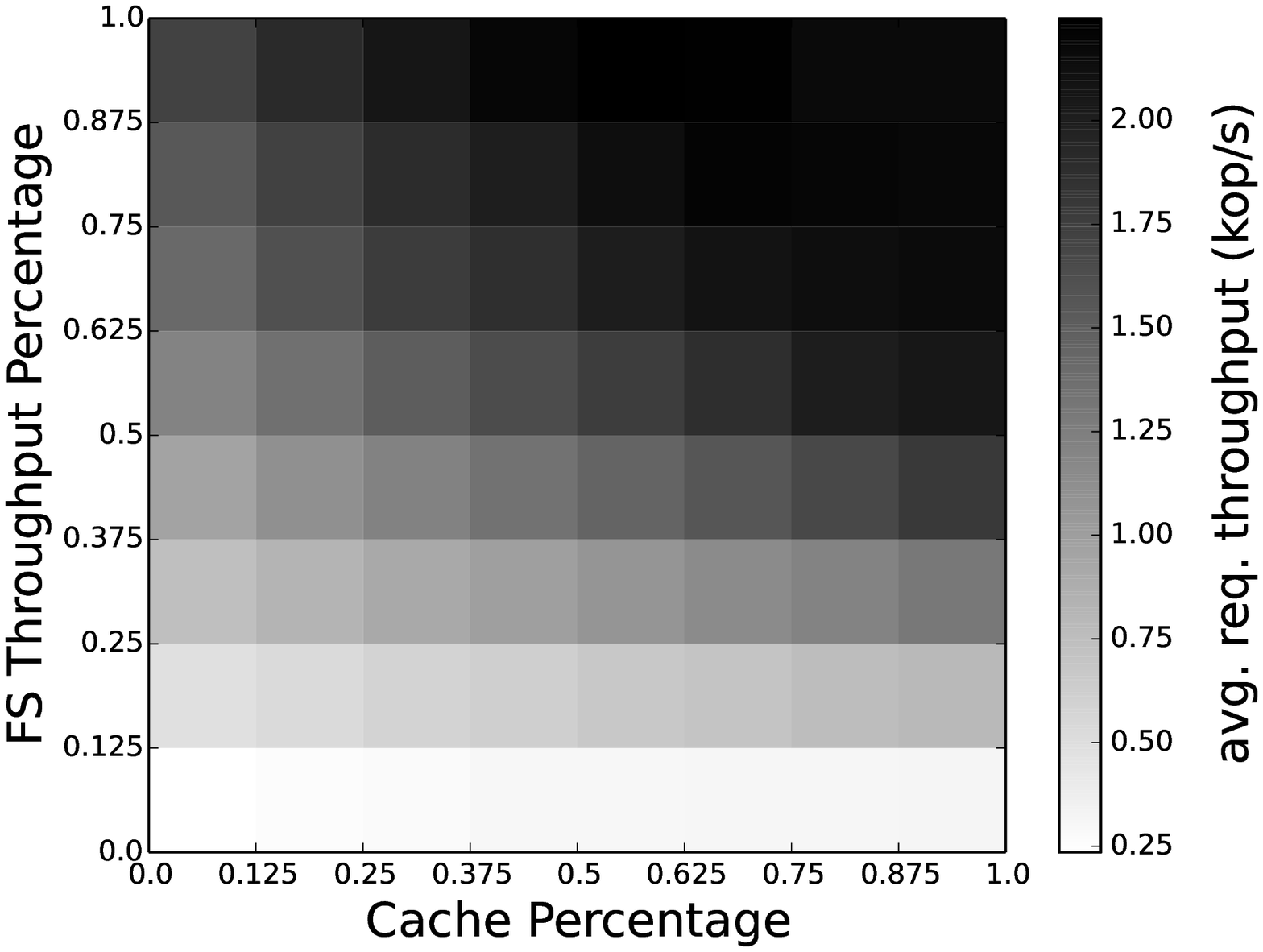}
        \caption{Regular hotspot workload.}
        \label{subfig:hot-heat}
    \end{subfigure}
    ~
    \begin{subfigure}[b]{0.32\textwidth}
        \includegraphics[scale=0.25]{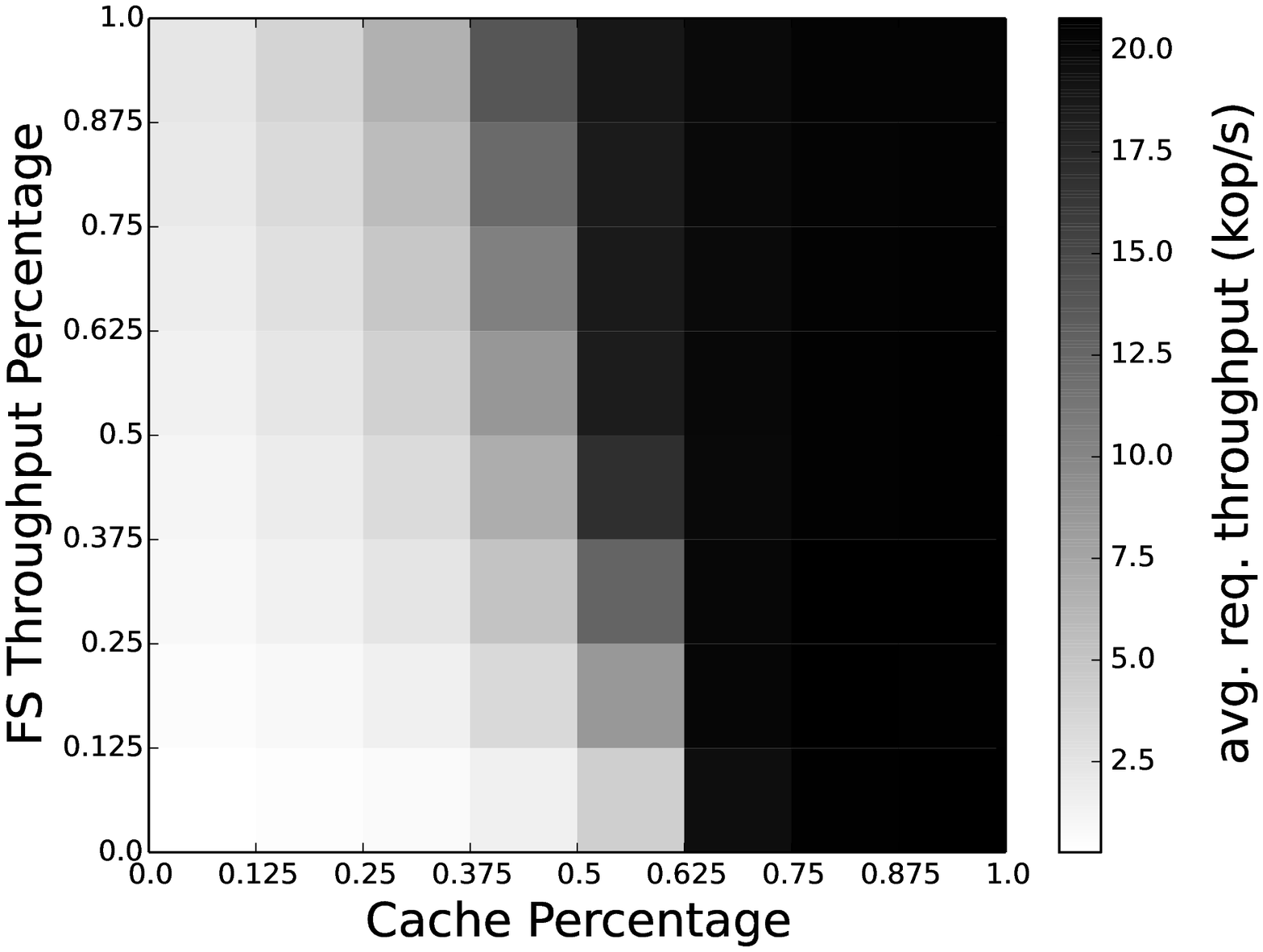}
        \caption{Extreme hotspot workload.}
        \label{subfig:exhot-heat}
    \end{subfigure}
    \caption[Throughput varies with different resource reservations.]{Throughput varies with different resource reservations. Each heat map shows experiments on a range of cache reservation percentage (x-axis) and HDFS throughput reservation percentage (y-axis). The magnitude of the color is shown in the legend on the right.}
    \label{fig:heatmap}
\end{figure*}

Setting the reservation evenly among tenants may not fully utilize the resources. We experimentally demonstrate that the same reservation could yield a different throughput when used for different workloads. We run the three workloads defined in Section \ref{subtitle:interference-analysis}. We vary the percentage of cache reservation and HDFS throughput reservation in a range from 0.125 to 1.0 in interval of 0.125. For example, 0.5 of cache and 0.25 of HDFS throughput means we reserve 50\% total cache space and 25\% total credits of HDFS. Each experiment only runs one workload in the cluster. Figure \ref{fig:heatmap} plots the results.

In Figure \ref{subfig:reg-heat}, the uniform workload is disk sensitive. Increasing HDFS throughput reservation increases its throughput significantly while increasing cache reservation does not. In Figure \ref{subfig:hot-heat}, the regular hotspot workload needs both disk and cache. It achieves maximum throughput when the reservations of HDFS throughput and cache are close to 1. In Figure \ref{subfig:exhot-heat}, the extreme hotspot workload is cache sensitive. Increasing cache reservation significantly improves throughput. In summary, results in Figure \ref{fig:heatmap} not only validate the reservation mechanisms presented in Section \ref{subtitle:enforcement} but also motivate the need of workload-aware reservation planning. For example, in the case of multiple uniform workloads, it is better to reserve resources equally among tenants. While in the case of uniform workloads mixed with extreme hotspot workloads, it is better to reserve more cache space for extreme hotspot workloads and more HDFS throughput for uniform workloads.

The resource reservation planning in Argus is done by the decision maker in the master node. In the following discussion, we first formalize the decision problem, and then present a solution based on the hill climbing algorithm with offline training models. Notations in Table \ref{tab:notations} are used for discussion.
\begin{table}[h]
    \centering
    \caption{Notations.}
    \label{tab:notations}
    \begin{tabular}{|l|l|}
      \hline
      \textbf{Notations} & \textbf{Description} \\
      \hline
      $n$ & The number of tenants \\
      \hline
      $m$ & The number of resources \\
      \hline
      $b_{i}$ & The baseline throughput for tenant $i$ \\
      \hline
      $t_{i}$ & The actual throughput for tenant $i$ \\
      \hline
      $v_{i}$ & The throughput violation for tenant $i$ \\
      \hline
      $r_{ij}$ & The amount of resource $j$ reserved for tenant $i$ \\
      \hline
      $M_{j}$ & The total amount of resource $j$  \\
      \hline
    \end{tabular}
\end{table}

The goal of the reservation planning is to let tenants have fairness in terms of throughput violation and keep the efficiency as much as possible. We use equation \ref{eq:j-index} to describe fairness and equation \ref{eq:efficiency} to represent efficiency. We need to maximize $J$ to achieve similar throughput violation among tenants as well as $E$ to improve efficiency. Combining equation \ref{eq:j-index} and \ref{eq:efficiency}, we can express the optimization objective in \ref{eq:objective}. The definitions of $J$ and $E$ are also reminded here.
\begin{eqnarray} \label{eq:objective}
D(v_{1}, v_{2}, \dots, v_{n}) = \alpha \times J + (1-\alpha) \times E \\
J(v_{1}, v_{2}, \dots, v_{n}) = \frac{(\sum_{1 \leq i \leq n}{v_{i}})^{2}}{n \times \sum_{1 \leq i \leq n}{v_{i}^{2}}} \nonumber \\
E(v_{1}, v_{2}, \dots, v_{n}) = 1 - \frac{\sum_{1 \leq i \leq n}{v_{i}}}{n} \nonumber
\end{eqnarray}
$\alpha$ is a variable between 0 and 1. It indicates how much impact $J$ and $E$ have in the decision procedure. In the current prototype, we set it to 0.5. Function $D$ is the objective that the decision maker needs to maximize.

With the baseline throughput and resources reserved, we express the resource reservation planning problem as a constrained optimization below.
\begin{equation} \label{eq:optimization}
    \begin{aligned}
        \max_{(r_{11}, \dots, r_{nm})} \quad
        D(v_{1}, \dots, v_{n}) & \\
        \text{s.t.} \quad
        \sum_{1 \leq i \leq n}r_{ij} &= M_{j} \\
        v_{i} &= (b_{i} - f_{i}(r_{i1},\ldots,r_{im}))/b_{i} \\
        i &= 1,2,\ldots,n \\
        j &= 1,2,\ldots,m
    \end{aligned}
\end{equation}
$f_{i}$ is the performance function that represents the throughput for tenant $i$'s workload, given a set of resource reservations $(r_{i1},\ldots,r_{im})$. Figure \ref{fig:heatmap} indicates that different workloads will have different performance functions. The solution for the above problem is a list of resource reservations $(r_{i1},\ldots,r_{im})$ for each tenant that maximizes the value of $D$. In this chapter, we only consider cache and HDFS throughput. So there are two resources in equation \ref{eq:optimization}, i.e., $m=2$.

\subsection{Solution}
Solving the problem above requires the knowledge of various performance functions. Instead of inferring to an analytic form of the function, which is difficult and error prone \cite{Multi-Resource}, we simply use regression to interpolate the function on some sample data collected by running the workload offline with different cache and HDFS throughput reservation percentages. Figure \ref{fig:heatmap} shows the mapping between resource allocations and throughput demonstrates linearity thus we use linear regression for the interpolation. The profiler in Figure \ref{fig:architecture-argus} conducts the interpolation and generates the performance function. The key repeat ratio is used to characterise a workload and is calculated as the number of keys repeatedly accessed divided by the total keys accessed within a certain period. For an incoming workload, Argus associates it with a performance function which has the closest key repeat ratio to the workload. If a workload changes its access pattern, say from uniform to hotspot, the decision maker is able to detect the change and adjust the performance function associated to the workload accordingly. The resource reservation may be changed consequently.

In more complicated scenarios, a workload may be characterised by more than one attribute e.g. read/write/scan percentages, priority, etc. The corresponding performance function may also have more resources involved e.g. \textit{MemStore} size in memory, write ahead log size, etc. In this chapter, we focus on the hotspot access pattern and use the key repeat ratio to characterise a workload and utilize block cache as well as HDFS throughput as input resources for the performance function. The modeling of workload and its performance function in more complicated scenarios are left for future work.

To find an optimum solution for equation \ref{eq:optimization}, we use the stochastic hill climbing algorithm to search the feasible space. The basic idea is to let the searching algorithm start from a potential solution point and pick a neighbor according to the probability distribution of all the neighbors. The distribution is based on the value of each neighbor state. Since the search space is infinite as both cache size and HDFS throughput are continuous variables, we discretize them into 20 equal pieces, i.e. the basic unit of share is 0.05 given the entire share is 1.0. Notice the number of pieces must be larger than the number of tenants in the current prototype to guarantee each tenant can get its share of the resources. Generally speaking, the finer the discretization is, the better result the searching algorithm can yield, but the longer it takes to search. Dynamically adjusting the granularity of discretization according to tenant number and accuracy is an ongoing work. The algorithm starts the searching from $(r_{1j},\ldots,r_{nj}), j= 1, \ldots, m$ where $r_{ij} = r_{kj}, i \neq k$. That is equal reservation. In addition, during the search, we only change one variable i.e., either cache size or HDFS throughput. This limits the number of neighbors to explore and makes the search tractable.

\subsection{Limitations} \label{subtitle:argus-limitation}
To simplify the scenario, the current prototype makes the following assumptions. First, we do not consider data locality and assume every byte read from HDFS consumes the same amount of resources. Reading from local disk is faster and consumes less resources than from remote nodes. Second, we do not address the read-write interference and assume both read and write consume the same amount of resources. HBase follows the LSM \cite{LSM} design which periodically flushes data from \textit{MemStore}, an in-memory structure, to fixed size files in HDFS and merges those files later in the \textit{compaction} procedure which incurs extra I/O in the background. Third, we assume both the data and the request are evenly distributed. Last but not least, we assume the access pattern does not change in a short time which allows sufficient time for Argus to detect and react.

\section{Evaluation} \label{subtitle:reservation-evaluation}
Argus is prototyped in HBase 0.94.21. We evaluate it in a 28-node cluster. Each node has a 2.0 GHz dual-core CPU with 4 GB memory and a 40 GB disk. Three nodes are setup as a Zookeeper ensemble, 1 node is setup up with both HDFS master and HBase master, and the other 24 nodes are setup as HRegion servers and HDFS data nodes. The block cache size is configured to be 1200 MB and the number of RPC handler threads in HBase is set to 30. The HDFS replication factor is set to 1 to conserve disk space. We use the YCSB benchmark \cite{YCSB} to populate the data and generate the workload. After all the data is pre-loaded, we run major compactions to compact the store files. We have HBase to balance the number of regions across nodes. The YCSB clients are run in additional nodes to simulate multiple tenants accessing the system simultaneously. Running YCSB clients on separated nodes can avoid interference on the client side. Each tenant has its own data set in HBase. We use both micro-benchmark and macro-benchmark to evaluate the performance of Argus. The throughput on the client side, i.e., operations per second (ops/sec), is used as a measurement to reflect a tenant's performance on the system. We first present the micro evaluation which mainly focuses on the reservation enforcement, then present the macro evaluation which studies the overall performance in various scenarios.

\subsection{Micro Evaluation} \label{subtitle:micro-argus}
This section presents micro-benchmark results of resource enforcement as it is the fundation of resource reservation. Specifically, we conduct an in-depth analysis of disk reservation to study its impact on fairness and efficiency. Then, we study the stability of resource enforcement in complex scenarios. In addition, we evaluate the effectiveness of elastic reservation. Finally, we assess the overhead introduced by the enforcement approaches.
%This section presents micro-benchmark results of the cache reservation strategies and an in-depth analysis of disk reservation using DRR as they are both the fundamental techniques of resource reservation enforcement. Specifically, these experiments investigate the adaptability of different cache reservation strategies and the overhead as well as the performance tradeoff between fairness and efficiency for disk reservation. Micro-benchmark experiments are run on the same setting mentioned above. For cache reservation experiments, we turn the disk reservation off to avoid its impacts. Similarly, for disk reservation experiments, we disable block cache in HBase to concentrate solely on HDFS usage.

\subsubsection{A. Disk Reservation}
We adapt the deficit round robin algorithm as the request scheduler to enforce disk reservation. The total number of credits per node is a parameter set by the system admin and has an impact on the performance of the scheduler. We start by studying its impact on throughput as well latency, and on fairness as well as efficiency. The block cache in HBase is disabled so that we can concentrate on HDFS usage.

\begin{figure*}[!htbp]
    \centering
    \begin{subfigure}[b]{0.45\textwidth}
        \includegraphics[scale=0.3]{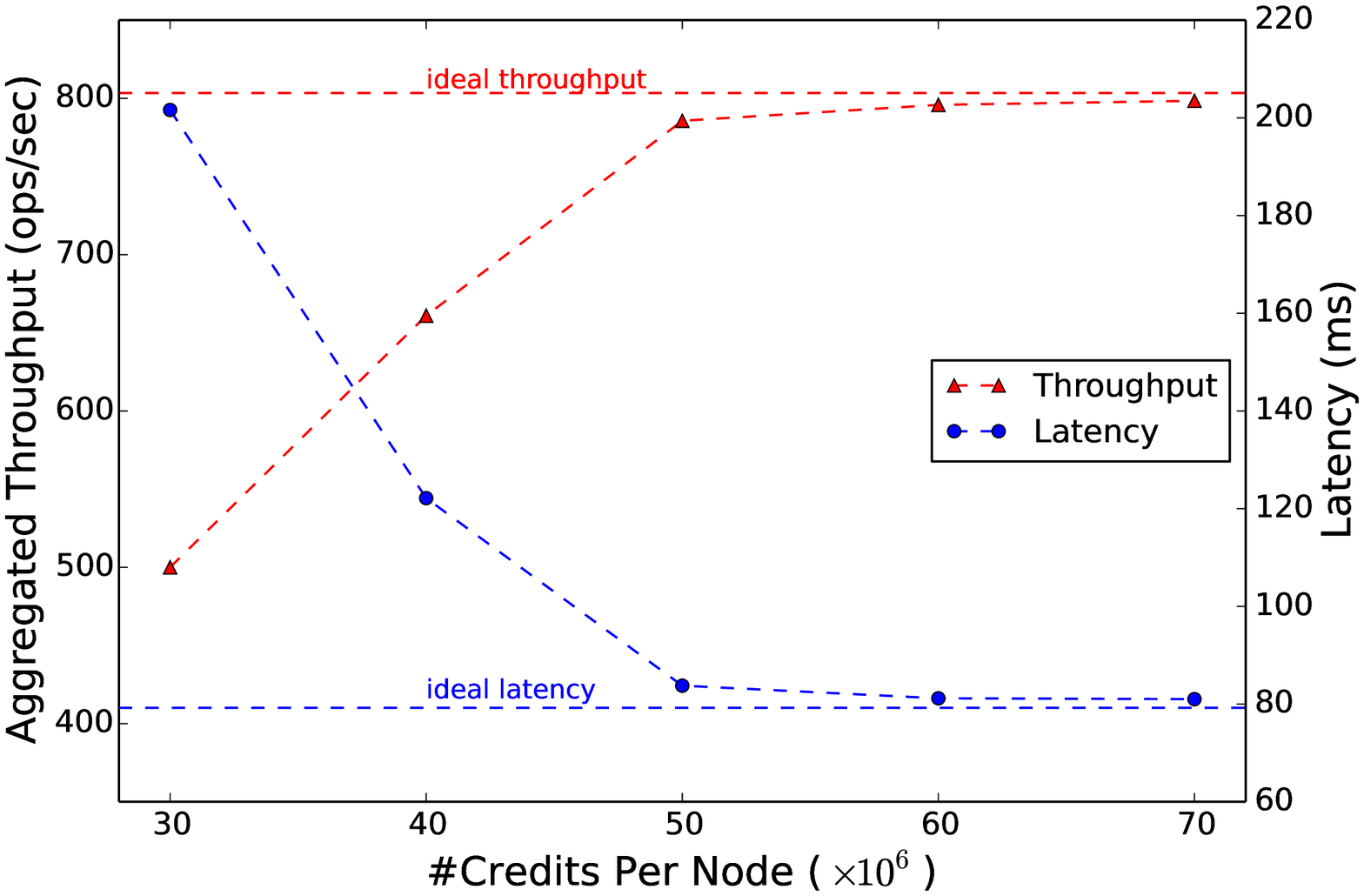}
        \caption{Impact of \#credits on throughput and latency.}
        \label{fig:drr-overhead}
    \end{subfigure}
    ~
    \begin{subfigure}[b]{0.45\textwidth}
        \includegraphics[scale=0.3]{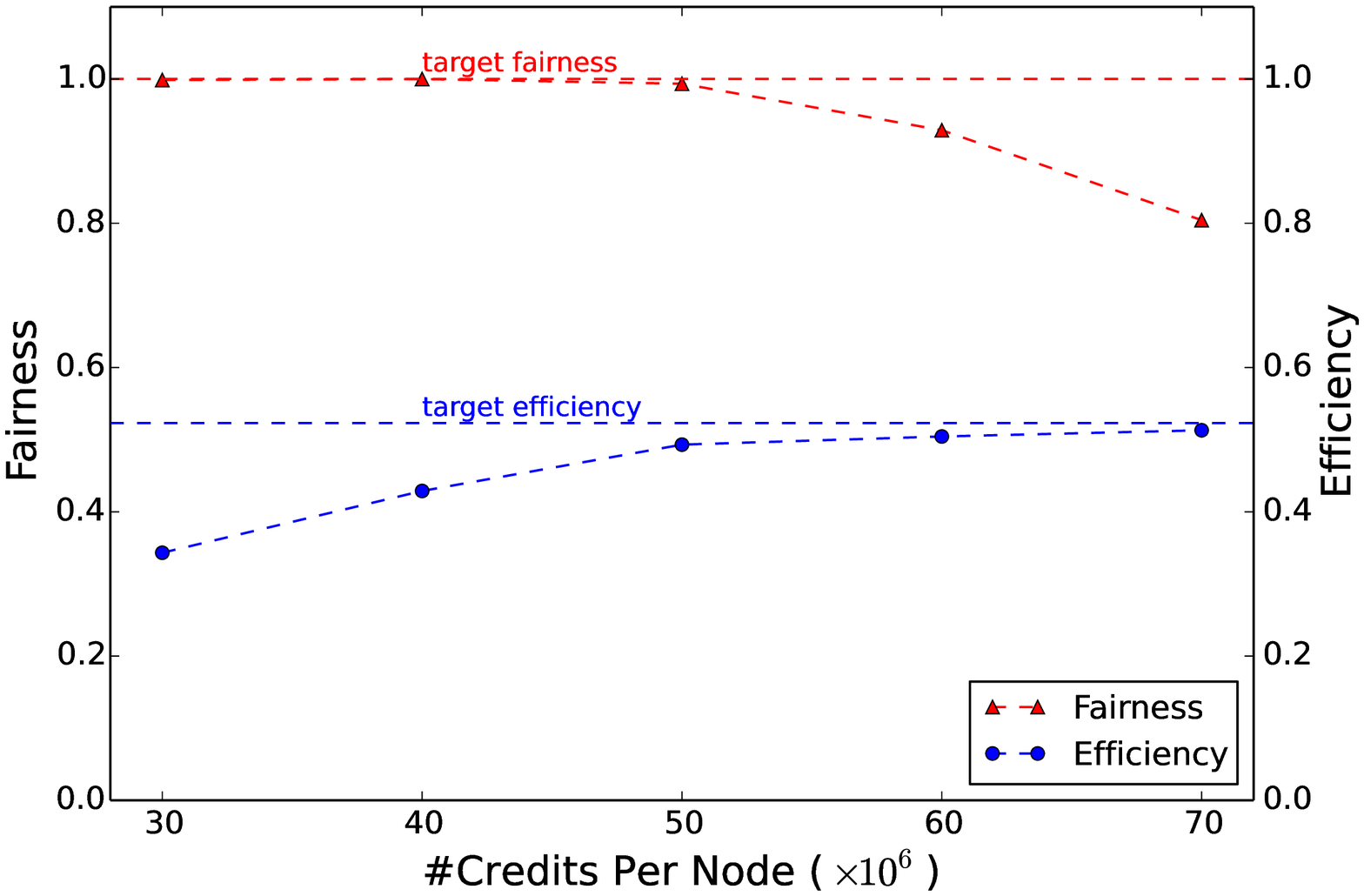}
        \caption{Impact of \#credits on fairness and efficiency.}
        \label{fig:drr-fairness-efficiency}
    \end{subfigure}
    \caption[Impacts of \#credits on disk reservation.]{Impacts of \#credits on disk reservation. The x-axis represents the number of credits allocated in every refill period.}
    \label{fig:micro-disk}
\end{figure*}

\paragraph{Impact on throughput and latency:} We have two tenants with 50 threads to carry out uniform read-only workloads respectively. We report the throughput as well as latency. Throughput is measured as aggregated throughput of tenants and latency is measured as average latency. Figure \ref{fig:drr-overhead} shows the result. The x axis indicates the number of credits allocated to tenants in every refill period (2 seconds in the current prototype) for each node. The ideal throughput and ideal latency are obtained by running the workloads against vanilla HBase. The throughput increases as the number of credits increases from 30 million to 50 million. Afterwards, the throughput gets close to the ideal throughput. Latency has a similar trend. As the number of credits increases, latency decreases until it gets close to the ideal one. In DRR, the number of credits is used to throttle requests sent to HDFS.

On one hand, the larger the number of credits is, the closer the throughput can get to the ideal one. On the other hand, the smaller it is, the more constraints are added to the scheduling. As a result, tenants' throughput degrade dramatically. From Figure \ref{fig:drr-overhead} we can see that both the throughput and latency tend to be stable when the number of credits exceeds 50 million.

\paragraph{Fairness and efficiency tradeoff:} To study the impact of the number of credits on fairness and efficiency, we have two tenants run the uniform read-only workload. One uses 50 threads while the other one uses 200 threads. Figure \ref{fig:drr-fairness-efficiency} displays the result. The fairness is quantified with the throughput violation in equation \ref{eq:j-index}, and the efficiency is measured as the average throughput violation in equation \ref{eq:efficiency}. The target fairness and efficiency are observed by running vanilla HBase and having both tenants use 50 threads. On one hand, when the amount of credits is small (e.g. less than 50 million), the scheduler can achieve target fairness. But at the same time the efficiency is smaller than the target one. On the other hand, when the credits go above 50 million, fairness drops but efficiency increases because the constraints the number of credits imposes become less. In a word, fewer credits tend to have better fairness but lower efficiency, while more credits have worse fairness but higher efficiency. Thus setting the amount of credits is a tradeoff between fairness and efficiency. In the current prototype, we set it as 50 million which sacrifices efficiency a little bit but gives good fairness as the experiments shown above. A more advanced option is to dynamically adjust the credits according to the workload characteristics suggested in \cite{Argon}. We leave that in the future work.

\subsubsection{B. Stability of Reservation Enforcement}
\begin{figure}[!htbp]
  \centering
  \includegraphics[scale=0.34]{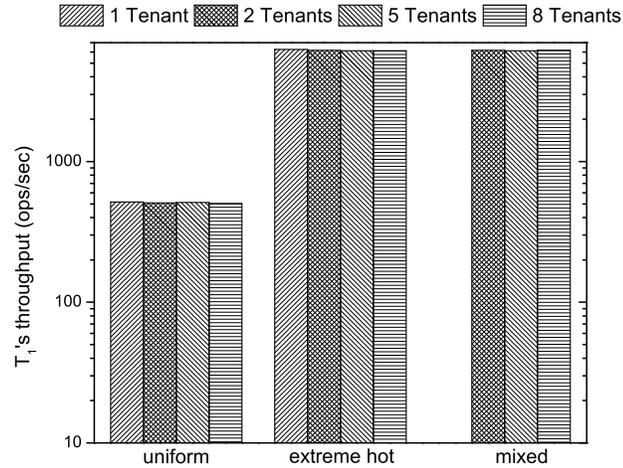}
  \caption{$T_{1}$'s throughput under different workloads and different number of tenants.}
  \label{fig:stability}
\end{figure}

Figure \ref{fig:heatmap} in Section \ref{subtitle:planning} already shows that the resource reservation is able to reserve given HDFS throughput and cache size. To take this one step further, we evaluate its stability in more complex scenarios. We have 3 groups of workloads: uniform, extreme hotspot, and mixed. There are 1, 2, 5 and 8 tenants respectively to run the workloads. Tenants run the same workload in the uniform and extreme hotspot groups while some tenants run uniform workload and others run extreme hotspot workload in the mixed group. In the cases of 1 and 2 tenants, each tenant uses 50 threads. In the cases of 5 and 8 tenants, the thread counts are 50, 50, 100, 200, and 300, and 50, 50, 100, 100, 200, 200, 300, and 300 respectively. The reservation planning is turned off in this evaluation to avoid resource re-allocation. We measure tenant \#1's throughput (represented as $T_{1}$) to see how much it changes in different settings. $T_{1}$ runs the extreme hotspot workload in the mixed workload group. $T_{1}$'s HDFS throughput and cache occupancy percentages are set to $0.5$. The remaining HDFS throughput and cache occupancy percentage, i.e. 0.5, are distributed evenly across other tenants. Figure \ref{fig:stability} displays the results. Notice that for the mixed group, we only report $T_{1}$'s throughput when the number of tenants is at least 2. Each bar represents the throughput of $T_{1}$. Bars with different stripe patterns mean the throughput observed under different tenant number settings. We can see that for all different workload groups, $T_{1}$ achieves consistent performance even when the number of tenants increases from 1 to 8 and $T_{1}$ is mixed with different workloads. Thus Argus is able to preserve throughput in a multi-tenant environment by enforcing resource reservations.

\subsubsection{C. Elastic Reservation}
Real world workloads usually have dynamics and require the storage system to be able to automatically adapt. The kinds of dynamics include the change of requesting rate and access pattern e.g. hotspot, uniform, etc. The elastic reservation approach is applied to block cache and disk reservation to dynamically adjust the reservation when some tenants decrease their throughput. The reservation planning is used to plan the reservation when tenants' workload have different resource demands.

\begin{figure*}[!htbp]
    \centering
    \begin{subfigure}[b]{0.48\textwidth}
        \includegraphics[scale=0.32]{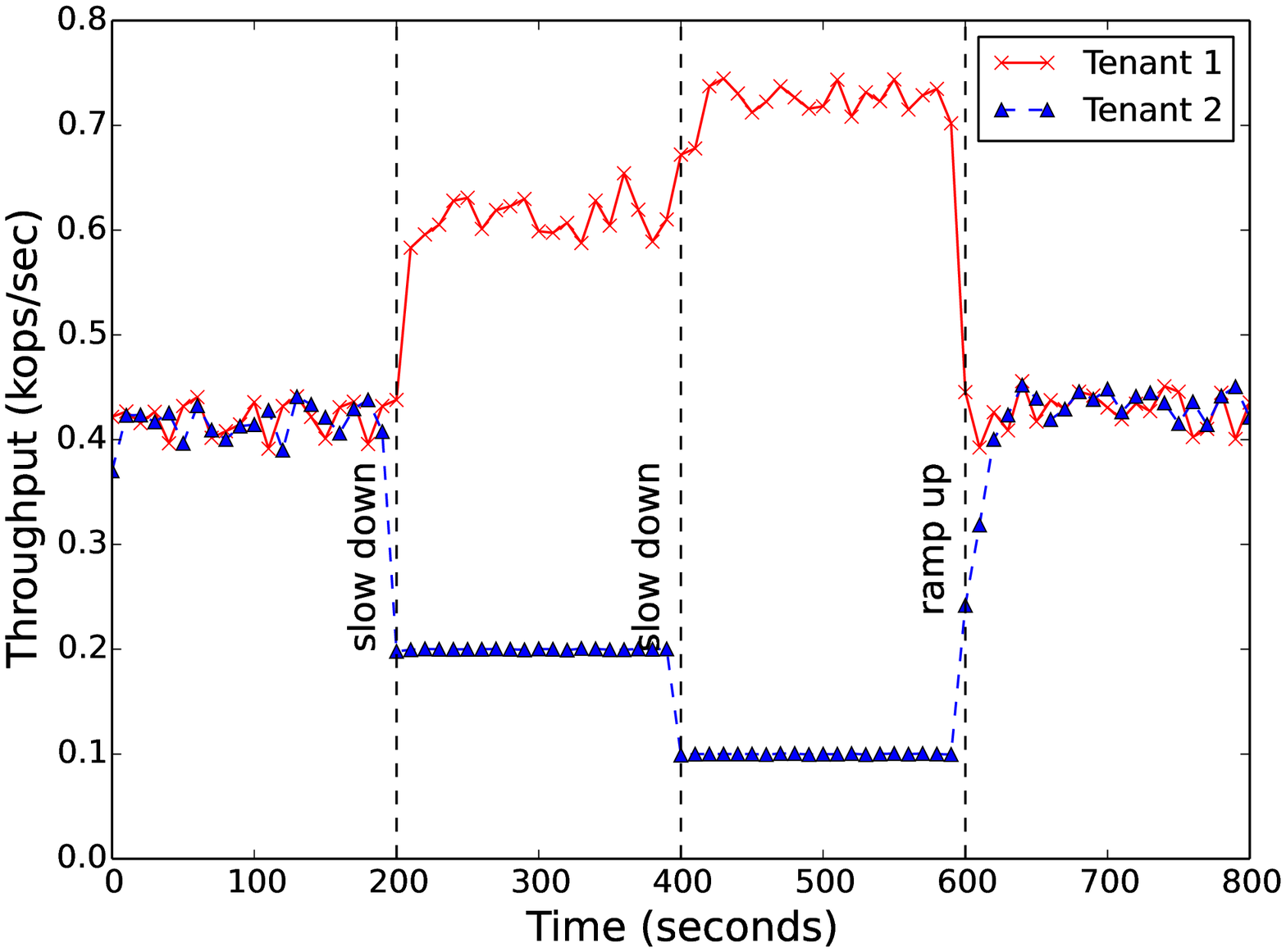}
        \caption{Tenant \#1 is allowed to have higher throughput when tenant \#2 slows down.}
        \label{fig:drr-slowtenants-throughputs}
    \end{subfigure}
    ~
    \begin{subfigure}[b]{0.48\textwidth}
        \includegraphics[scale=0.32]{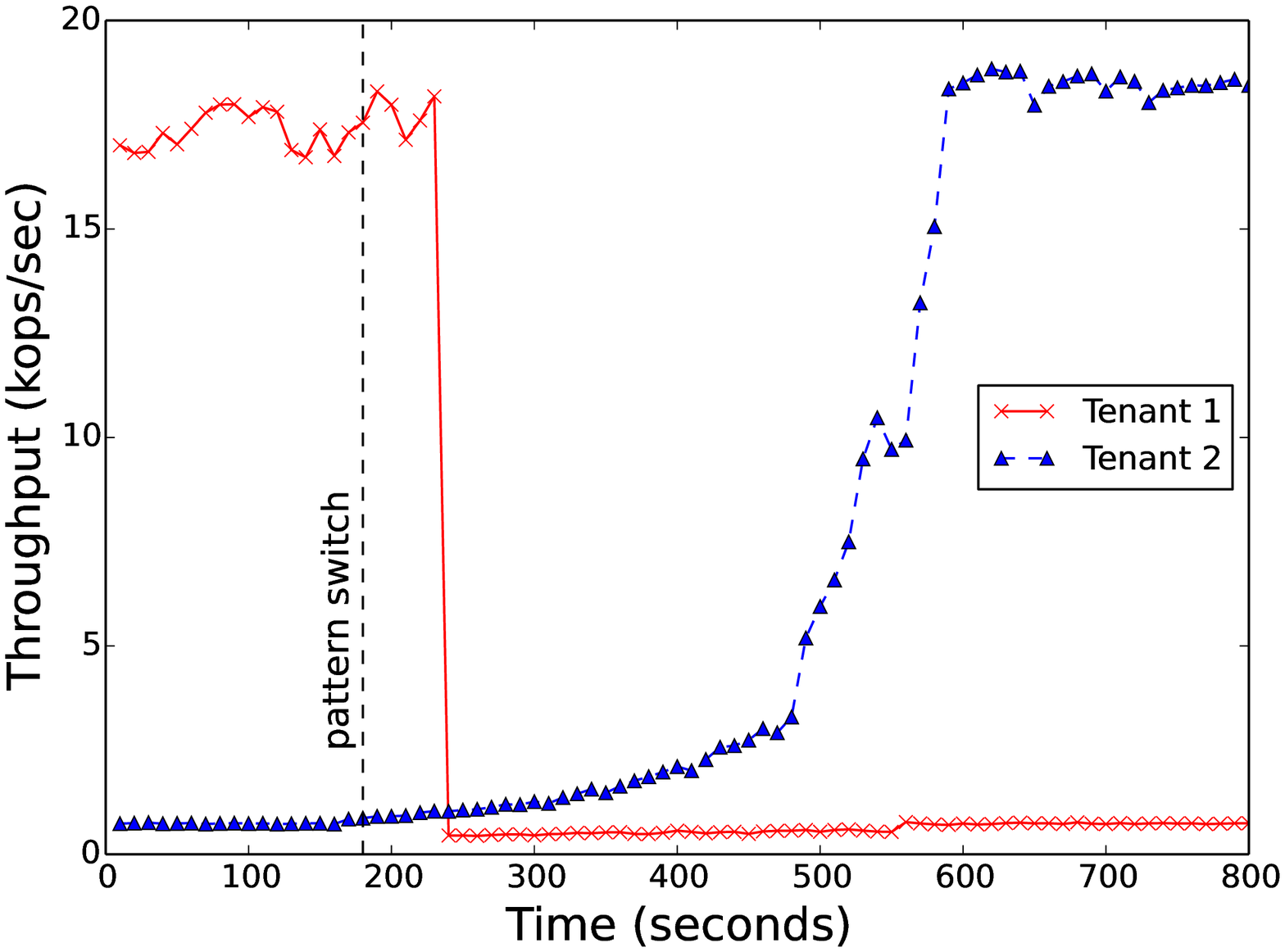}
        \caption{Argus reacts when workload changes its access pattern over time.}
        \label{fig:slowtenants-planning}
    \end{subfigure}
    \caption{Argus dynamically adjusts its reservations.}
    \label{fig:argus-dynamic}
\end{figure*}

We evaluate the elastic reservation by having two tenants run the uniform workloads and one of them depresses its request rate. They both use 50 threads. Figure \ref{fig:drr-slowtenants-throughputs} shows the throughput as a function of time. Both tenants fair share the system in the first 200 seconds. Between the 200th second and the 400th second, tenant \#2 decreases its throughput to 200 ops/sec. During that period, Argus is able to raise the throughput of tenant \#1 to about 600 ops/sec. Then tenant \#2 further decreases its throughput to 100 ops/sec in the next 200 seconds. Because of the elastic reservation, Argus can further increases tenant \#1's throughput to about 700 ops/sec. Finally, tenant \#2 increases its throughput and both tenants start seeing similar throughput after the 600th second. There are three throughput changes.

Next, we examine Argus' capability of dealing with workloads changing access patterns. Two tenants switch their access patterns at some point to simulate access pattern change in Figure \ref{fig:slowtenants-planning}. Tenant \#1 starts with an extreme hotspot workload and tenant \#2 runs an uniform workload. At the 180th second, they switch the access pattern i.e., tenant \#1 now runs uniform workload and tenant \#2 runs extreme hotspot workload. It takes the decision maker about 60 seconds to realize the change because it operates every 60 seconds. Once the decision maker adjusts the resource reservation for both tenants, tenant \#2's throughput increases gradually. Finally at about 580th second, tenant \#2 achieves its maximum throughput. Tenant \#2 takes around 300 seconds to get to the maximum because it needs to replace most of the cache items in the block cache. The reservation planning plays an important role to adjust the reservation according to workloads' demands. The results above show that Argus can handle workloads varying in throughput and access pattern efficiently.

\subsubsection{D. Overhead}
Finally, we study the overhead introduced by resource reservation. For the disk reservation, there are two sources where overhead comes from. One is the implementation of the DRR algorithm and the queues associated with it. Figure \ref{fig:drr-overhead} shows that when the number of credits per node is 70 million, which does not impose any constraints to the request scheduling, about 2\% overhead is observed for throughput and about 1\% for latency. We attribute that to the implementation of DRR and queues. The other source of overhead is the number of credits used in the DRR algorithm. For 50 million credits in Figure \ref{fig:drr-overhead}, we observe around 3\% overhead for throughput and 2\% for latency. To study the overhead of the cache reservation, we disable the disk reservation and have two tenant run the uniform, hotspot and extreme hotspot workloads respectively with the same number of threads. The overhead is ignorable for all three workloads (less than 1\%).

To get the total overhead of a fully functioning system, we enable both the cache and the disk reservations in Argus. We have two tenants run the same workloads above with 50 threads, and compare the aggregated throughput with the ones generated from vanilla HBase. We observe approximately a 5\% throughput decrease for the uniform workload, a 4\% drop for the hotspot workload, and ignorable overhead for the extreme hotspot workload. Compared with the overhead obtained solely from disk reservation, the overhead from cache and disk reservation increases. We think the usage of cache magnifies the throughput overhead.

\subsection{Macro Evaluation} \label{subtitle:macro-argus}
We study the overall performance of Argus in more complex scenarios in this section. We first present the performance of Argus by running the three workloads used in our interference analysis in Section \ref{subtitle:interference-analysis}. Then we further assess Argus in versatile workloads with scan operations as well as write operations. Next, we investigate the effectiveness of reservation planning and its capability of dynamic workload handling. Finally, we compare Argus with A-Cache \cite{A-Cache}, a HBase based system that aims at preventing cache interference.

\subsubsection{A. Overall Performance} \label{subtitle:overall-argus}
\begin{figure*}[!htbp]
   \begin{subfigure}[b]{0.3\textwidth}
        \includegraphics[scale=0.22]{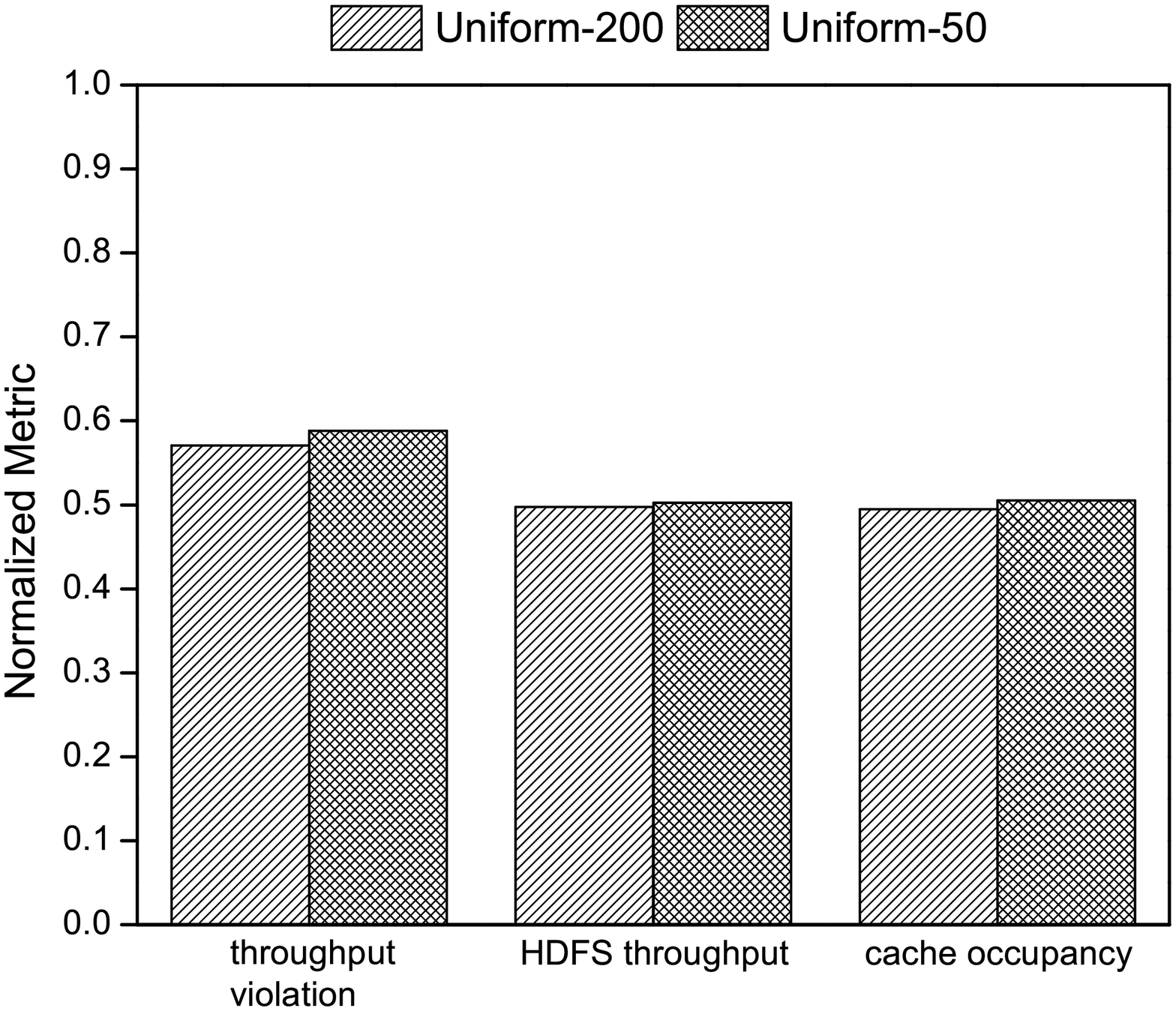}
        \caption{Two uniform workloads with different number of threads.}
        \label{subfig:overall-uniform}
    \end{subfigure}
    ~
    \begin{subfigure}[b]{0.3\textwidth}
        \includegraphics[scale=0.22]{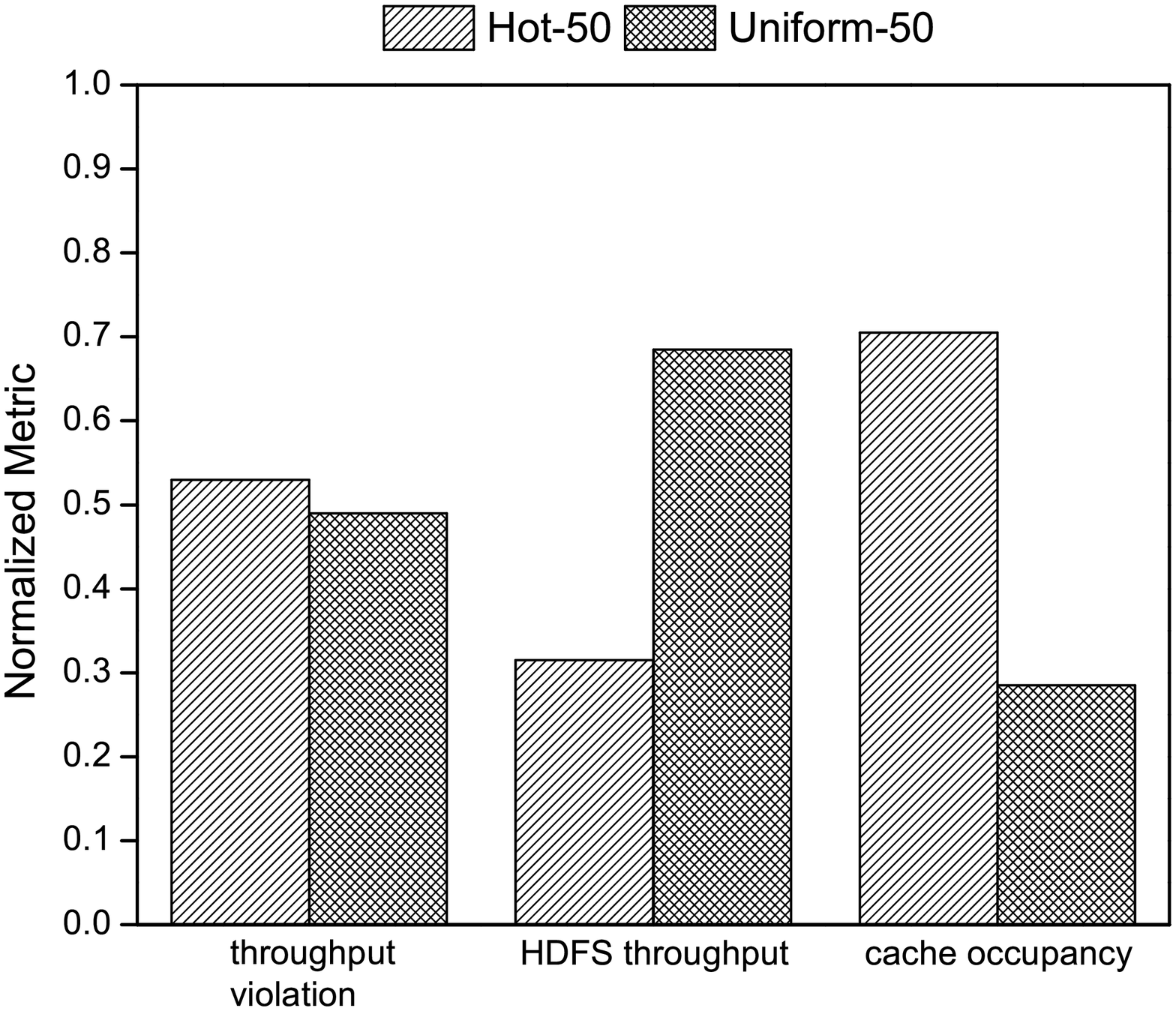}
        \caption{Regular hotspot workload mixed with uniform workload.}
        \label{subfig:overall-hot}
    \end{subfigure}
    ~
    \begin{subfigure}[b]{0.3\textwidth}
        \includegraphics[scale=0.22]{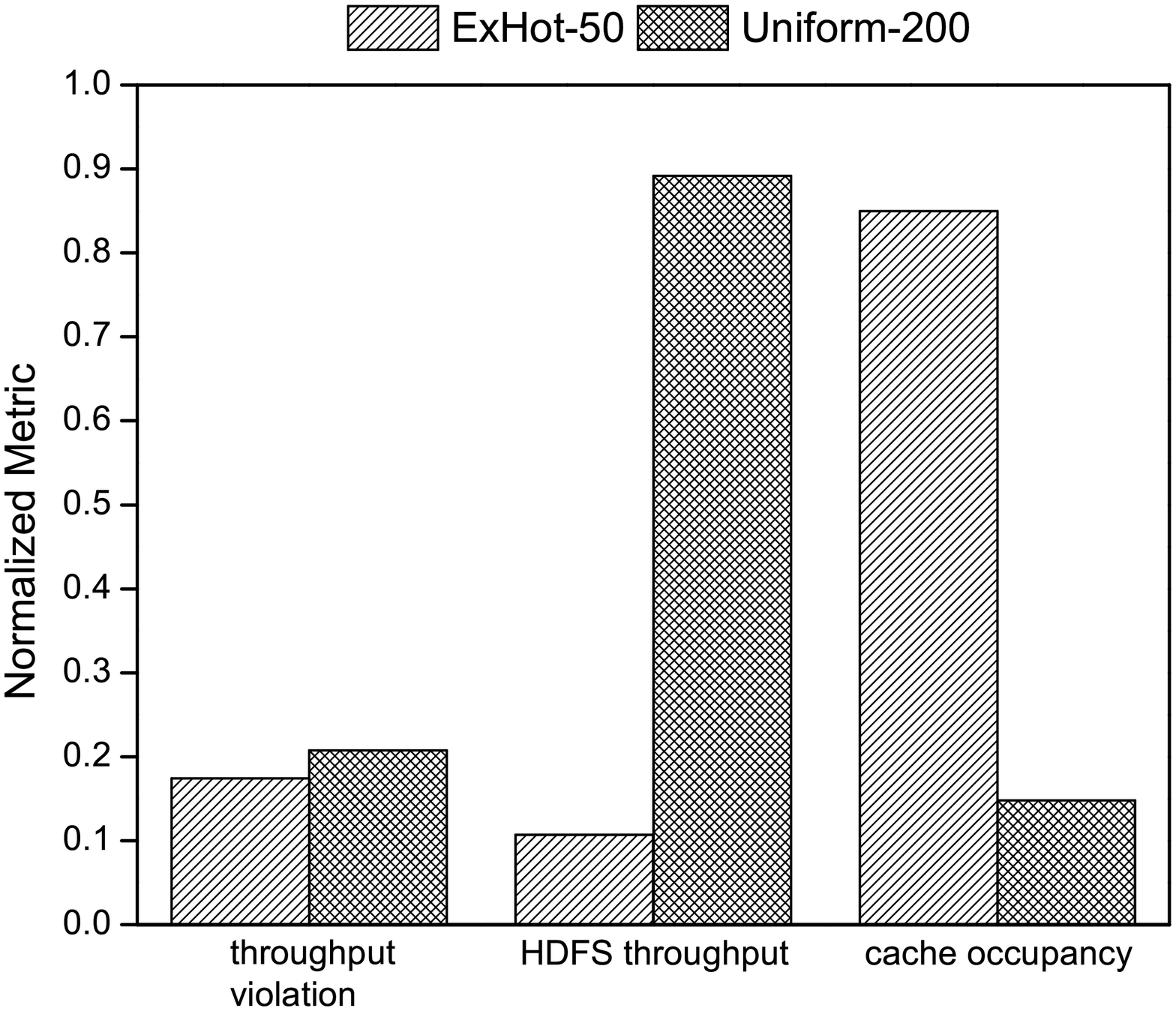}
        \caption{Extreme hotspot workload mixed with uniform workload.}
        \label{subfig:overall-exhot}
    \end{subfigure}
    \caption{Overall performance of different workloads.}
    \label{fig:overallperf-argus}
\end{figure*}

The throughput (ops/sec) is measured as an average over an 800 second period. We use the J-index in equation \ref{eq:j-index} to measure the fairness in terms of throughput violation and the value of $D$ (D-score) in equation \ref{eq:objective} to measure the improvement. To get a stable output, we report the results after a ramp-up time of 300 seconds. We clear the file system cache after each run. We evaluate whether Argus can prevent interference in cases where workloads with different access patterns are mixed together. We run the interference experiments described in Section \ref{subtitle:interference-analysis}. The purpose is threefold. First, we test if Argus is able to handle tenants with different thread numbers. Second, we want to see if Argus can differentiate cache and disk demands from different workloads. Third, we evaluate how Argus reacts when tenants with various number of threads and resource demands coexist. The workload is labeled with the same convention used in Figure \ref{fig:interference}. We pre-load 80,000,000 rows to each tenant. Each row is about 1.2 KB. Figure \ref{fig:overallperf-argus} plots the throughput violation, normalized HDFS throughput and cache occupancy. Table \ref{tab:improvement-reservation} summaries the J-index, D-score and compares them with the ones obtained from vanilla HBase.
\begin{table}[h]
    \centering
    \caption{Performance interference compared with vanilla HBase. Values in parentheses are the performance numbers from vanilla HBase.}
    \label{tab:improvement-reservation}
    \begin{tabular}{|l|c|c|}
      \hline
      \textbf{Workload} & \textbf{J-index} & \textbf{D-score} \\
      \hline
      Uniform-50 and Uniform-200 & 0.999 (0.746) & 0.715 (0.620) \\
      \hline
      Hot-50 and Uniform-50 & 0.997 (0.968) & 0.741 (0.704) \\
      \hline
      ExHot-50 and Uniform-200 & 0.995 (0.662) & 0.909 (0.656) \\
      \hline
    \end{tabular}
\end{table}

For throughput violation, both tenants see roughly the same in all experiments which indicates that Argus is able to prevent interference by proper resource reservations. The extreme hotspot workload mixed with the uniform workload in Figure \ref{subfig:overall-exhot} experiences the least violation because the two workloads do not compete for the same resource i.e., tenant \#1 mainly needs cache and tenant \#2 mainly wants disk. The scheduler realizes the resource demands and reallocates resources based on needs. For HDFS throughput and cache occupancy, both tenants have similar share in Figure \ref{subfig:overall-uniform}. As tenant \#1 becomes more hotspot oriented (i.e. Hot-50 and ExHot-50), tenant \#2 running the uniform workload takes a larger share of HDFS throughput and gives away more share in cache occupancy to tenant \#1. This further verifies Argus's capability of identifying workloads' resource demands and adjusting resource reservation. In Figure \ref{subfig:overall-hot} and \ref{subfig:overall-exhot}, the regular hotspot workload takes a larger share of HDFS throughput and less of a share of cache occupancy than the extreme hotspot workload does which matches our expectation because the regular hotspot workload demands more disk accesses and less cache visits. From Table \ref{tab:improvement-reservation}, it is clear that Argus outperforms vanilla HBase in terms of J-index and D-score. Notice that for regular hotspot workload mixed with the uniform workload, the J-index from vanilla HBase is close to the one in Argus. We think it is because the block cache replacement algorithm with priority in vanilla HBase is able to identify popular data and avoid evicting them too early, which protects the hotspot workload. But due to the lack of resource reservation, the throughput violation from vanilla HBase is larger (smaller D-score) than Argus' as the D-score indicates.

We next evaluate Argus in versatile workloads by running two additional workloads with 200 threads: scan and read/write workloads, against the read workload. The scan workload reads 100 rows per request while read/write workload sends read and write requests together. The keys accessed in scan and read/write workload follow the uniform distribution. Notice that Argus is the most effective in read operations because it deals with the cache and disk usage, and the write operation does not consume cache in HBase. Table \ref{tab:versatile-reservation} summarizes the results.

Similar to Table \ref{tab:improvement}, Argus achieves better interference isolation than vanilla HBase does. However, the values of J-index and D-score drop for both scan mixed and read/write mixed scenarios. This is especially true for scan mixed workloads, where the J-index values drop over 20\%, implying that throughput violation becomes more serious. We attribute the decrease to the inability of system to isolate tenant resource demands in HDFS. First, a scan request holds resources in HDFS for a relatively long period of time and prevents \textit{Get} requests' access. Because Argus works on top of HDFS, it has no direct control of the resource in HDFS. \cite{Cake} suggests a two-level scheduler that works both at the HBase level and HDFS level to deal with get and scan mixed workloads. Second, writes increase the size of \textit{MemTable} as well as the number of \textit{SSTable} which force reads to consult more SSTable. Additionally, writes will trigger a background procedure called \textit{compaction} that merges multiple SSTables into a single one. Thus the compaction procedure may compete HDFS resources with reads and writes from clients.

It is challenging to deal with the interferences incurred internally e.g. the increase of SSTables and compaction discussed above. It requires a model that can represent the I/O behavior of the system with background procedures running, which is very difficult to infer even in a single node storage system \cite{Libra}. Similar to our offline modeling approach, \cite{Libra} derives a non-linear function that transforms a request to underlying I/O cost by running workload offline. In a word, we feel Argus can be extended to use the approaches in \cite{Cake,Libra} although there are challenges from aggregating additional resources in the workload model. We leave such extensions for future work.
\begin{table}[h]
    \centering
    \caption{Evaluate Argus in versatile workloads. Values in parentheses are the performance numbers from vanilla HBase.}
    \label{tab:versatile-reservation}
    \begin{tabular}{|l|c|c|c|}
      \hline
      \textbf{Workload} & \textbf{J-index} & \textbf{D-score} \\
      \hline
      Uniform-50 and Scan-200 & 0.750 (0.513) & 0.651 (0.589) \\
      \hline
      ExHot-50 and Scan-200 & 0.686 (0.650) & 0.544 (0.447) \\
      \hline
      Uniform-50 and RW-200 & 0.926 (0.894) & 0.592 (0.469) \\
      \hline
      ExHot-50 and RW-200 & 0.995 (0.698) & 0.654 (0.507) \\
      \hline
    \end{tabular}
\end{table}

\subsubsection{B. Resource Reservation Planning}
\begin{figure}[!htbp]
  \centering
  \includegraphics[scale=0.36]{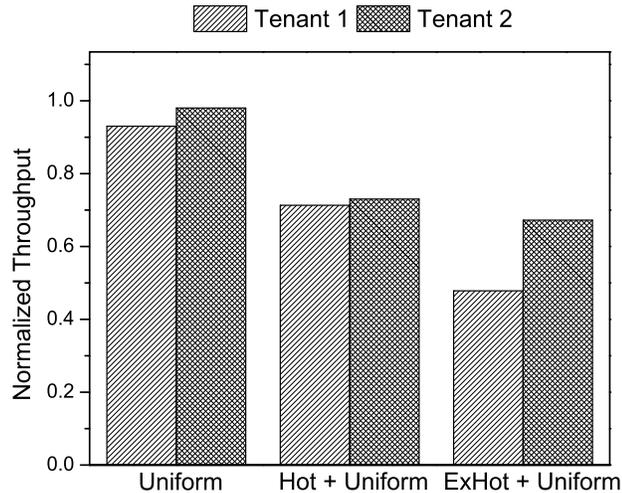}
  \caption{Throughput without planning normalized to throughput with planning.}
  \label{fig:reservation-planning}
\end{figure}

Argus relies on the reservation planning to decide how much resource to reserve for each tenant. We investigate the effectiveness of planning by rerunning the interference experiments in \ref{subtitle:overall-argus} with the planning turned off (i.e. even reservation for HDFS throughput and cache occupancy) We compare the throughput with the corresponding run where the planning is on. Figure \ref{fig:reservation-planning} displays the normalized results. Overall, without planning, the throughput in the uniform workload experiment is similar to the one with planning but it falls behind in the other two experiments. We think it is because both tenants run uniform workloads and the planning does not need to reallocate cache and HDFS throughput. So tenants see similar throughput with the planning turned off.

In the Hot+Uniform and ExHot+Uniform experiments, the throughput of uniform workloads i.e. tenant \#2's workloads, drops when compared with the one in the Uniform experiment. We attribute that to the lack of reservation planning. Tenant \#2 is not able to use some of the HDFS throughput reservation of tenant \#1. Instead, it only relies on the reallocation approach described in Section \ref{subtitle:enforcement} to ``steal'' HDFS throughput from tenant \#1, which is very limited in such cases. On the other hand, tenant \#1 is not able to use the redundant cache space held by tenant \#2. In summary, Argus' resource reservation planning is necessary and effective when workloads with different resource demands coexist in a shared system.

\subsubsection{C. Comparison with A-Cache}
Lastly, we compare Argus with A-Cache \cite{A-Cache} in terms of fairness and efficiency. A-Cache is also developed on top of HBase to prevent performance interference among workloads. It focuses on preventing the cache interference. It uses the cache reuse ratio to represent the cache utilization for each tenant. The reuse ratio is calculated as the ratio of the number of cache blocks visited at least twice to the total number of cache blocks. It estimates how much cache space this workload may need because a cache block only becomes useful when it is visited at least twice. The intuition is a tenant should only have what is needed. Workloads with large reuse ratios deserve more cache space while workloads with small reuse ratios need less.
%A-Cache uses the reuse ratio multiplied by a margin factor as the portion of cache size a tenant needs. For example, a tenant with a 50\% reuse ratio will get 60\% of the cache size it occupies given the margin factor is 1.2.
\begin{table}[h]
    \centering
    \caption{Comparison with A-Cache. Values in parentheses are the performance numbers from A-Cache.}
    \label{tab:acache-comparison}
    \begin{tabular}{|l|c|c|}
      \hline
      \textbf{Workload} & \textbf{J-index} & \textbf{D-score} \\
      \hline
      Uniform-50 and Uniform-200 & 0.999 (0.733) & 0.715 (0.691) \\
      \hline
      Hot-50 and Uniform-50 & 0.997 (0.963) & 0.741 (0.745) \\
      \hline
      ExHot-50 and Uniform-200 & 0.995 (0.691) & 0.909 (0.667) \\
      \hline
    \end{tabular}
\end{table}

We implemented the A-Cache approach and evaluate its fairness as well as improvement with the three workloads used in Section \ref{subtitle:interference-analysis}. Table \ref{tab:acache-comparison} shows the results. From the J-index and D-score values, it is clear that A-Cache is not able to provide the same fairness and efficiency as Argus does. For the extreme hotspot and uniform mix workload and uniform workload with different threads, the J-index values of A-Cache are close to the ones of vanilla HBase in table \ref{tab:improvement-reservation}, which means A-Cache fails to prevent interference. It is because A-Cache only focuses on cache interference and does not take disk usage into account. Therefore, compared with A-Cache, Argus can prevent interference by providing stronger performance isolation.

\section{Real World Applications}\label{title:realworld}
To see how Argus could work for real world applications, we draw on big data text mining of the HathiTrust Research Center (HTRC) \cite{HTRC,DataCapsules}. HTRC provisions for community research text mining of the nearly 14 million digital documents (books, serials, government documents) of the HathiTrust digital repository \cite{HT}. HTRC manages different types of data objects: raw text data, metadata about the books, and derived data \emph{e.g.} term frequency count. This is managed through a single key-value store. All of these data objects are slow changing so workloads against all three are largely read-only, matching the observation in \cite{fb-workload} capturing realistic workloads in NoSQL stores.

The access patterns are different amongst workloads and could thus result in performance interference. Reading from raw text and metadata has evidenced locality as parts of the corpus are more interesting than other parts. Reading from derived data, however, is likely done using a linear scan, which may interfere with the cache contents and access patterns used by workloads over raw text and metadata. Additionally, some of the text analysis may run on the same data set repeatedly \emph{e.g.} topic modeling analysis, advanced machine learning for classification \cite{IJCAI15,AAAI15,IJAMS}, while some of the others may just run for one time, \emph{e.g.} a tag cloud generator may fetch the term frequency from the derived data set. The cache in the topic modeling workload may be interfered by the other one-time workloads where requests are mostly random. Argus can protect the cache by enforcing reservation and allocating more disk resource to the random workload accordingly. Another interference scenario is the number of requests made to the metadata is a lot higher than the number made to the raw text as users may want to investigate enough metadata before studying the text. Argus can prevent the interference from high volume of requests on metadata to workloads on raw text.

\section{Summary}
In this chapter, we characterize multi-tenancy interference in the context of NoSQL data stores. We present Argus, a workload-aware resource reservation framework that prevents interference by enforcing reservation on cache and disk usage. Furthermore, the resource reservation technique is workload-aware. Empirical results show that Argus is able to prevent interference across tenants and adapt to dynamic workloads accordingly.

Future work can go in several directions. We intend to quantify the impact of writes on reads and model the I/O behavior through offline sampling. We want to investigate another resource reservation, i.e. the memory usage for writes. Increasing the size of write buffer will boost the write performance but harms read performance as the size of block cache decreases. It is beneficial to set the sizes of cache and write buffer according to different workload characteristics. Furthermore, we want to extend Argus to other NoSQL solutions beyone HBase. Last but not least, it is interesting to study Argus in a heterogenous environment.

\chapter{A Lightweight Key-Value Store for Distributed Access} \label{title:kvlight}
Previous chapters present the performance isolation mechanisms we propose for the non-shared data on local file system case. In this chapter, we study multi-tenancy in the case where tenants share the same data set through a parallel file system. Specifically, we study the key-value store (KVS), a special form of NoSQL data store. KVS offers flexible data model, high scalability, as well as many other attracting features, and thus becomes increasingly popular. Various key-value stores \cite{DynamoDB,Cassandra,Voldemort} have been developed to facilitate analysis on social media feeds, web logs, and etc. With the advent of cloud computing, users are willing to move their data infrastructures to the cloud. They set up the KVS across a set of virtual machines (VMs) billed by a flexible price model, i.e. ``pay-as-you-go'' model.

KVS is usually architected as a layer over local file system. Single node KVS (S-KVS), such as Berkeley DB \cite{BDB} and Level DB \cite{leveldb}, targets a single node environment. It allows direct access to the local file system by embedding to the applications. In contrast, multi-node KVS (M-KVS), constructs network connected nodes as a cluster and provides a unified interface for applications through the network. Examples include Cassandra \cite{Cassandra}, HBase \cite{HBase}, and etc. It usually stores the data on the local file system in a cluster of VMs and runs daemon services on each individual node to delegate application access. In general, S-KVS is much more lighter-weight and has better performance in the sense that it is embedded as a library in applications and allows direct file system access without going through daemon services. However, S-KVS suffers from data loss and scalability issues due to the limit of a local file system. Besides, it does not allow concurrent writes because of file system locking, which is not desirable in a cloud environment where the store may be accessed concurrently. On the contrary, M-KVS distributes data across different nodes, supports concurrent access, and provides data replication as well as fail over. In this chapter, we intend to retain the high performance access of S-KVS, but extend its capability in a distributed environment, \emph{e.g.} cloud, which M-KVS is good at, with the help of a parallel file system.

Parallel file system (PFS) has begun seeing usage in the cloud in both industry \cite{Lustre-AWS} and academia \cite{Lustre-OpenStack,Lustre-S3,IUJetStream}. Originating from the high performance computing (HPC) platform, PFS is a type of clustered file system that spreads data in a dedicated storage node cluster \cite{PFS-Wiki}. PFS can be mounted to multiple VMs and provides the same interfaces and semantics as local file systems. Essentially, PFS decouples data storing from VM's local disk to a dedicate storage system and provides a hybrid storage solution along with the local file system in the cloud. However, there are some challenges to run KVS over PFS. On one hand, although PFS can resolve the data reliability and scalability issues, S-KVS over PFS is still subject to the exclusive writes constraint. On the other hand, M-KVS over PFS introduces overheads owing to its unawareness of PFS. For example, data may be unnecessarily replicated; extra network trips may be needed to access the PFS because the daemon service delegates all the requests to the back-end file system; overheads may also come from the data replication and failover protocols, both of which are taken care of by PFS. Additionally, most M-KVS withhold resources to have persistent running services even if no request comes, which is not cost effective. Recently, Greenberg et al. also point out the burden and inefficiency of running persistent KVS service in the HPC environment \cite{MDHIM}.

Therefore, we propose a lightweight and distributed KVS, KVLight, over a PFS to better utilize the sharing and reliability nature of PFS. The design, presented in a poster \cite{KVLight-Poster}, is further developed in this chapter. Similar to S-KVS, KVLight is implemented as a library embedded in applications for high performance. It makes use of the log structure merge tree (LSM) \cite{LSM} structure to support concurrent writes and uses a novel tree based compaction strategy to support concurrent reads efficiently. In summary, this chapter makes the following contributions:
\begin{itemize}
  \item A LSM based framework with asynchronous mechanisms to support concurrent writes and reads;
  \item A tree based compaction equipped with parallel processing to improve read performance;
  \item Experimental results that show KVLight has better performance than other M-KVS including Cassandra and Voldemort \cite{Voldemort} in several different workloads including two real world applications.
\end{itemize}

\section{Background and Motivation} \label{title:kvlight-motivation}
\subsection{Background}
Parallel file system (PFS) is designed for parallel and high performance access. It allows concurrent access from a number of clients and operates over high-speed networks. Below we summarize several PFSs, mainly from the architecture and data distribution prospects.

The Parallel Virtual File System (PVFS) \cite{PVFS} runs its server processes, i.e. \textit{pvfs2-server}, over a cluster of nodes. The pvfs2-server process stores data locally. Data is stored in files and metadata is stored in Berkeley DB. PVFS provides two sets of APIs: the UNIX API backed by the \textit{pvfs2-client}, a user-space process, and the MPI-IO API which can bypass the pvfs2-client and be more efficient. File is striped across all available servers in a round robin fashion. The striping can be tuned with various parameters \cite{PVFS-Striping} according to data size and access patterns. As its development continues, PVFS is now known as OrangeFS \cite{OrangeFS}. Some company has provided OrangeFS as a service through Amazon AWS \cite{OrangeFS-AWS}. However, PVFS does not support locking which prevents its API from conforming to POSIX semantics.

The General Purpose File System (GPFS) \cite{GPFS}, developed in IBM, bears a similar architecture with PVFS. It runs the servers on a number of dedicated storage nodes called file system nodes. The file system nodes are connected to a set of disks through switching fabric. The disks are set up in RAID to provide reliable storage. A file in GPFS is divided into blocks (256 KB by default) and distributed evenly across the disks. Reads and writes can be served in parallel by multiple file system nodes and disks. Unlike the Hadoop File System \cite{HDFS} which stores all the metadata in a single server, GPFS distributes the metadata e.g. directory tree in its servers. Due to this kind of distribution, GPFS does not have limits on the number of files a directory could have, which is often 65536 in many file systems. Additionally, GPFS introduces a distributed locking mechanism which allows it to support full POSIX file system semantics.

\begin{figure}[h]
  \centering
  \includegraphics[scale=0.45]{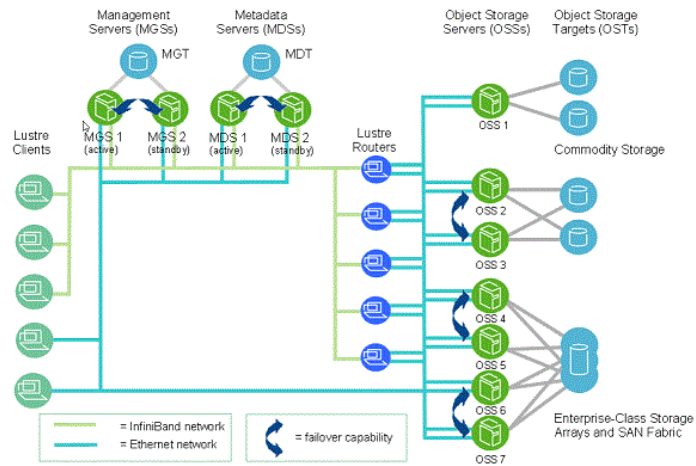}
  %\caption{Lustre Architecture.}
  \caption[Lustre Architecture.]{Lustre Architecture. Source: \url{http://lustre.org/about/}}
  \label{fig:kvlight-lustre-architecture}
\end{figure}

Among various PFS e.g. PVFS, GPFS, and etc., Lustre \cite{Lustre} is the most widely used PFS nowadays. Figure \ref{fig:kvlight-lustre-architecture} displays its architecture. It has a set of metadata servers (MDS) to host file system metadata and a set of object storage servers (OSS) to interact with clients. Behind each OSS, there are many object storage targets (OST) that store the data in a redundant fashion. A file is striped into several pieces and stored across different OST so that read/write operations can be performed in parallel. To read/write from/to Lustre, a client will first consult the MDSs to get the locations of OSSs. Afterwards, the data transfer is between the client and OSSs. With the separation among MDSs, OSSs and OSTs, Lustre is able to provide highly reliable and scalable data access. Lustre also uses a distributed locking mechanism and is able to support full POSIX filesystem semantics. In practice, Lustre is often mounted to other compute nodes that provide computation power. The access to Lustre in compute nodes is just like it is a local file system. The complexities are completely hidden from users.

We prototype KVLight over Lustre. Next, we motivate the design of KVLight by comparing a single node KVS (S-KVS) against a multi-node KVS (M-KVS) on Lustre.

\subsection{KVS on Parallel File System}
\begin{figure}[h]
  \centering
  \includegraphics[scale=0.3]{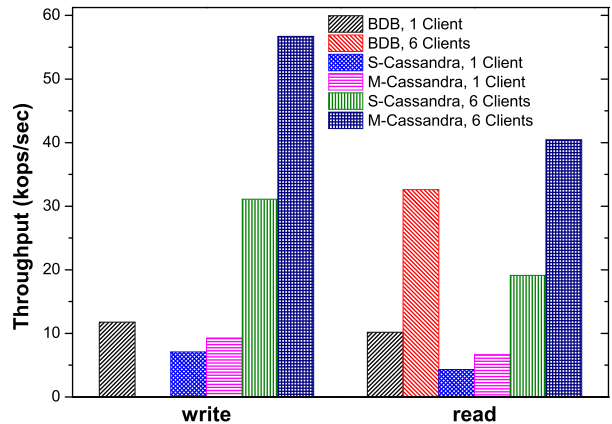}
  \caption[Read/write performance for different KVS in different setup.]{Read/write performance for different KVS in different setup. ``S-Cassandra'' means a single node Cassandra instance while ``M-Cassandra'' represents a multi-node Cassandra instance. We do not report the throughput of 6-client write workload in BDB as it does not support concurrent write.}
  \label{fig:kvlight-motivation}
\end{figure}

We run the experiments on Lustre 2.1.6 in Data Capacitor 2 at Indiana University \cite{DC2}. We use Berkeley DB Java Edition (BDB) version 6.2.3 \cite{BDBJE} as the S-KVS and Cassandra version 2.0.14 \cite{Cassandra-URL} as the M-KVS. Yahoo Cloud Storage Benchmark (YCSB) \cite{YCSB} is used to generate the workloads. For more details about the setup, please refer to Section \ref{title:kvlight-evaluation}. We run write-only and read-only workloads. There are two setup variants for Cassandra: single node Cassandra instance (S-Cassandra) and 15-node Cassandra cluster (M-Cassandra). We use one client and six clients to access the BDB and Cassandra instances respectively. A client runs on a separate node to carry out the workloads. We report the aggregated throughput i.e. operations per second (ops/sec) on client side. Figure \ref{fig:kvlight-motivation} displays the results.

In the one client case, BDB outperforms S-Cassandra and M-Cassandra. It is because the access to BDB is lightweight -- direct file system access, which avoids the overheads Cassandra imposes. When there are 6 clients, the aggregated throughput of Cassandra is much higher than the one of BDB with one client access. M-Cassandra delivers the highest throughput among all KVSs in Figure \ref{fig:kvlight-motivation}. Although BDB does not support concurrent writes from different clients, it does allow multiple clients to read from and yields a higher throughput than S-Cassandra does.

We conclude from the above discussions that 1) S-KVS has better performance than M-KVS in a single node due to its lightweight access, but S-KVS does not support concurrent writes; 2) M-KVS is much better than S-KVS when it is deployed and accessed in a distributed environment because it well supports concurrent writes and reads. Motivated by such experiments, we extend S-KVS to support concurrent writes and reads in a distributed environment while retaining its lightweight access as much as possible.

\section{The KVLight Structure} \label{title:kvlight-structure}
\subsection{System Model}
\begin{figure}[h]
  \centering
  \includegraphics[scale=0.85]{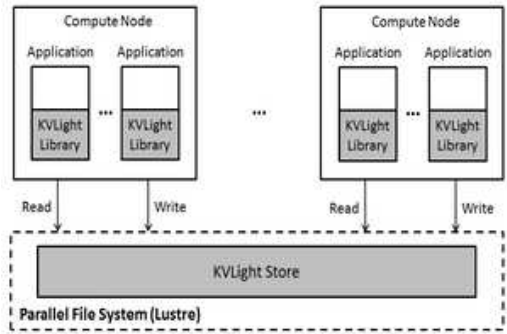}
  \caption{KVLight architecture.}
  \label{fig:kvlight-scenario}
\end{figure}

In a large scale compute environment, e.g. HPC and cloud, the system can usually be organized into a 2-layer architecture, consisting of application and storage layer. The application layer generates queries to the storage layer and processes the query results, while the storage layer stores the data and handles queries. We design KVLight as a middleware that stays between the application and storage layers in Figure \ref{fig:kvlight-scenario}. The KVLight library provides basic key-value store APIs including \textit{Get(key)}, \textit{Put(key, value)}, and \textit{Delete(key)}. The KVLight store is a list of files in Lustre. The files contain metadata \emph{e.g.} KVLight status and data i.e. the key-value pairs. Lustre is used as the underlying storage system and a communication media among nodes.

\subsection{Design Choices}
\begin{figure}[h]
  \centering
  \includegraphics[scale=0.85]{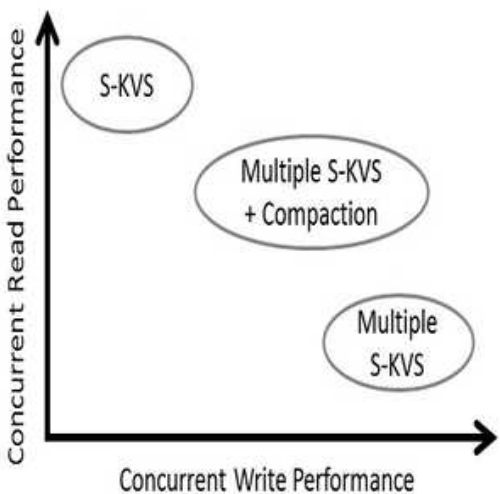}
  \caption{Design space of KVLight. The arrow points to high performance.}
  \label{fig:kvlight-design}
\end{figure}

We address the problem of building a lightweight KVS in a distributed environment. The heavyweight mechanisms \emph{e.g.} data replication, fail over, nodes coordination and etc. are shifted from KVS itself to the underlying file system i.e. Lustre. We explore the design space in Figure \ref{fig:kvlight-design}. A S-KVS has low concurrent write performance as it only supports exclusive writes. A simple solution is to have multiple processes write to independent S-KVSs and let a read search all the existing S-KVSs. However, the read performance will deteriorate as the number of S-KVS grows because a read has to consult more BDBs to get the data. To remedy the read deterioration, compaction can be used to merge multiple S-KVSs into one to reduce the number in the system. Therefore, to support concurrent writes without sacrificing too much read performance, we design KVLight in the ``Multiple S-KVS + Compaction'' category.

\section{Design Details} \label{title:kvlight-design}
KVLight uses Berkeley DB (BDB), a widely used S-KVS, to store key-value pairs. In this section, we present the log structure merge tree (LSM) design that allows KVLight to support concurrent writes and the asynchronous mechanisms that hide the overheads introduced by LSM. Then we describe two different compaction approaches i.e. size based compaction and tree based compaction, and their parallel implementations. Finally, we present the consistency model used in KVLight.

\subsection{Concurrent Write} \label{subtitle:concurrent-write}
\begin{figure}[h]
  \centering
  \includegraphics[scale=0.8]{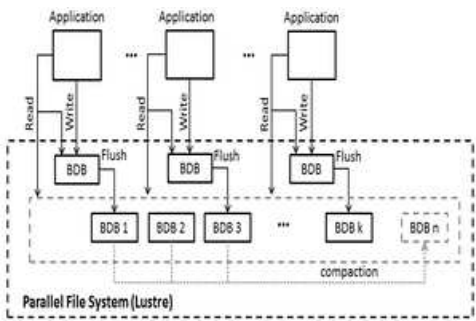}
  \caption{The structure supports concurrent write.}
  \label{fig:kvlight-lsm}
\end{figure}

To support concurrent writes, KVLight has each application write to a dedicated BDB (called write BDB) which is not shared with other applications. Figure \ref{fig:kvlight-lsm} shows the details. All the writes of an application go to one write BDB or multiple write BDBs. A write BDB is flushed as an immutable BDB and shared by other applications according to a certain policy. Sample policies are when the size of the write BDB exceeds a threshold or when the application wants to. After a write BDB is flushed, it becomes accessible by other applications. Any new write requests will be added to a new write BDB. To read a key-value pair, KVLight consults the write BDB first and then the immutable BDBs. To delete a key-value pair, it marks the key as deleted by updating the key with a special value. KVLight will report ``key not found'' if an application intends to read the key marked as deleted. The key-value pair will be removed during the compaction.

To further improve performance, we introduce an asynchronous mechanism that runs the flushing in the background as a separate thread and keeps admitting new write requests at the same time. When a write BDB is closed for flushing, it will invoke a number of procedures such as syncing buffer to disk, reclaiming unused disk space, and etc. These procedures result in blocking for coming writes. The asynchronous mechanism can overlap the flushing with write admission and thus hides the flushing overhead from applications as much as possible.

Organizing BDBs in this way has some advantages. Using different ``types'' of BDB allows KVLight to support both concurrent write and concurrent read. Thus we enact dedicated BDB for write and immutable BDB for read. Having different applications to write to different BDBs can better utilize the parallelism Lustre provides.

\begin{figure}[h]
  \centering
  \includegraphics[scale=0.26]{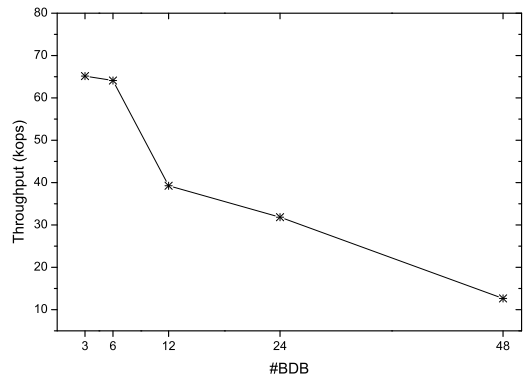}
  \caption{The aggregated throughput of read decreases as the number of BDBs increases.}
  \label{fig:kvlight-bdbs}
\end{figure}

However, there are some caveats about such an organization. First, the read performance will degrade as the system keeps admitting new write requests. Figure \ref{fig:kvlight-bdbs} shows the read throughput degradation as a result of the number of BDB increases. That is because the number of immutable BDBs keeps increasing and forces a read to consult more BDBs than before. Second, there will be some versioning issues as different values associated with the same key may exist in different BDBs. Third, the consistency is weakened as a write BDB is visible to other applications until it is flushed and becomes immutable. We address the read performance degradation and versioning issues through the compaction procedure discussed in Section \ref{subtitle:kvlight-compaction}. We present the consistency model used for KVLight in Section \ref{subtitle:kvlight-consistency}.

\subsection{Compaction} \label{subtitle:kvlight-compaction}
To serve reads, all the immutable BDBs have to be searched so as to return the values associated with the given key. The lookup cost is $O(n \times k)$ where $n$ is the number of BDBs, $k$ is the cost of fetching data from BDB, and $n$ keeps increasing as the system admits new writes and flushes write BDBs. Compaction is a background procedure that merges multiple BDBs into one to reduce the number of BDBs. Once a new BDB is generated, old BDBs will be deleted. If there are multiple values associated with the same key, KVLight only returns the one with the latest timestamp during read and discards other values during compaction. We first present a straightforward approach to compact BDBs based on their sizes. Then we propose a tree based compaction approach.

\subsubsection{Size Based Compaction}
A compaction will be triggered if the number of immutable BDBs exceeds a threshold. The compaction picks a subset of BDBs with the smallest sizes to merge. This is to avoid merging large BDBs repeatedly, which is similar to the tier compaction used in HBase and Cassandra. To better utilize the parallism of Lustre and speed up the compaction, we introduce parallel compaction to KVLight. Specifically, KVLight launches several procedures that compact BDBs in parallel. Each of these procedures merges a separated set of BDBs. Algorithm \ref{alg:kvlight-sizecompaction} describes the details. The \textit{Compaction} procedure monitors the number of BDBs and dispatches them to different workers for compaction in a round robin fashion if needed. Workers may be run on different nodes instead of the ones KVLight applications run to avoid I/O competition. The \textit{CompactionWorker} procedure merges the set of BDBs assigned.
\begin{algorithm}
    \caption{Size Based Compaction Algorithm}\label{alg:kvlight-sizecompaction}
    \begin{algorithmic}[1]
        \State $worker\_list$ is a list of workers where compaction can be performed.
        \State $bdb$ is a set of BDBs to be compacted.

        \Procedure{Compaction}{}
            \If{\#BDBs $\geqslant$ threshold}
                \If{$worker\_list$ is not empty}
                    \State $bdb \leftarrow$ Pick $(\#BDBs - threshold + 2)$ BDBs
                    \State Update \#BDBs
                    \State Remove $worker_{j}$ from $worker\_list$
                    \State Schedule $worker_{j}$ to work on $bdb$
                \EndIf
            \EndIf
        \EndProcedure

        \Procedure{CompactionWorker}{$bdb$, $worker_{j}$}
            \State Compact($bdb$) on $worker_{j}$
            \State Add $worker_{j}$ back to $worker\_list$
        \EndProcedure
    \end{algorithmic}
\end{algorithm}

%A compaction runs slower than writes admission as it needs to read several BDBs and write to a new BDB. Therefore, a read request may still need to consult multiple BDBs.
To further boost the read performance, KVLight uses bloom filter, an in-memory hash structure, to quickly locate the BDBs that may have the data. A bloom filter can test if a given key is in a BDB in \textit{O(1)} with some false positive. This is much faster than using the indices of BDB to test. Every BDB is associated with a bloom filter generated during the flush of a write BDB. Given a key, KVLight tests the key against the bloom filters and finds a subset of the BDBs that might have the associated values. Afterwards, KVLight linearly searches the subset of the BDBs with their indices.

\subsubsection{Tree Based Compaction}
\begin{figure}[htbp]
    %\centering
    \begin{subfigure}[b]{0.3\textwidth}
        \includegraphics[scale=0.37]{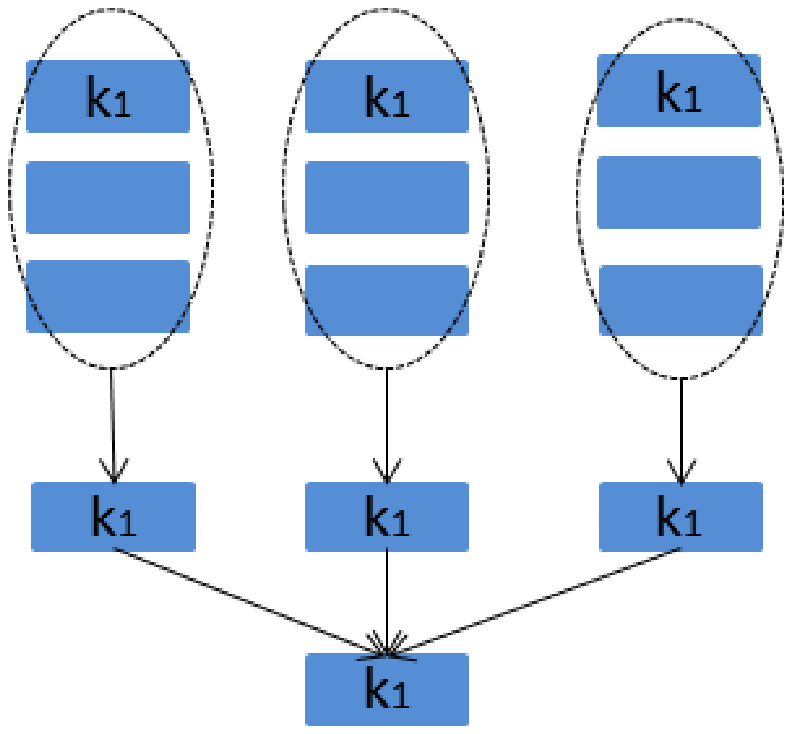}
        \caption{}
        \label{fig:kvlight-compaction-sizebased-example}
    \end{subfigure}
    ~
    \begin{subfigure}[b]{0.3\textwidth}
        \includegraphics[scale=0.37]{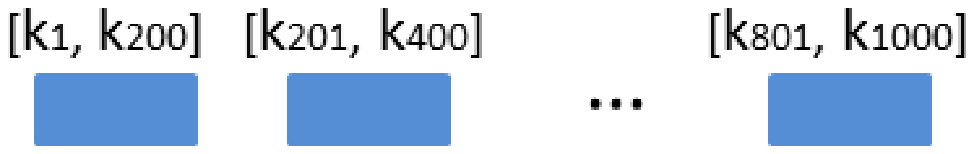}
        \caption{}
        \label{fig:kvlight-compaction-ideal-example}
    \end{subfigure}
    ~~
    \begin{subfigure}[b]{0.3\textwidth}
        \includegraphics[scale=0.37]{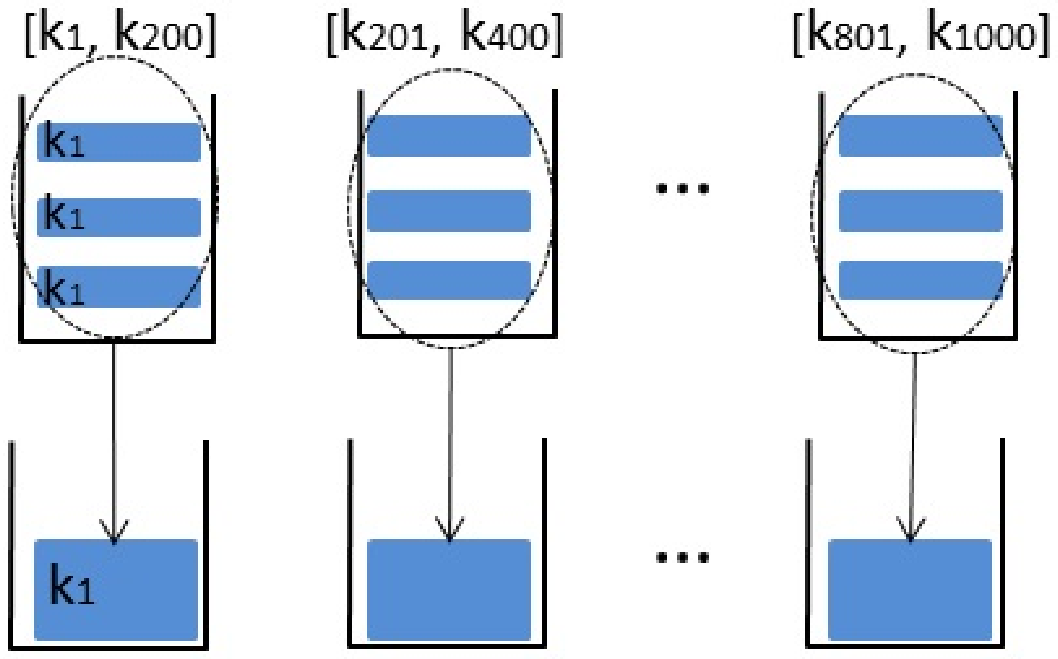}
        \caption{}
        \label{fig:kvlight-compaction-treebased-example}
    \end{subfigure}
    \caption[Different key range organizations across BDBs.]{Different key range organizations across BDBs. A blue stripe represents a BDB. The number in the key range is used as illustrations. (a) Overlapped key ranges; (b) Non-overlapped key ranges; (c) Non-overlapped key ranges across partition, overlapped key ranges within the same partition.}
\end{figure}

The problem of the size based compaction is a read may still have to lookup a large number of BDBs even after a compaction. Consider the following example in Figure \ref{fig:kvlight-compaction-sizebased-example}. \textit{Key1} spreads across 3 BDBs. After the first compaction, it is still in 3 BDBs. Only after a second compaction it stays in 1 BDB. It is worse if the threshold is larger than 3 because there will not be a second compaction and \textit{Key1} stays across three BDBs.

The cause of the scenario described above is the key ranges in different BDBs overlap with each other. Ideally, if the key ranges of all the BDBs are disjointed, then a read request only needs to lookup one BDB as shown in Figure \ref{fig:kvlight-compaction-ideal-example}. However, implementing such a model will be very inefficient in KVLight because concurrent writes from different applications may write to the same BDB which only supports exclusive write. Therefore, we propose another model that approximates the ideal one. The entire key space is divided into a few partitions. A partition has a disjoint key range and is associated with a list of BDBs. BDBs falling into different partitions will have non-overlapped key ranges but BDBs in the same partition may have overlapped ones. A compaction may only be applied within a partition. Figure \ref{fig:kvlight-compaction-treebased-example} shows an example. \textit{Key1} spreads across 3 BDBs in the same partition. After a compaction, \textit{Key1} stays in 1 BDB instead of 3 compared with Figure \ref{fig:kvlight-compaction-sizebased-example}. Thus after one compaction, KVLight is able to read \textit{Key1} by searching only one BDB. BDBs in the same partition can be further partitioned into multiple sub-partitions during the compaction, which makes the organization similar to a tree.

\begin{figure}[h]
  \centering
  \includegraphics[scale=0.85]{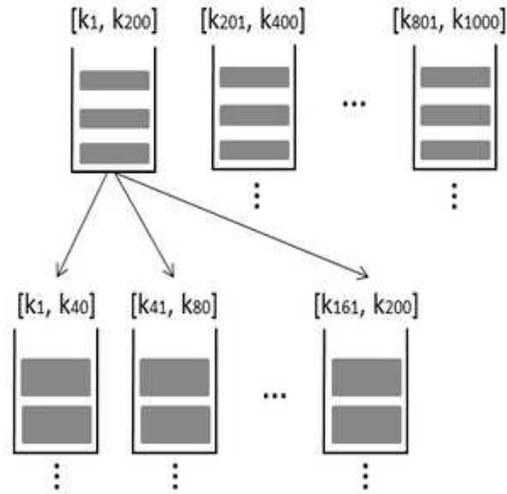}
  \caption{Tree based compaction design.}
  \label{fig:kvlight-treecompaction}
\end{figure}

Motivated by the example in Figure \ref{fig:kvlight-compaction-treebased-example}, we organize BDBs as a tree for read. Figure \ref{fig:kvlight-treecompaction} describes the design. A node in the tree has a key range and is associated with a list of BDBs whose keys fall within the node's key range. Its children nodes further partition the parent node's key range into disjoint ranges. BDBs belonging to different nodes have non-overlapped ranges while BDBs within the same node have overlapped key ranges. The root of the tree is a node with the entire key range and an empty list of BDBs. Writes will go the children of the root node. There are two types of compaction over the tree: one is to push the key-value pairs down to the next level by reading the BDBs in the parent node and writing into its children nodes; the other one is to merge several BDBs under the same node into one. The former one further divides the key ranges while the latter one reduces the number of BDBs. In addition, the height of the tree and the number of children of a node can be dynamically adjusted. These two parameters determine the granularity of the key range partition. This is especially helpful when dealing with skew data. If a node has much more data than other nodes have, we may increase the height and the number of children nodes to refine the partition of the key range.

\begin{figure}[h]
  \centering
  \includegraphics[scale=0.85]{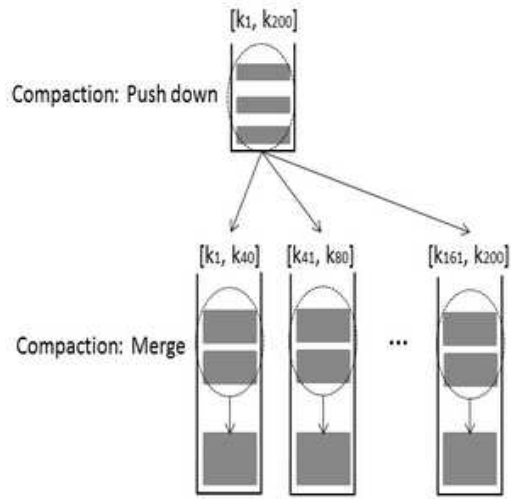}
  \caption[Compactions run in nodes from different levels of the tree.]{Compactions run in nodes from different levels of the tree. For the node in the top, the compaction pushes keys down to the next level by further partitioning the key range. For the nodes in the bottom, the compaction merges BDBs of the same node into one.}
  \label{fig:kvlight-running-compaction-in-tree}
\end{figure}

In the current prototype of KVLight, the maximum level of the tree and the maximum number of children of a node, i.e. the fan out, are set as fixed numbers. We leave the dynamical adjustment of the parameters in future work. The KVLight library maintains an index that describes the tree structure of BDBs. The key range is decided by hashing. A portion of the key is extracted for hashing in different levels. Specifically, we have two levels (the root is not counted in the level hierarchy). In the 2nd level, we hash the last 16 bytes of the key to decide which child it goes to. In the 3rd level, we hash the first 16 bytes. The compaction procedure is triggered when the number of BDBs under a node exceeds a threshold. BDBs in a non-leaf node will be pushed down to the leaf node and the BDBs in the leaf node are merged together. Figure \ref{fig:kvlight-running-compaction-in-tree} shows a running example of compaction in the tree structure.

To parallelize tree based compaction, we allow multiple compactions to run on different branches of the tree independently. Algorithm \ref{alg:kvlight-treecompaction} describes the procedure. The \textit{Compact} procedure dispatches nodes to compaction workers. The $Update$ function traverses the tree level by level and updates a list of nodes to be compacted. The \textit{CompactionWorker} procedure either pushes the BDBs to the next level for non-leaf nodes or merges the BDBs for leaf nodes.
\begin{algorithm}
    \caption{Tree Based Compaction Algorithm}\label{alg:kvlight-treecompaction}
    \begin{algorithmic}[1]
        \State $node\_list$ is a list of nodes to be compacted.
        \State $worker\_list$ is a list of workers where compaction can be performed.

        \Procedure{Compaction}{}
            \State Update($node\_list$)
            \For{each $node_{i}$ in $node\_list$}
                \If{$worker\_list$ is not empty}
                    \State Remove $worker_{j}$ from $worker\_list$
                    \State Schedule $worker_{j}$ to work on $node_{i}$
                \EndIf
            \EndFor
        \EndProcedure

        \Procedure{CompactionWorker}{$node_{i}$, $worker_{j}$}
            \State Compact($node_{i}$) on $worker_{j}$
            \State Add $worker_{j}$ back to $worker\_list$
        \EndProcedure
    \end{algorithmic}
\end{algorithm}

To read a key-value pair, KVLight searches down the tree and locates the node containing the key. Then it linearly searches all the BDBs under the node to get the values. Bloom filter is also used here to quickly detect if a BDB contains the key or not. Compared with the size based compaction, the tree based compaction can quickly focus on a subset of BDBs containing the key and provides a flexible structure that can dynamically adjust according to different key distribution.

\subsection{Consistency} \label{subtitle:kvlight-consistency}
Although Lustre, where KVLight is built, provides strong consistency for concurrent access, KVLight introduces data inconsistency because it needs to support concurrent writes. There are two scenarios related to consistency: consistency within a process and consistency among processes. For consistency within a process, according to \cite{distbook}, KVLight follows the read-your-write consistency. The effect of a write can always be seen by following reads in the same process. That is because a write will either stay in the write BDB or one of the immutable BDBs. Successive reads in the same application can read the value of write from either the write BDB or immutable BDBs immediately. For consistency among processes, a process might not read the latest value written by another process in advance, but it will eventually. The reason is that a key-value pair is admitted to the write BDB which only becomes visible to other applications until flushed. The time took for the key-value pair becomes visible depends on the flushing policy discussed in Section \ref{subtitle:concurrent-write}. To support strong consistency in the inter-processs case, the KVLight library can search all the existing BDBs to serve reads. Such an operation is very expensive in terms of performance because KVLight has to reopen the write BDBs every time it tries to read a key-value pair in order to get the newly admitted data. KVLight follows an eventual consistency model by default, but allows a client to specify strong consistency as an option in the API.

\subsection{Limitations}
To simplify the situations, the current KVLight prototype is limited to simple key-value pair lookup. It does not support transaction and atomic operation. Both require a central mechanism that can coordinate among applications. Additionally, it does not support advanced operations like scan and join. The scan operation requires a global order of keys across different BDBs while the join requires extra indices. Furthermore, it assumes the back-end PFS can accommodate reads and writes from both applications and internal procedures i.e. compaction without creating contentions.

\subsection{Applications}
The aforementioned consistency model and limitations impose constraints to the applications that can use KVLight. For applications that involve operations across multiple processes, KVLight is not suitable. Examples include the strong consistency case where a process reads from the writes made by another process, the transaction case where a process wants to commit a series of operations as a transaction, and the atomic operation case where a counter is maintained among applications. Despite of these constraints, KVLight is suitable to a wide range of applications. KVLight can be used in situations where eventual consistency can be tolerated. An example is the advertisement listing application. Many users (applications) post advertisements. It is OK that some advertisements do not get to the reader immediately. KVLight can also be used in situations where write and read are separate. An example is log processing \cite{Pipeline}. Log data is injected as key-value pairs to KVLight by multiple processes (potentially distributed). Once the injection is done, several other processes read from KVLight to process the log data. Another usage scenario is to persistently store data in HPC environment. The access to compute nodes is granted as the job is scheduled and revoked as the job is terminated, which makes traditional M-KVS like Cassandra unable to persist data as it requires a long running service on each compute node \cite{MDHIM}. KVLight does not require persistent running services and thus allows on-demand access.

\section{Implementation} \label{title:kvlight-implementation}
\begin{figure}[h]
  \centering
  \includegraphics[scale=0.85]{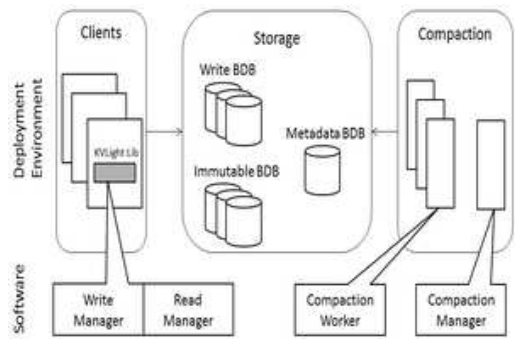}
  \caption{KVLight implementation and deployment.}
  \label{fig:kvlight-impl}
\end{figure}

We implemented KVLight in Java 1.7 with Berkeley DB Java Edition 6.2.3. The KVLight library shown in Figure \ref{fig:kvlight-impl} consists of a write manager and a read manager. The write manager dispatches the key-value pairs to different BDBs based on the hash partition. It also implements an asynchronous mechanism to avoid blocking when flushing a write BDB. The read manager maintains the tree structure of various BDBs and is responsible for handling reads. The compaction manager is a separate process triggered by the library. It launches serval compactions to compact BDBs in parallel. Only one compaction manager is allowed to run at a time. Both the compaction manager and the workers are run through \textit{ssh} on additional nodes. KVLight has to maintain system status. We implement the system status e.g. paths of the immutable BDBs in several metadata tables backed by a BDB in Lustre. Retry logic is applied to update operation on the metadata table in case of failure caused by concurrent write. The read manager updates the tree model every 1 second by reading the paths from the metadata table.

\section{Evaluation} \label{title:kvlight-evaluation}
We evaluate the performance of KVLight in a cluster environment and compare against Cassandra version 2.0.14 and Voldemort version 1.9.17, both of which are state-of-the-art KVS. Voldemort is a KVS using Berkeley DB Java Edition as the default backend store and used widely in LinkedIn \cite{Voldemort}. The experiments are carried out in a distribute environment. Each node has 2 Intel Xeon E5-2650 v2 8-core processors and 32 GB memory. Both Cassandra and Voldemort are setup as a 15-nodes cluster. We use Data Capacitor II (DC2) as the parallel file system \cite{DC2}. DC2 runs Lustre 2.1.6 with 26 storage nodes connected with 56-Gb FDR InfiniBand and provides 3.5 PB storage capacity. It is mounted to all the nodes as a shared file system. The client uses YCSB \cite{YCSB} to generate the data as well as workloads and runs on additional nodes. To use YCSB with KVLight, we develop a KVLight plugin for it.

For KVLight, it uses the tree based compaction by default. A write BDB will be flushed when its size exceeds 256 MB. The maximum level is set to 2. The maximum children per node is 4. The compaction threshold for each node is set to 3. Therefore, the maximum number of BDBs is 48 ($4\times4\times3$). The maximum number of compaction procedures running in parallel is 6. For Cassandra, we keep its default settings, and configure the Java heap size to 4GB so as to leave most of the memory to OS as recommended in \cite{Cassandra-Book}. The consistency level defaults to one which enables eventual consistency. We configure BDB with 5 GB cache size and 256 MB log file size for both KVLight and Voldemort according to the suggestions from \cite{VoldemortConfig}. For the data stored in Lustre, we set the stripe count to 1 and stripe size to 4 MB. An evaluation of impacts of different stripe counts and sizes is left in future work. The workloads used in evaluation include write-only, read-only and write-read mixed. Requests are drawn from 2,500,000 randomly generated key-value pairs. Each pair is about 1.2 KB. The write-only workload sends write requests, while the read-only workload sends read requests. 50\% of the requests in the write-read workload is writes while the other 50\% is reads. We use throughput, i.e. operations per second (ops/sec), as the performance measurement. We report the aggregated throughput from all clients to reflect the throughput of a KVS.

We first evaluate the overall performance of KVLight and quantify the impact of compaction to workloads. Then we investigate the effectiveness of different compaction strategies. Finally, we examine KVLight's performance under two real world applications.
\begin{figure}[th]
  \centering
  \includegraphics[scale=0.3]{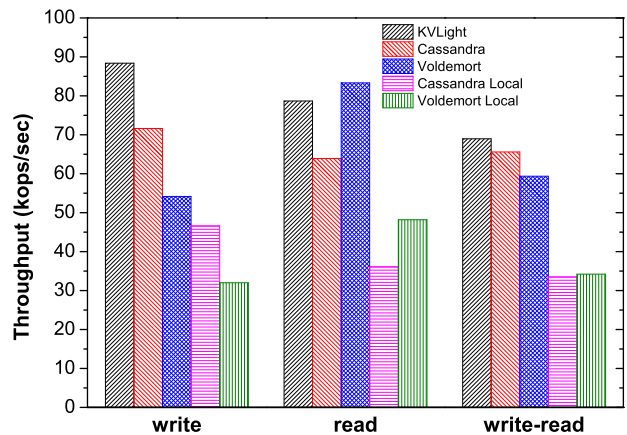}
  \caption{Aggregated throughput for various workloads on different KVS.}
  \label{fig:kvlight-overall}
\end{figure}

\subsection{Overall Performance}
\begin{figure*}[th]
   \begin{subfigure}[b]{0.3\textwidth}
        \centering
        \includegraphics[scale=0.18,angle=-90]{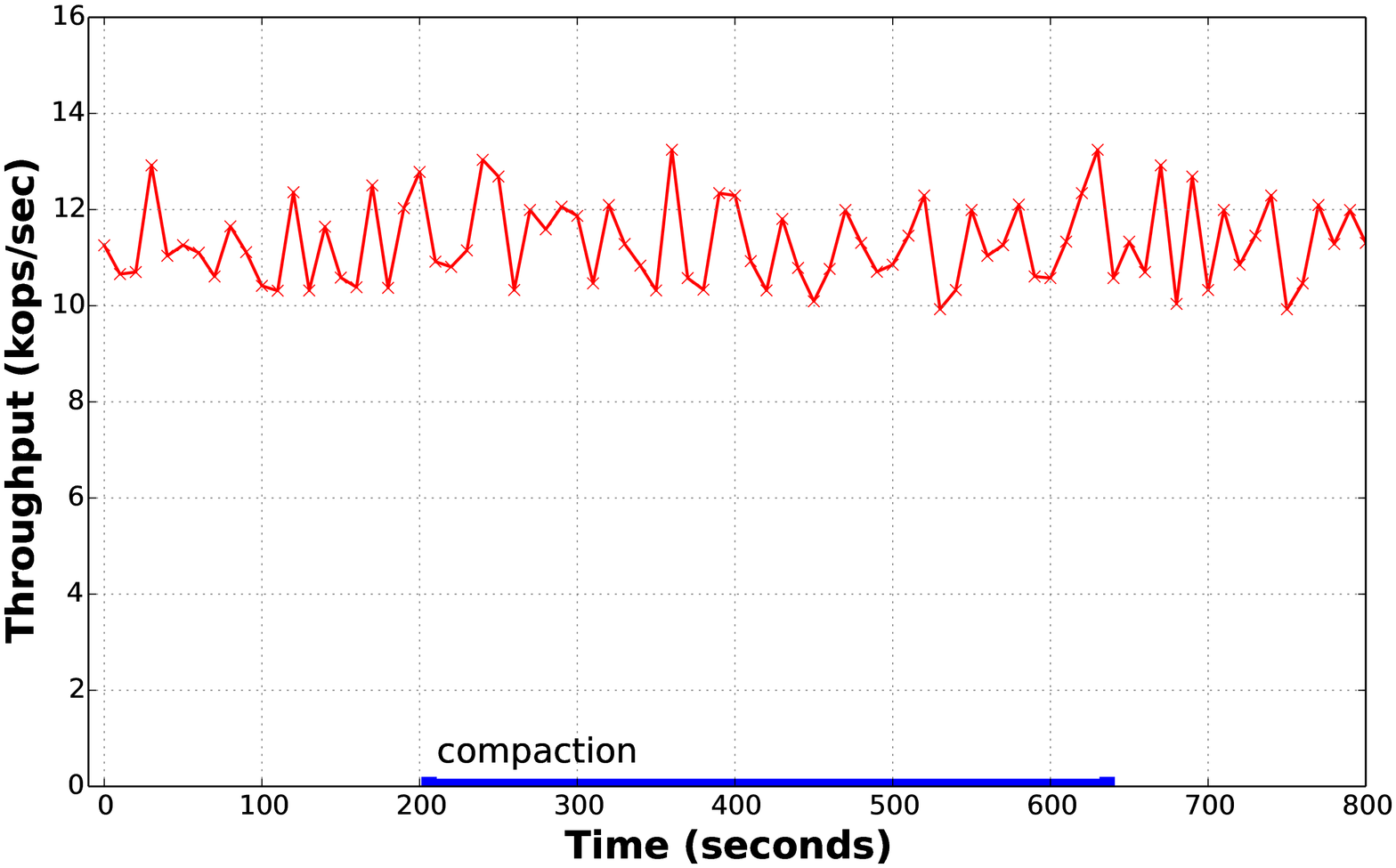}
        \caption{Write-only workload}
        \label{fig:kvlight-compaction-impact-w}
    \end{subfigure}
    ~
    \begin{subfigure}[b]{0.3\textwidth}
        \centering
        \includegraphics[scale=0.18,angle=-90]{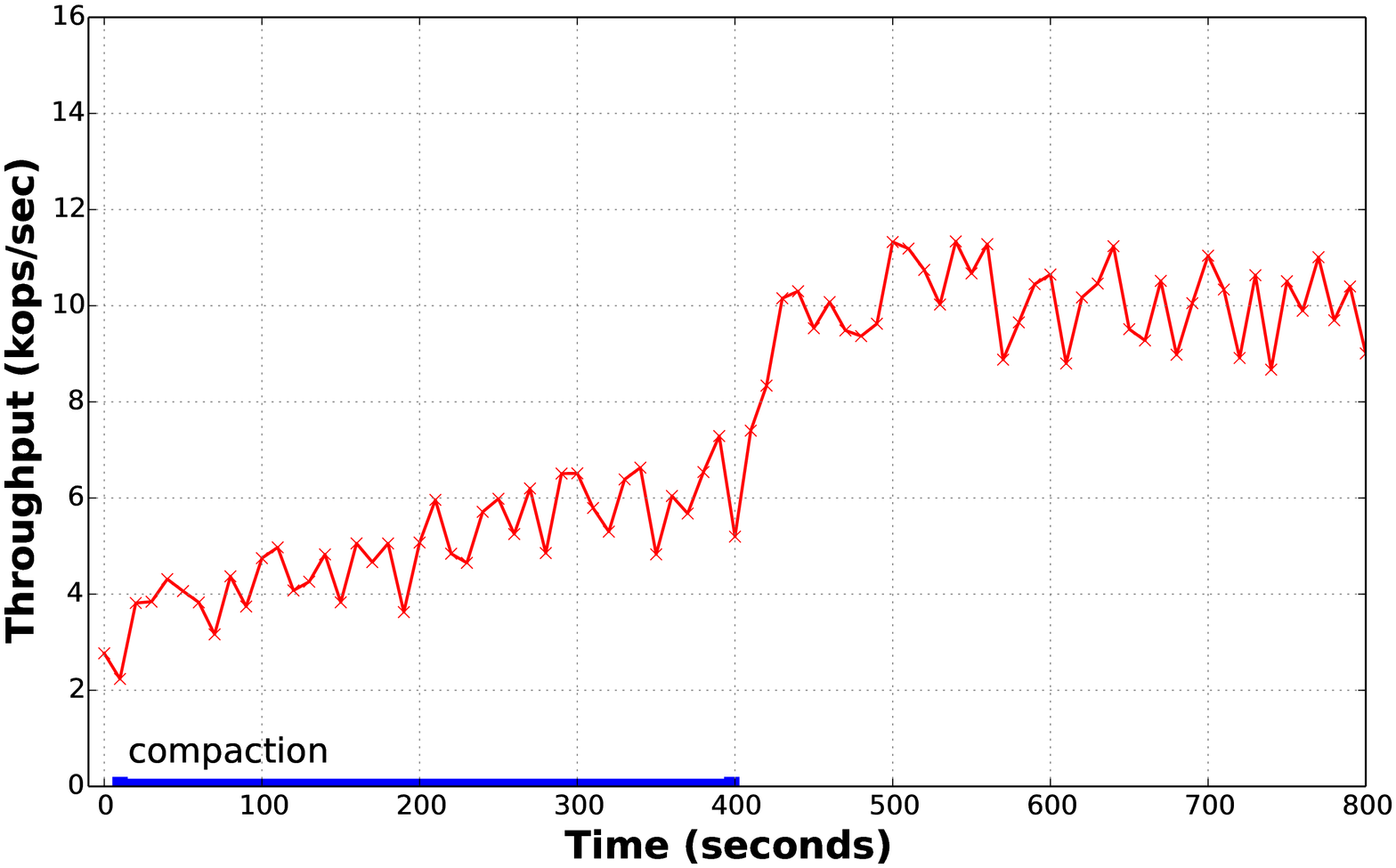}
        \caption{Read-only workload}
        \label{fig:kvlight-compaction-impact-r}
    \end{subfigure}
    ~
    \begin{subfigure}[b]{0.3\textwidth}
        \centering
        \includegraphics[scale=0.18,angle=-90]{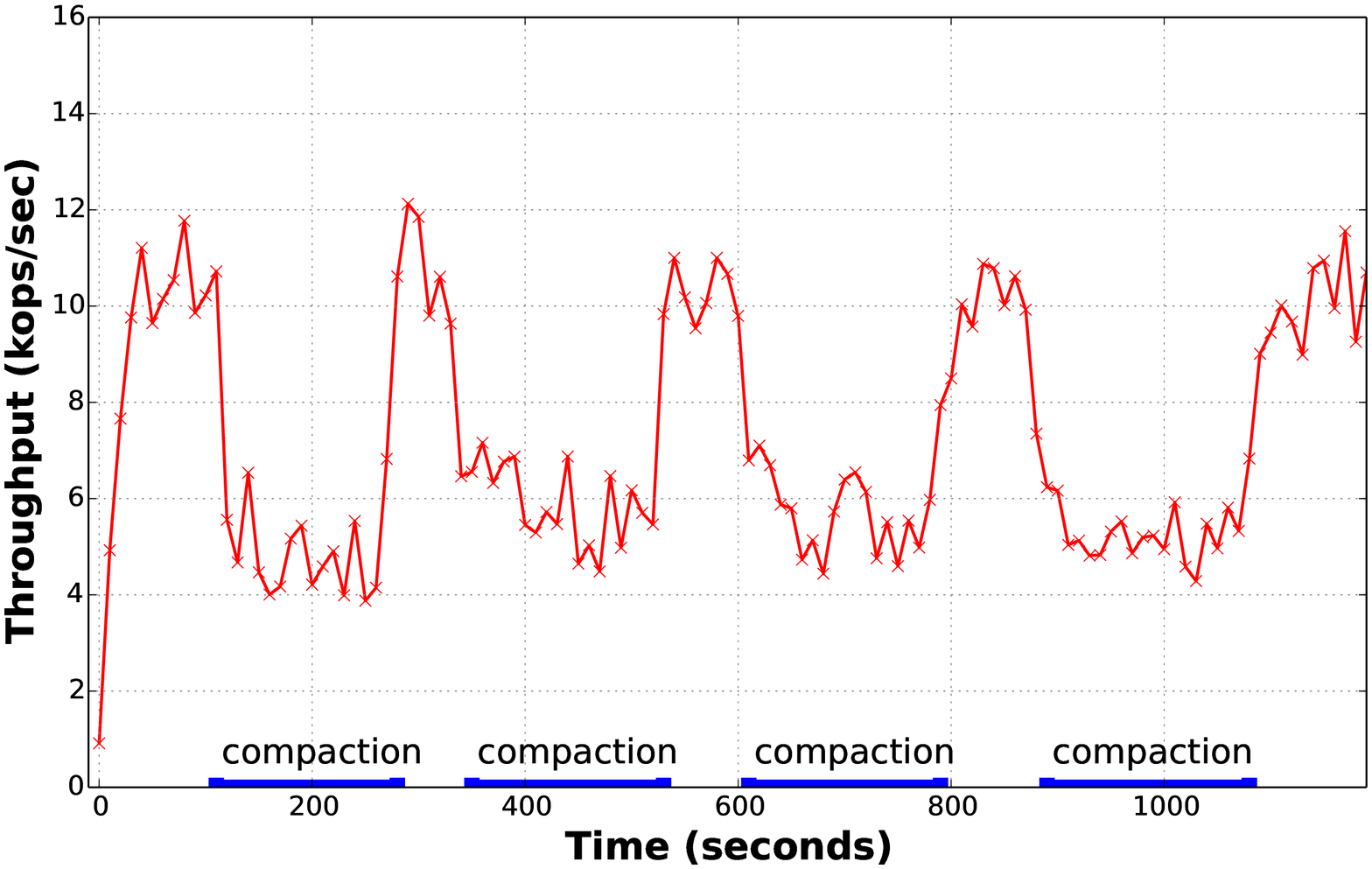}
        \caption{Write-read workload}
        \label{fig:kvlight-compaction-impact-rw}
    \end{subfigure}
    \caption[Impact of compaction on different workloads in terms of throughput.]{Impact of compaction on different workloads in terms of throughput. (a) There is no observable impact for the write-only workload. (b) Throughput improves as compaction goes on for the read-only workload. (c) Throughput drops during compaction and improves afterwards for the write-read workload.}
    \label{fig:kvlight-compaction-impact}
\end{figure*}

We start by running the workloads against KVLight, Cassandra and Voldemort. The write workload tests if KVLight can handle concurrent writes. The read workload evaluates if the organization of BDBs can serve concurrent reads efficiently. The write-read workload assesses the impact of compaction when writes and reads are present. To thoroughly test the KVS, Cassandra and Voldemort are set up on two different file systems: a parallel file system i.e. Lustre and a local file system backed by the local disk in the compute node. We have 8 clients run simultaneously on separate nodes. We preload 2,500,000 key-value pairs for the read-only workload and the write-read workload. Both the read-only and write-read workloads are launched after the store has been compacted for Cassandra and KVLight. Cassandra and Voldemort are restarted to clear the cache before serving reads. Figure \ref{fig:kvlight-overall} displays aggregated throughput of different KVS. We first compare three KVSs when they store data in Lustre. For the write workload, KVLight outperforms Cassandra and Voldemort by 23\% and 62\% respectively. We think it is because KVLight avoids the protocol overhead and redundant network trip mentioned in Section \ref{title:kvlight-motivation}. For the read workload, KVLight's throughput is about 26\% higher than Cassandra's which reflects the effectiveness of tree organization of BDBs in KVLight. Compared with Voldemort, KVLight's throughput is about 5\% less. We attribute that to the overhead of locating the BDBs. To serve a read request, KVLight searches down the tree to locate the node whose key range contains the target key. Then it linearly searches the BDBs that belong to the node to find the value. In contrast, Voldemort divides the key space into non-overlapped partitions (one BDB per partition) through consistent hashing and serves a read request by just looking up one BDB. For the write-read workload, KVLight's throughput is about 5\% and 16\% higher than the ones of Cassandra and Voldemort respectively. Such results show KVLight is able to run compaction efficiently to mitigate the write impact. Compared with Cassandra, KVLight runs multiple compactions in parallel instead of one compaction at a time. Furthermore, KVLight runs compactions on additional nodes rather than the nodes hosting the applications to avoid I/O contentions.

When the data is stored in the local disk, both Cassandra and Voldemort perform worse, evidenced by the throughput drop in Figure \ref{fig:kvlight-overall}. After consulting to the system admin of DC 2, we conclude that it is because the local file system is slower than Lustre in the kind of workloads generated by KVS. The local disk is a 7200 RPM SATA drive while a OST i.e. a Lustre data node is composed from (10) 7200 RPM SATA drives configured in Raid-6. The nodes running the Cassandra and Voldemort instances have 10Gb Ethernet connectivity to Lustre nodes. A 7200 RPM SATA drive typically yields about 90 MB/s bandwidth while an OST yields about 300 MB/s. Because the network bandwidth is 1.2 GB/s which is larger than an OST can sustain, data transferred to or from Lustre is subjected to the bandwidth of OSTs rather than to the network.
\begin{figure}[th]
  \centering
  \includegraphics[scale=0.3]{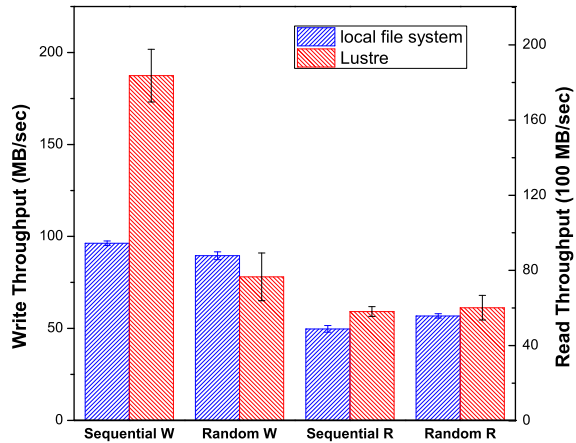}
  \caption{I/O performance from Iozone.}
  \label{fig:kvlight-iobenchmark}
\end{figure}

We run an I/O benchmark, Iozone \cite{iozone} to report writes and reads throughput with respect to sequential and random patterns on the local file system and on Lustre. The size of data written and read is set to 30 GB. Figure \ref{fig:kvlight-iobenchmark} plots the average throughput of 10 runs. For sequential write, Lustre outperforms the local file system about 2 times, which matches the expectation based on the hardware spec mentioned. That also explains why Cassandra and Voldemort are worse in local file system because both systems append data to log files which are sequential writes. However, Lustre is about 15\% worse than the local file system in the random write workload. For sequential read and random read, both local file system and Lustre obtain higher throughput than the disk can generate. We think it is due to the impact of cache including Linux VFS cache, disk cache, and Lustre cache. Iozone does not support direct I/O and we do not have the privilege to reset the cache in the system. For reads, Lustre's throughput is about 20\% higher than the local file system in the sequential read and 8\% higher in the random read. In a word, Lustre outperforms the local file system in sequential write as well as read and random read, all of which are main workloads Cassandra and Voldemort impose to the file system.

Next, we investigate the impact of compaction on aforementioned workloads. We use a YCSB client to carry out the workloads and plot the throughput as well as compaction span as functions of time in Figure \ref{fig:kvlight-compaction-impact}. We preload the data for the read-only and write-read workloads. We disable the compaction when preloading data for the read-only workload so that compaction will be triggered during reads. Figure \ref{fig:kvlight-compaction-impact} shows the result. For the write-only workload, although the compaction lasts about 400 seconds, there is no observable impact to the throughput as displayed in Figure \ref{fig:kvlight-compaction-impact-w}. For writes, a compaction is just another batch of writes and the reads in compaction do not interfere with writes from applications. For the read-only workload in Figure \ref{fig:kvlight-compaction-impact-r}, the compaction starts shortly after the workload begins and lasts about 400 seconds. The throughput gradually improves during the compaction which shows the compaction indeed can improve read performance by reducing the number of BDBs and organizing them in a tree structure. For the write-read workload in Figure \ref{fig:kvlight-compaction-impact-rw}, its throughput stays about 10,000 ops/sec but drops to around 5,500 ops/sec when the compaction runs. There are several reasons behind that. First, some of the slowdown may come from the frequent update of the tree structure of BDBs because obsolete BDBs are deleted and new BDBs are generated. Second, a BDB being read by a KVLight library may be deleted by the compaction procedure which cause exception. The KVLight library ignores such types of exceptions, updates its view of available BDBs and retries. All these steps contribute to the slowdown.

Finally, we evaluate the overhead KVLight adds to BDB by running 1-client workloads against them. Table \ref{tab:kvlight-overhead} shows the results. The overhead of writes in KVLight is ignorable owning to the asynchronous write mechanism in Section \ref{title:kvlight-design}. KVLight suffers about 5\% degradation in the read workload and 9\% degradation in the write-read workload. The overhead mainly comes from the tree based organization of BDBs and compaction. However, KVLight stands out in concurrent access situation where BDB struggles.
\begin{table}[h]
    \centering
    \caption{Overhead evaluation of KVLight.}
    \label{tab:kvlight-overhead}
    \begin{tabular}{|l|c|c|}
      \hline
      \textbf{Workload} & \textbf{Berkeley DB (ops/sec)} & \textbf{KVLight (ops/sec)} \\
      \hline
      Write & 11763 & 11045 \\
      \hline
      Read & 10323 & 9832 \\
      \hline
      Write-read & 9425 & 8545 \\
      \hline
    \end{tabular}
\end{table}

\subsection{Effectiveness of Compaction}
In this section, we investigate the effectiveness of different compaction strategies and the speedup of parallel compaction.

\subsubsection{Compaction Strategies}
\begin{figure}[h]
  \centering
  \includegraphics[scale=0.3]{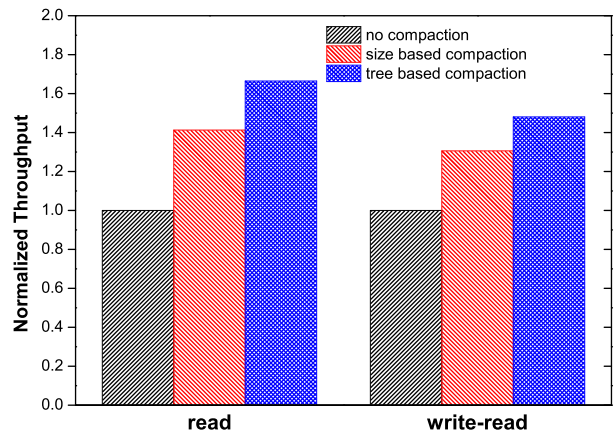}
  \caption{Throughput comparison among different compaction strategies.}
  \label{fig:kvlight-compacteffort}
\end{figure}

To study different compaction strategies in KVLight, we use the size based compaction and the tree based compaction to merge the data respectively, and afterwards run workloads against KVLight. The compaction threshold for the size based compaction is set to 48 which is the same as the maximum number of BDBs in tree based compaction. We use a read-only workload and a write-read workload in this experiment. Throughputs are normalized to the one obtained without any compaction and reported in Figure \ref{fig:kvlight-compacteffort}. With compaction, throughput is improved by around 40\% and 65\% in the read-only workload, and around 30\% and 50\% in the write-read workload. That is because without compaction, KVLight has to consult many BDBs with overlapping key ranges to get a key-value pair, which is time consuming. The tree based compaction is better than the size based compaction because it further reduces the number of BDBs searched during reads by partitioning the key space into disjoin sets. To verify such an statement, we report the average number of BDBs read per request for the read-only workload. The tree based approach visited 1.3 BDBs in average while the size based approach accessed 8.7 BDBs in average which is about 7 times higher. Therefore we can conclude from the above results that compaction can improve read performance significantly and the tree based compaction strategy can further boost performance by efficiently organizing the BDBs.

\subsubsection{Parallel Compaction}
\begin{figure}[h]
  \centering
  \includegraphics[scale=0.3]{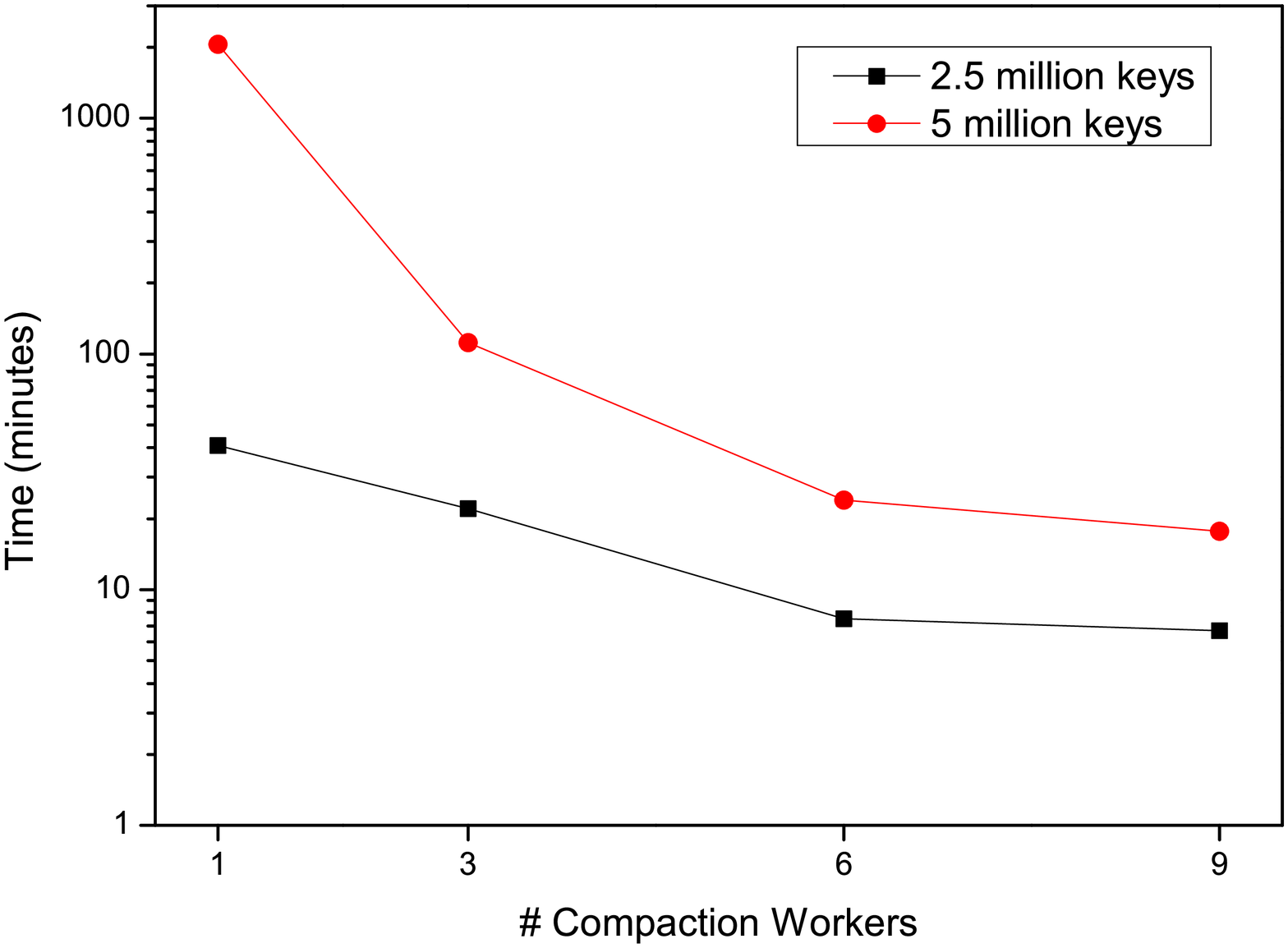}
  \caption{Speedup tests over various data sizes. Time is reported in log scale.}
  \label{fig:kvlight-speedup}
\end{figure}

KVLight runs multiple compactions in parallel to reduce the total time spent on compaction. We evaluate the speedup by varying the total number of compaction workers that run in parallel. We have a YCSB client to inject 2,500,000 and 5,000,000 key-value pairs respectively. Figure \ref{fig:kvlight-speedup} displays the time spent in compaction corresponding to different numbers of compaction workers. In the 2,500,000 case, when the number of workers increases from 1 to 3, we obtain near ideal speedup. The speedup begins to deteriorate when the number of workers goes beyond 6. We think it is because the compaction rate has already matched the write BDB flush rate. In the 5,000,000 case, we observe a similar trend. In addition, the performance gain in terms of time saved in the 5,000,000 case is much more significant than the one in the 2,500,000 case. That implies parallel compaction becomes more effective when the size of data increases.

%\subsection{Impact of Consistency on Performance}
%Although KVLight is not designed for applications with a strong consistency requirement, we investigate the impact of maintaining strong consistency in KVLight. We have 2, 4, 6, and 8 clients run the write-read workload and compare the throughput when the strong consistency option is turned on and off. Figure \ref{fig:kvlight-consistency} displays the results. As the number of clients increases, the throughput decreases significantly. This is because to maintain strong consistency, KVLight has to reopen every write BDB in the system to get the latest value, which turns out to be very expensive. A possible approach to strengthen the consistency is to have a middleware between the applications and KVLight to record consistency dependency information and enforce consistency \emph{e.g.} causal consistency \cite{bolt-on-consistency}. We leave such an extension in the future work.
%\begin{figure}[h]
%  \centering
%  \includegraphics[scale=0.23]{./graph/kvlight-consistency.eps}
%  \caption{To maintain strong consistency, throughput decreases as the number of clients increases.}
%  \label{fig:kvlight-consistency}
%\end{figure}

\subsection{Real World Applications}
To further evaluate KVLight's performance, we apply two real applications. The first one is a Facebook key-value pair access application i.e. ETC \cite{fb-workload}. We generate 25 million key-values pairs based on the key-size and value-size distributions specified in \cite{fb-workload}. We use the power law distribution with the shape parameter set as 3.2 to approximate the key access sequence in \cite{fb-workload}. There are three workloads on this data set: a write-only workload (W) that loads the data into the store; a read-write workload (R) whose read-write ratio is 30:1; a read-write-delete workload (R/W/D) whose read-write-delete ratio is 30:1:15. These ratios are also specified in \cite{fb-workload}. The second application is from the file I/O trace of Los Alamos national laboratory Anonymous App1 application (LANL) \cite{lanl-trace}. Yin et al. interpret each write in the trace as a key-value pair whose key size is fixed and value size is the number of bytes written \cite{pfs-kvs}. We adopt the same interpretation but repeat the trace 5 times to have a larger data set with 860390 key-value pairs. We only evaluate the write-only (W) and read-only (R) workloads for the LANL data set as the trace does not reveal the read-write ratio. To generate aforementioned workloads, we extend YCSB to support customized key size and value size distribution, as well as the delete operation. Eight clients are used to carry out each workload. We report the aggregated throughput.
\begin{figure}[htbp]
    %\centering
    \begin{subfigure}[b]{0.55\textwidth}
        \includegraphics[scale=0.3]{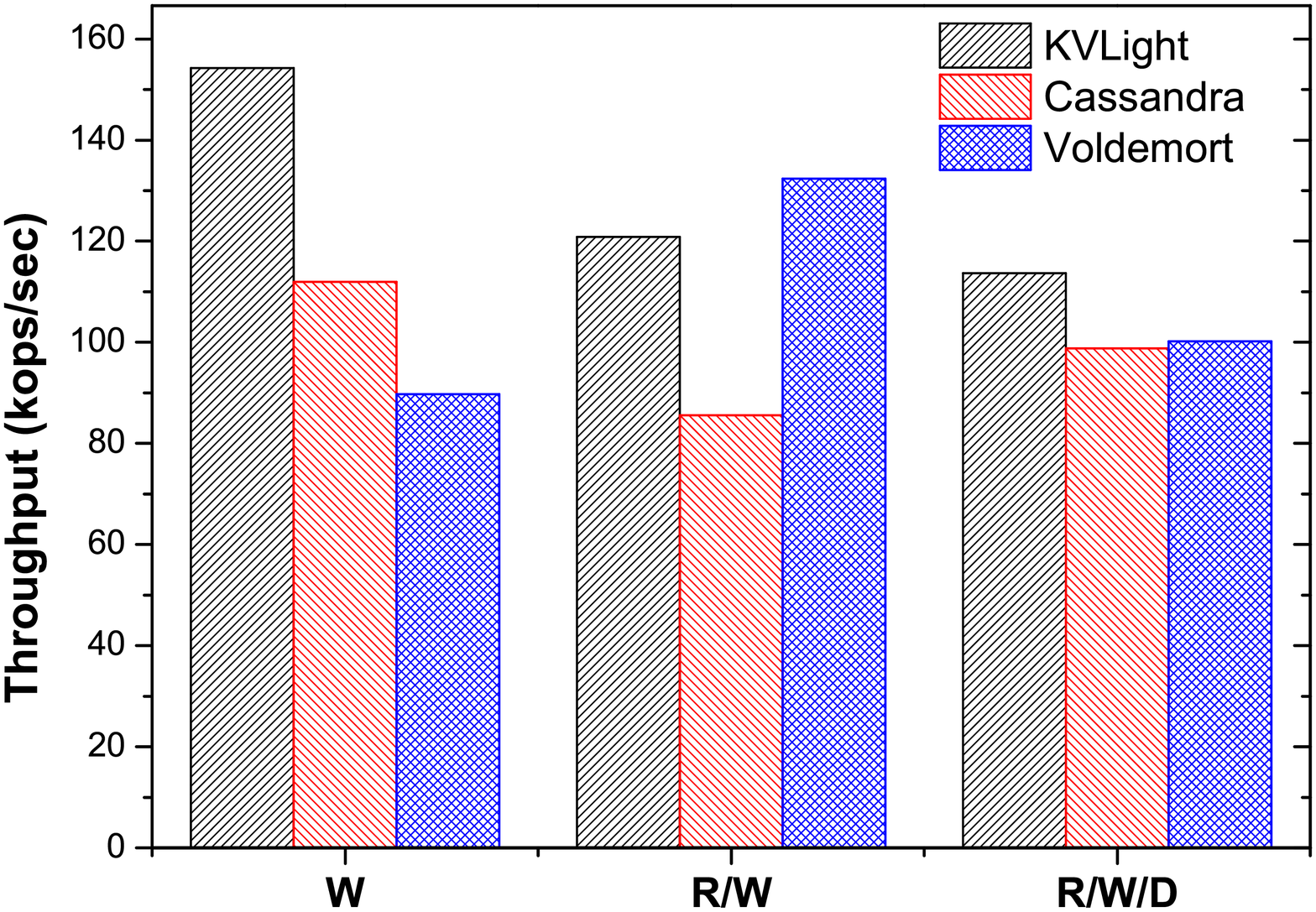}
        \caption{Facebook application.}
        \label{fig:kvlight-fbworkload}
    \end{subfigure}
    ~
    \begin{subfigure}[b]{0.35\textwidth}
        \centering
        \includegraphics[scale=0.3]{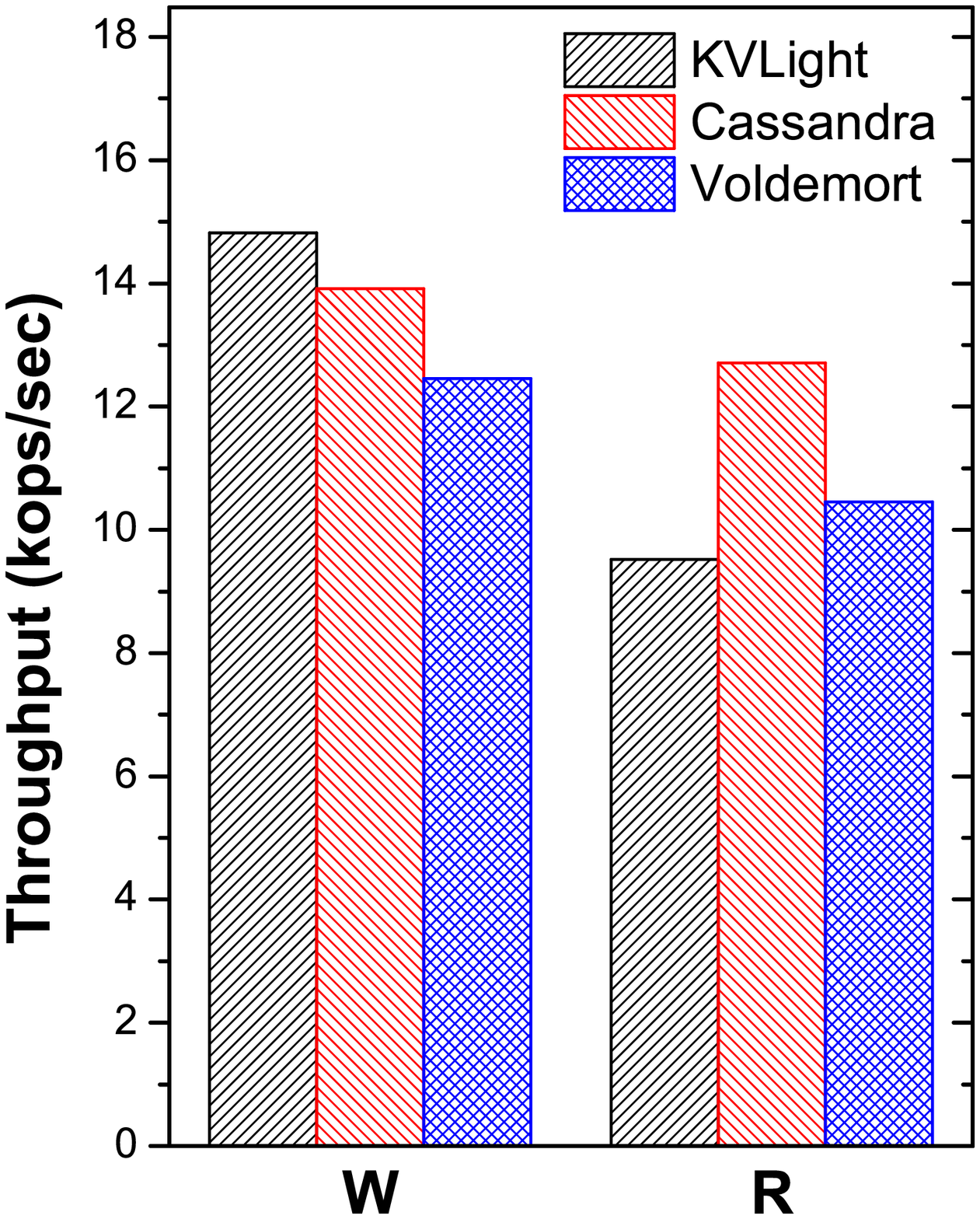}
        \caption{LANL App1 application.}
        \label{fig:kvlight-lanlworkload}
    \end{subfigure}
    \caption{Performance comparison in real world applications.}
    \label{fig:kvlight-realworkloads}
\end{figure}

Figure \ref{fig:kvlight-fbworkload} displays the results of the Facebook application. We can observe that KVLight yields 38\% and 70\% higher throughput than Cassandra and Voldemort do respectively. In the R/W workload, KVLight's throughput is 49\% higher than Cassandra's but is 8\% less than Voldemort's. In the R/W/D workload, KVLight achieves about 13\% higher throughput than the other two systems. Figure \ref{fig:kvlight-lanlworkload} shows the results of the LANL App1 application. KVLight outperforms both Cassandra and Voldemort in the write-only workload, although the improvement is not as significant as the one in the Facebook workload. However, KVLight is the worst in the read-only workload. Voldemort is also worse than Cassandra in the read-only workload. It is probably because Berkeley DB does not handle key-value pairs with large value size well. About 2/3 of the key-value pairs in the LANL App1 application have value size over 100 KB which is significantly larger than the value size in the synthetic experiments and the Facebook application.

\section{Summary} \label{title:kvlight-conclusion}
This chapter describes KVLight, a lightweight key-value store in a distributed environment. KVLight uses Berkeley DB for the lightweight access and extends it with a parallel file system for data reliability, fail over and concurrent access. The core design behind KVLight is a novel tree based organization of data with parallel compaction. Empirical results show that KVLight is able to outperform Cassandra and Voldemort in most of the workloads.

There are a number of future directions for this work. First, it is useful to adjust the tree structure (i.e. height and fan out) dynamically to handle different data distributions. For key ranges with many keys, it is better to further partition such ranges to avoid consulting too many BDBs in read, which results in the increase of height or fan out. For key ranges with few keys, it is reasonable to keep the height as well as fan out low to avoid overheads in compaction. Second, we plan to further optimize the metadata structure. Right now, KVLight has to maintain a bloom filter for each BDB, which puts pressures on the use of memory. Last but not least, we intend to investigate the impact of strip count and strip size setting of Lustre on the BDB performance.

\chapter{Conclusions and Future Work} \label{title:conclusion}

\section{Conclusions}
Multi-tenancy in cloud hosted NoSQL data stores is favored by providers as it allows effective resource sharing amongst different tenants and thus lowers operating cost. For tenants working on non-shared data sets, performance isolation becomes critical as it provides predictable performance and prevents interference. For tenants sharing the same data set, cost effectiveness is the main concern. As we describe and experimentally show in chapter \ref{title:fairshare}, \ref{title:reservation}, and \ref{title:kvlight}, traditional distributed NoSQL stores do not support performance isolation well and are not cost effective while running over parallel file system. This dissertation proposes several approaches in an attempt to address the isolation and efficiency issues. Empirical results show that our system allows tenants to share system throughputs fairly and the performance of read operations is protected.

In chapter \ref{title:fairshare}, we propose a system that targets fair share across tenants by throughput regulation. We first show that interference can occur when tenants use different thread numbers to run workloads against Cassandra. Then we propose an architecture based on feedback control to enforce fair share. Essentially the system puts requests into queues and has a scheduler to schedule them. When a response returns, the system collects some metrics as feedbacks to the scheduler for adaptive control. The scheduler adapts the deficit round robin algorithm with a linear programming model for credit refill. The system is further enhanced with several adaptive control approaches. We first experimentally demonstrate that single node fairness (called local fairness) is unnecessary to achieve system-wide fairness (called global fairness) and lower throughput. Then we propose a mechanism to dynamically adjust the credit allocation of tenants for each node to accomplish global fairness without achieving local fairness. Additionally, the system splits a scan operation into small chunks to avoid head-of-line blocking and allows the overlapping of processing between scan operations and get operations.

The aforementioned approach uses the number of bytes delivered from the store to represent the actual resource consumption in the system. It works in situations where workloads have the same access patterns but fails in cases when they do not. For example, bytes of a workload with random access pattern mainly come from or go to disk while the ones of a workload with a hotspot pattern rely on cache heavily, as discussed in chapter \ref{title:reservation}. Therefore, we propose a workload-aware resource reservation approach which targets multiple resources to prevent interference in chapter \ref{title:reservation}. We first conduct a set of experiments on a state-of-the-art NoSQL store, i.e. HBase, and reveal that interference could be triggered by tenants using different thread numbers and access patterns. Also, interference could occur in block cache (a cache layer maintained by HBase), disk, or both. We present Argus, a workload-aware resource reservation framework that prevents interference by enforcing reservation on cache and disk usage. We divide the block cache space into partitions and limit a tenant's activities to the cache partition it is assigned. We approximate the disk usage by the HDFS throughput and design a scheduler in HBase to limit the number of requests sent to HDFS. The reservation is elastic that it can adapt to workload changes on the fly. Furthermore, the resource reservation technique is workload-aware. We have a reservation planning engine that decides how much resource to reserve according to the resource demands of workloads. The engine models the problem as a constrained optimization and relies on the performance functions of various workloads. The performance function of a workload is approximated by using linear interpolation over sample date collected offline. Evaluation results show that Argus is able to prevent interference across tenants and adapt to dynamic workloads accordingly.

Chapter \ref{title:kvlight} investigates the multi-tenancy in the case that tenants share the same data set over a parallel file system. We particularly target the key-value store (KVS), a special instance of NoSQL, over parallel file system (PFS). Traditional KVS is inefficient in the sense that it requires long running daemon services which prohibits resource reuse or repurpose and also introduces overheads while it runs over PFS. We explore the opportunity of building a lightweight, high performant and distributed KVS, called KVLight, on PFS. KVLight uses an embedded KVS, i.e. Berkeley DB, for the lightweight access but extends it with a parallel file system for data reliability, fail over and concurrent access. To overcome the limit of exclusive writes in most embedded KVS, KVLight proposes a novel tree based organization of data with parallel compaction. KVLight employs the log structure merge tree design and has each application write to a dedicated BDB to support concurrent writes. The dedicated BDB for write does not become visible by other applications until it is flushed as an immutable BDB. To improve the read performance, KVLight divides the key space into disjoint partitions and employs the tree structure to organize BDBs. The operation of reading a key-value pair only needs to search a portion of the BDBs. Compaction, a procedure that merges different BDBs into a single one to reduce the number of BDBs, is used to further improve performance. A parallel mechanism is used to run multiple compactions in parallel to speed up the process. Empirical results show that KVLight is able to outperform Cassandra and Voldemort in most of the workloads, including two real world application workloads.

\section{Future Directions}
There are several directions for future work.

For performance isolation in the non-shared data with local file system setting, it is important to isolate reads and writes. Most of the NoSQL stores are log-structured stores which append writes in log files and have internal data structures to organize the log files. Our evaluations \cite{Argus, Zeng} show that writes from client requests may trigger extra writes to the internal data structures and influence reads. Modeling the extra I/O cost from writes is non-trivial and requires further study. In addition, as mentioned in chapter \ref{title:reservation}, there are multiple resources involved beyond just block cache and disk in serving requests, e.g. the memory used to buffer writes. It is beneficial to incorporate additional resources to extend the capability of reservation. Incorporating additional resources requires extensions of the resource model in chapter \ref{title:reservation}. For example, the increase of write buffer size may decrease the block cache size whose impact needs to be considered. And a workload needs to be identified by multiple factors including the read/write ratio, key repeat ratio, etc. It is also interesting to apply the NoSQL store equipped with multi-tenant support in the MapReduce framework \cite{Yuan12, Yuan12-CPE} to support multi-tenancy. Another direction for exploration is to apply the isolation mechanisms to cloud environments with security requirements \cite{secure-1,secure-2,secure-3,secure-4}. We envision the performance isolation can further improve the security by eliminating possible covert channels.

For cost effective access in the shared data with parallel file system setting, the prototype of KVLight organizes all the immutable Berkeley DBs as a static tree structure and can be extended to a dynamic one. A dynamic tree structure can address uneven data distribution i.e. data skew. For a tree node with many keys, it is better to push down the BDBs to the next level through compaction so as to further partition the key range. Such an operation results in the increase of tree height. For a tree node with few keys, it is reasonable to keep the height as well as fan out small to avoid overheads over compaction. Additionally, KVLight is built over a parallel file system without fine tuning its parameters. The strip count and strip size play an important role in data access performance. It is worthwhile to quantify the impact of those two factors and has the compaction procedure dynamically change the two factors according to the workload. Last but not least, it is interesting to apply the performance isolation mechanisms in Chapter \ref{title:fairshare} and \ref{title:reservation} to KVLight to support the case where tenants are independent. It is also interesting to explore the application space for KVLight in a digital library setting \cite{DataCapsules} and a finance setting \cite{AAAI14}.

Systems proposed in this dissertation assume the environment is homogeneous. Such assumptions limit the applicability of our results as the cluster environment may be heterogeneous. Handling the uneven distribution and heterogeneous environment requires each node in the cluster has its own policy to deal with multi-tenant access. A more complicated global coordination among nodes than the one used in chapter \ref{title:fairshare} is needed.

Finally, this dissertation only considers simple key-value pair query. Advanced queries like join, filter, and etc. have not been investigated. Different queries may demand different resources from the NoSQL data store. Some of them may hold the resources for a long time like the scan query. We plan to categorize the queries and refine or propose new resource models to accommodate advanced queries.

\addcontentsline{toc}{chapter}{Bibliography}
\bibliographystyle{plain}
\bibliography{chapters/reference}

\newpage
\renewcommand\baselinestretch{1.45} 
  \section*{Curriculum Vitae}
  \addcontentslinewithoutnumber{toc}{chapter}{Curriculum Vitae}{}
  \thispagestyle{empty}
  \pagestyle{empty}
  \begin{center}
\begin{minipage}{.2\textwidth}
Jiaan Zeng
\end{minipage}
\end{center}

\noindent{Education}
\begin{itemize}
  \item Doctor of Philosophy, 2009 - 2015\\
        Computer Science, Indiana University Bloomington, USA
  \item Master of Science, 2006 - 2009\\
        Computer Science and Engineering, South China University of Technology, China        
  \item Bachelor of Science, 2002 - 2006\\
        Software and Applied Mathematics, South China University of Technology, China                
\end{itemize}

\noindent{Publications}
\begin{itemize}
  \item Jiaan Zeng, Beth Plale, Workload-Aware Resource Reservation for Multi-Tenant \\
        NoSQL, IEEE International Conference on Cluster (CLUSTER), 2015
  \item Jiaan Zeng, Beth Plale, Multi-Tenant Fair Share in NoSQL Data Stores, IEEE International Conference on Cluster (CLUSTER), 2014
  \item Jiaan Zeng, Guangchen Ruan, Alexander Crowell, Atul Prakash, Beth Plale, Cloud Computing Data Capsules for Non-Consumptive Use of Texts, the 5th Workshop on Scientific Cloud Computing (ScienceCloud), 2014
  \item Jiaan Zeng, Beth Plale, Data pipeline in MapReduce, the 9th IEEE Conference on e-Science (eScience), 2013
  \item Jiaan Zeng, Yinghua Han, A New Approach for Value Function Approximation Based on Automatic State Partition, Proceedings of the International MultiConference of Engineers and Computer Scientists, 2009
  \item Huaqing Min, Jiaan Zeng, Ronghua Luo, Fuzzy CMAC with Automatic State Partition for Reinforcement Learning, the 2009 World Summit on Genetic and Evolutionary Computation
  \item Huaqing Min, Jiaan Zeng, Jian Chen, Jinhui Zhu.  A Study of Reinforcement Learning in a New Multi-agent Domain, IEEE/ACM International Conference on Intelligent Agent Technology (IAT' 08), 2008
\end{itemize}
\vfill\newpage

\end{document}